\documentclass[aps,prd,onecolumn,groupedaddress,nofootinbib,preprintnumbers,superscriptaddress]{revtex4-1}  
\usepackage{graphicx}
\usepackage{amsmath}
\usepackage{amsfonts}
\usepackage[colorlinks=true,citecolor=red,linkcolor=blue,breaklinks=true]{hyperref}
\usepackage[figure, table]{hypcap}
\usepackage{amssymb}
\usepackage{xcolor}
\usepackage{subfigure}
\usepackage{setspace}
\usepackage{footnote}
\usepackage{multirow}
\usepackage[normalem]{ulem}
\usepackage[utf8]{inputenc}
\usepackage{mathrsfs}
\usepackage{slashed}
\usepackage{longtable}
\usepackage{enumitem}

\allowdisplaybreaks

\definecolor{tabred}{rgb}{0.8392156862745098, 0.15294117647058825, 0.1568627450980392}
\definecolor{tabblue}{rgb}{0.12156862745098039, 0.4666666666666667, 0.7058823529411765}
\hypersetup{colorlinks=true, linkcolor={tabred}, citecolor={tabblue}, urlcolor={tabblue}} 

\usepackage{orcidlink}

\usepackage[skins,theorems]{tcolorbox}
\tcbset{highlight math style={enhanced,
  colframe=red,colback=white,arc=0pt,boxrule=1pt}}

\newcommand{\orcid}[1]{\,\orcidlink{#1}}

\newcommand{\dd}{\mathrm{d}}

\begin{document}


\title{
Radiative corrections to double-$\beta$ decay and their impact on experimental analyses
}

\author{Jordy de Vries\orcid{0000-0001-5037-5108}}
\affiliation{Institute for Theoretical Physics Amsterdam and Delta Institute for Theoretical Physics, University
of Amsterdam, Science Park 904, 1098 XH Amsterdam, The Netherlands}
\affiliation{Nikhef, Theory Group, Science Park 105, 1098 XG, Amsterdam, The Netherlands}
\author{Emanuele Mereghetti\orcid{0000-0002-8623-5796}}
\affiliation{Theoretical Division, Los Alamos National Laboratory, Los Alamos, NM 87545, USA}
\author{Saad el Morabit\orcid{0009-0000-0193-8891}}
\affiliation{Institute for Theoretical Physics Amsterdam and Delta Institute for Theoretical Physics, University
of Amsterdam, Science Park 904, 1098 XH Amsterdam, The Netherlands}
\affiliation{Nikhef, Theory Group, Science Park 105, 1098 XG, Amsterdam, The Netherlands}
\author{Stefan Sandner\orcid{0000-0002-1802-9018}}
\affiliation{Theoretical Division, Los Alamos National Laboratory, Los Alamos, NM 87545, USA}

\begin{abstract}
\noindent
We present a comprehensive study of the $\mathcal{O}(\alpha)$ radiative corrections to two-neutrino and neutrinoless double-$\beta$ decay within a heavy-nucleus effective field theory. We compute all virtual and real-emission corrections for intermediate nuclear states of arbitrary excitation energy, including new closed-form expressions for the required one-loop integrals with heavy-nucleus propagators. For excitation energies large compared with the lepton energies, the corrections factorize into the tree-level rate times a universal double-weak Sirlin function, which differs substantially from the sum of two single-$\beta$ Sirlin functions used so far in the literature. We discuss the impact of the radiative corrections on the analysis of measured spectra in terms of ratios of nuclear matrix elements, $\xi_{31}$, and show that neglecting radiative corrections can significantly bias the extraction. Real-photon emission, not considered in previous analyses, plays a central role and the impact of radiative corrections depends crucially on the detector setup. We provide the photon spectrum, fully differential in the photon energy and direction, and recommend that experiments simulate the photon explicitly. For $0\nu\beta\beta$ decay, the radiative corrections increase the inclusive rate by about $5\%$, independently of the nuclear matrix elements or the neutrino Majorana mass.
\end{abstract}

\maketitle

{
\hypersetup{linkcolor=black}
\tableofcontents
}

\noindent\makebox[\linewidth]{\rule{0.8 \paperwidth}{0.2pt}}
\vspace{1 mm}
\vspace{-4.5mm}

\section{Introduction}
\label{sec:intro}

Two-neutrino double-$\beta$ decay ($2\nu\beta\beta$) occupies an interesting position at the intersection of nuclear and particle physics. As the rarest Standard Model (SM) process observed in the laboratory, it provides a direct window into nuclear weak interactions, and its precise theoretical description is increasingly important for the broader program of double-$\beta$ decay experiments. The ongoing and next-generation tonne-scale experiments, including LEGEND, nEXO, CUPID, NEXT, and others~\cite{Ray:2024pse, LEGEND:2025jwu, Zhao:2025iyx, NEXT:2025yqw,Adams:2022jwx,Agostini:2022zub}, aim to probe neutrinoless double-$\beta$ decay ($0\nu\beta\beta$) at half-life sensitivities of $10^{28}$ yr, while recording millions of $2\nu\beta\beta$ events. Large samples of $^{136}$Xe $2\nu\beta\beta$ decays are also collected by dark-matter experiments with xenon targets~\cite{PandaX:2025yly}. This statistical power elevates $2\nu\beta\beta$ from a background to a precision observable in its own right: one that can constrain nuclear matrix elements (NMEs) relevant for $0\nu\beta\beta$ searches, test the SM in the electroweak sector, and probe physics beyond the SM through distortions of the electron energy and angular spectra, such as those induced by the emission of sterile neutrinos, light scalars, or other new particles coupling to neutrinos~\cite{Deppisch:2020mxv,Cepedello:2018zvr,deVries:2025hqa,Boudjema:2025okq,PandaX:2025tls}.

Exploiting this potential requires a systematic treatment, at the sub-percent level, of all corrections to the $2\nu\beta\beta$  spectral shape. The standard theoretical description organizes the spectrum as an expansion in the lepton energies over typical nuclear excitation energies, which factorizes the rate into NMEs and leptonic phase-space functions~\cite{Simkovic:2018rdz}. At next-to-leading order in this expansion, the spectrum becomes sensitive to the NME ratio $\xi_{31} = M_{GT}^{(-3)}/M_{GT}^{(-1)}$ (defined in detail below), which encodes information about the intermediate nuclear states and has been the target of recent experimental extractions~\cite{KamLAND-Zen:2019imh, CUPID-Mo:2023lru, CUORE:2025xue, PandaX:2025yly}. Subleading nuclear effects, including weak magnetism and pion-exchange two-nucleon currents, induce spectral distortions of similar size~\cite{Morabit:2024sms}.

Electromagnetic radiative corrections constitute another class of corrections that must be addressed at the sub-percent level. In single-$\beta$ decay, the structure of radiative corrections is well understood. The Sirlin function~\cite{Sirlin:1967zza, Sirlin:2012mh} captures the universal, nuclear-structure-independent contribution from ultrasoft photons, while hard and potential photon modes renormalize the single-nucleon couplings and generate finite-size corrections. For $2\nu\beta\beta$, however, the situation is considerably more complex. The process involves two weak decay vertices, two outgoing electrons, and a tower of intermediate nuclear states, introducing new kinematic scales in the form of the two-electron invariant mass and the nuclear excitation energies as well as new diagram topologies with no analogue in single-$\beta$ decay. Prior treatments of radiative corrections to $2\nu\beta\beta$ approximated the correction as the sum of two independent Sirlin functions, one for each outgoing electron~\cite{Nitescu:2024tvj}. This approximation ignores correlations between the two electrons, the spectrum of intermediate nuclear excitations, and is not independent of the renormalization scale.

In a companion Letter~\cite{deVries:2026udq}, we reported the first explicit calculation of ultrasoft radiative corrections to $2\nu\beta\beta$, working within a heavy-nucleus effective field theory (EFT) that includes fields for the initial and final nuclei, as well as for each relevant intermediate nuclear excitation. The main result of that work is the derivation of a universal ``double-weak Sirlin function'', the analogue of the Sirlin function for double-weak processes, which depends on both electron energies and their relative angle, and differs significantly from the sum-of-two-Sirlin-functions approximation across most of the available phase space. We further demonstrated that the resulting spectral distortions are comparable in magnitude to the leading nuclear-structure effect governed by $\xi_{31}$, implying that precision extractions of $\xi_{31}$ and SM tests from $2\nu\beta\beta$ data must consistently include double-weak radiative corrections.

The present paper provides a comprehensive account of the methods and results summarized in the Letter, with several extensions and additional analyses. Specifically, this work contains:
\begin{itemize}
\item \textbf{Detailed derivation of the EFT framework.} We present the heavy-nucleus EFT used to compute the ultrasoft radiative corrections: the Lagrangian, the degrees of freedom,  scalar fields for the $0^+$ initial and final nuclei and vector fields for the $1^+$ intermediate states, with their minimal electromagnetic couplings, and the power counting that organizes the expansion in $E_e/k_F$.
\item \textbf{Computation of the Feynman diagrams.} 
We evaluate all virtual and real-emission diagrams at $\mathcal{O}(\alpha)$, including the topologies unique to the double-weak process, such as photons coupling to the intermediate nucleus and photon exchange between the two outgoing electrons.  We present new closed-form expressions for all scalar one-loop integrals with a heavy-nucleus propagator needed for radiative corrections to single- and double-$\beta$ decay, with infrared divergences handled in dimensional regularization. They are obtained with an auxiliary-mass representation of the heavy propagator, which maps the integrals onto standard relativistic ones~\cite{Saad}. General solutions for such scalar integrals were given in Ref.~\cite{Zupan:2002je}, but in an implicit form, in which the roots of quadratic equations and the branches of logarithms and dilogarithms must be selected case by case, and without infrared divergences regulated in dimensional regularization. The loop integrals can be used directly in other one-loop calculations with heavy-baryon propagators, such as electric dipole moments~\cite{Dekens:2025skl} or radiative corrections to lepton-proton scattering~\cite{Goswami:2025zoe}. We describe the treatment of ultraviolet divergences, the cancellation of infrared divergences, and the subtraction method used for the real emission.

\item \textbf{The large-$\omega_n$ EFT.} For excitation energies $\omega_n$ large compared with the lepton energies, we compute the corrections in two independent ways: by expanding the full result, and by integrating out the intermediate states and matching onto an EFT that contains only the ground states of the initial and final nuclei. The agreement at fixed order in $1/\omega_n$ provides a nontrivial check of the calculation. At leading order, the corrections factorize into the tree-level rate times the double-weak Sirlin function. We assess the accuracy of this expansion for realistic nuclear spectra.

\item \textbf{Real-photon emission.} We derive the rates of the radiative modes $2\nu\beta\beta+\gamma$ and $0\nu\beta\beta+\gamma$, differential in the photon energy and direction, and the spectra in the total visible energy which is relevant when the photon is absorbed in the detector. The photon distributions are needed to model the response of a given detector to radiative events.

\item \textbf{Radiative corrections to $0\nu\beta\beta$ decay.} With the same tools we compute the virtual and real-emission corrections to $0\nu\beta\beta$ decay, for which an EFT with only the ground-state nuclei is appropriate. The radiative corrections increase the inclusive rate by about $5\%$ independent of nuclear matrix elements or the value of the neutrino Majorana mass. We study the change in the $2\nu\beta\beta$ background in the endpoint region due to radiative corrections and the impact of real photon emission from $0\nu\beta\beta+\gamma$ events. 

\item \textbf{Phenomenology of $2\nu\beta\beta$.} We study the impact of the double-weak radiative corrections on the summed-energy spectra and angular distributions of the electrons and on the photon emission rates for various isotopes, and their interplay with the nuclear-structure effects parametrized by $\xi_{31}$ and $\xi_{51}$. We quantify the bias on $\xi_{31}$ when tree-level templates are fitted to radiatively corrected spectra and show that it depends on whether the energy of the real photon is measured, also for event selections that reject hard photons. We discuss the implications for existing measurements and recommend how radiative corrections, together with other known corrections to the spectral shape, should be included in experimental templates.

\end{itemize}

\paragraph*{Reader's guide.}
The paper consists of two largely independent parts. Sections~\ref{sec:EFT}--\ref{sec:0nu} are technical. They present the EFT, the evaluation of the one-loop diagrams and of the real emission, the closed-form loop integrals, the large-$\omega_n$ expansion, and the extension to $0\nu\beta\beta$ decay. They are mainly of interest to readers concerned with the theoretical framework. Readers primarily interested in the phenomenological consequences can proceed directly to Sec.~\ref{sec:results}, which is largely self-contained and refers back to earlier sections for the explicit expressions it uses. It discusses the accuracy of the large-$\omega_n$ expansion, the spectral and angular distortions, the photon emission rates, the bias on $\xi_{31}$ and its dependence on the detection of the photon, and the endpoint region and $0\nu\beta\beta$ searches. Sec.~\ref{sec:othercorr} combines the radiative corrections with other known corrections to the spectral shape. The main results are summarized in Sec.~\ref{sec:conclusions}.

\section{Heavy-nucleus Effective Field Theory}\label{sec:EFT}

The issue of radiative corrections to nuclear electroweak processes has a long history \cite{Brown:1966zza,Sirlin:1967zza,Abers:1968zz,Abers:1967zzb}, predating the formulation of the Standard Model. The problem is complicated by the fact that the exchanged virtual photons are sensitive to all scales relevant to the process, from the electroweak scale to the hadronization scale, from typical scales of nuclear physics, such as the nuclear radius, to the low-energy scales of the reaction $\mathcal{Q}$ value, and even down to the relevant scales of atomic physics. The formalism of Effective Field Theories provides a natural framework to split the problem into a set of simpler problems by focusing on a specific energy and momentum region at a time.
Recently, EFT frameworks have been proposed for the description of neutron and superallowed $\beta$ decays
\cite{Cirigliano:2022hob,Hill:2023acw,Hill:2023bfh,Cirigliano:2024msg,Cirigliano:2024rfk,Plestid:2025idc,Plestid:2025ojt,Borah:2024ghn},
where, in order to extract fundamental SM parameters such as the $V_{ud}$ element of the Cabibbo--Kobayashi--Maskawa (CKM) quark mixing matrix, it is important to control corrections to the total rate.
Electromagnetic corrections are also crucial for predicting $\beta$-decay spectra
with permille accuracy
\cite{Holstein:1974zf,Hayen:2017pwg},
allowing such observables to be used in searches for physics beyond the Standard Model \cite{Brodeur:2023eul,Sargsyan:2026ygi}.

In the case of pure Fermi or Gamow--Teller decays, energy-independent corrections to the leading-order Fermi or Gamow--Teller matrix elements can be absorbed into a normalization factor that can be fitted to the lifetime, while it is more important to control energy-dependent corrections to the spectral shape.
Similarly, in the case of $2\nu\beta\beta$, the large theoretical uncertainties in the leading-order nuclear matrix elements will mask any energy-independent electromagnetic
corrections, while it is important to control percent-level corrections to the  electron energy distributions.

The energy scales relevant to $2\nu\beta\beta$ are the electroweak scale, $M_W \sim 80$ GeV, the
hadronization scale $\Lambda_\chi \sim m_N \sim 1$ GeV, the typical scales of nuclear physics $M_\pi \sim k_F \sim R^{-1} \simeq \mathcal O(100\,\mathrm{MeV})$, and finally the scale of the nuclear excitation energies and of the reaction $\mathcal Q$ value,
$\omega \sim \mathcal Q \sim m_e \simeq  \mathcal O(\mathrm{MeV})$. We will exploit the separation between these energy scales, and start from a low-energy Lagrangian for nuclear degrees of freedom, which is formally obtained by integrating out quark and nucleon degrees of freedom. This theory is closely related to the Heavy Quark Effective Theory and Heavy Baryon Chiral Perturbation Theory Lagrangians \cite{Jenkins:1990jv,Georgi:1990um}.
The calculation is organized as an expansion in several small parameters. The nuclear-level EFT is an expansion in $E_e/\Lambda$, with $\Lambda \sim M_\pi \sim k_F$, which we truncate at leading order apart from the weak-magnetism and two-body corrections discussed below. We work at $\mathcal O(\alpha)$, but resum the Coulomb interaction between the electrons and the daughter nucleus, enhanced by $Z\alpha$, through the Fermi function, as is standard in single-$\beta$ decay~\cite{Hayen:2017pwg}. Finally, in Sec.~\ref{sec:largeomega} we expand in $E_e/\omega_n$, which defines the large-$\omega_n$ EFT. We will be mostly interested in the contribution of photon modes that are sensitive to the scale of the electron energy, $k \sim \mathcal Q \sim m_e$, which, adopting the language of nonrelativistic EFTs, we call ``ultrasoft''~\cite{Caswell:1985ui,Pineda:1997bj}.

We now specify our Lagrangian. We consider
\begin{equation}
    \mathcal L = \mathcal L_{\text{QED}}  + \mathcal L_{\text{EFT}}\,,
\end{equation}
where $\mathcal L_{\text{QED}}$ denotes the standard QED Lagrangian for photons, electrons, and neutrinos 
\begin{equation}\label{eq:QED}
    \mathcal L_{\text{QED}} = -\frac{1}{4} F^{\mu \nu} F_{\mu \nu}
 +   \bar{e} \left(i\slashed{D}-m_e\right) e + \bar{\nu}_L\left(i\slashed{\partial}\right) \nu_L,
\end{equation}
with $F_{\mu\nu}$ the photon field strength, $e$ and $\nu$ the electron and neutrino fields, and the covariant derivative $D$ given by
\begin{equation}\label{eq:cov}
    D_\mu = \partial_\mu  + i e Q A_\mu,
\end{equation}
where $Q = -1$ for the electron. $\mathcal L_{\text{EFT}}$ denotes the heavy-nucleus Lagrangian. 
For $2\nu\beta\beta$ decays to the ground state,  
the initial and final state nuclei both have $0^+$ quantum numbers. We will thus introduce two scalar fields
 $\mathcal{A}_i(0^+)$ and $\mathcal{A}_f(0^+)$ to describe these nuclei.
The intermediate nuclear states can have several possible values of angular momentum and parity. At lowest order in the multipole expansion, the weak charged current operator connects the initial and final state to $0^+$ or
$1^+$ intermediate states. In the isospin symmetric scenario, which nuclear calculations have found to be a good approximation, the Fermi contribution vanishes, leaving only $1^+$ states. 
We will thus introduce a vector field for each $1^+$ intermediate state,
denoted by  $\mathcal{B}_n(1^+)$.
At higher order in the multipole expansion and when including electromagnetic effects, it is also possible to connect to other intermediate states, but we will neglect this possibility in what follows.
We can split the heavy-nucleus Lagrangian into an electromagnetic and weak sector. In the electromagnetic sector, the Lagrangian is a generalization of the standard HQET Lagragian \cite{Manohar:1997qy}. At second order in the external momenta, the Lagrangian is given by \cite{Manohar:1997qy,Hill:2012rh}
\begin{align}
\label{eq:HEFT_em}
\mathcal L_{\mathrm{EFT}} &= \bar {\cal A}_i \left( i D_0 + \frac{\vec D^2}{2 m_{A_i}} \right) {\cal A}_i  + \bar {\cal A}_f \left( i  D_0 + \frac{\vec D^2}{2 m_{A_f}} - \Delta_{fi} \right)  {\cal A}_f  + \sum_n \bar{\mathcal B}^j_n  \left(i  D_0 + \frac{\vec D^2}{2 m_{B_n}} - \Delta_{ni}\right) \mathcal B^j_n  \nonumber \\
& - i  \sum_n \frac{e c_{F_n}}{m_{B_n}} \varepsilon^{i j k} {\vec B}^i \bar{\mathcal B}^j_n  \mathcal B^k_n + e c_{D_i}\,  \bar{\mathcal A}_i \frac{\left[\vec D, \vec E\right]}{8 m_{A_i}^2} \mathcal A_i + e c_{D_f}\,  \bar{\mathcal A}_f \frac{\left[\vec D, \vec E\right]}{8 m_{A_f}^2} \mathcal A_f +  \sum_n  e c_{D_n}\,  \bar{\mathcal B}^j_n \frac{\left[\vec D, \vec E\right]}{8 m_{B_n}^2} \mathcal B^j_n + \ldots .
\end{align}
The residual masses  are defined with respect to the initial nucleus,
\begin{equation}\label{eq:omegadef}
\Delta_{ni} = m_{B_n} - m_{A_i}\,,\qquad \Delta_{fi} = m_{A_f} - m_{A_i} = -\left(\mathcal Q + 2 m_e\right) \equiv -E_0\,,
\end{equation}
where $m_{A_{i,f}}$ and $m_{B_n}$ are nuclear masses and $\mathcal Q$ is the double beta $Q$ value defined with atomic masses. The energy denominators of the intermediate states depend on
\begin{equation}
\omega_n \equiv E_n - \frac{E_i + E_f}{2} = \Delta_{ni} - \frac{\Delta_{fi}}{2} = \Delta_{ni} + \frac{E_0}{2}\,.
\end{equation}
The superscript $j$ on the vector fields $\mathcal B$  in Eq.~\eqref{eq:HEFT_em} denotes their polarization.
At lowest order, the photon only couples to the nuclei via minimal coupling, with
the covariant derivative defined as in Eq. \eqref{eq:cov}, and $Q_i = Z$,
$Q_n = Z+1$, and $Q_f = Z+2$.
Beyond leading order, one gets additional effects from the coupling to the electric and magnetic fields
\begin{equation}
B^i = -\frac{1}{2}\varepsilon^{ijk} F_{jk}\,, \qquad E^{i} = F_{0i}\,. 
\end{equation}
In Eq. \eqref{eq:HEFT_em} we have included the magnetic term $c_{F_n}$, and the operators $c_{D_j}$, which receive contributions from the  charge radii of initial and final state nuclei. The dots denote other next-to- and next-to-next-to-leading order terms, including terms
that mediate magnetic transitions between the towers of excited states, the spin-orbit terms for the $1^+$ intermediate states, or lepton-nucleus interactions \cite{Manohar:1997qy,Hill:2012rh}.
Except for the coefficients fixed by Lorentz invariance,  the factors of nuclear masses are purely conventional,
and the couplings in Eq.~\eqref{eq:HEFT_em} need to be determined by matching either to experiment or to ab initio calculations with nucleon degrees of freedom. For example, the coupling $c_D$ contains information on the nuclei charge radii. They are given by
\cite{Hill:2012rh}
\begin{equation}
    c_{D_j} = Z_j \left(1 + \frac{4}{3} m^2_{A_j} \langle r_{A_j}^2 \rangle \right).
\end{equation}
The nuclear charge radii typically scale as $k_F^{-1}$, which determines the cut-off scale in the EFT expansion.  In this work we only need the leading-order Lagrangian. Nuclear recoil, which is suppressed by $E_e/m_{A} \sim 10^{-5}$, is neglected and the nuclear propagators become static propagators $i/(v\cdot k - \Delta + i0)$, with $v^\mu = (1,\vec 0)$ the nuclear velocity and $\Delta$ the residual mass of the propagating nucleus. The magnetic and charge-radius terms in Eq.~\eqref{eq:HEFT_em} contribute at higher order in $E_e/\Lambda$ and are not considered further. 

At lowest order, weak decays are mediated by the Lagrangian
\begin{equation}\label{eq:weak}
    \mathcal L_{\rm weak} = - \frac{2 G_F}{\sqrt{2}} V_{ud} \, g_V  \sum_n  \left\{g_{A} ^{i n}\, \bar{\mathcal  B}^j_n \mathcal A_i  + g_{A}^{ fn} \bar{\mathcal A}_f \mathcal B^j_n \right\} \bar e \gamma^j P_L\nu + \textrm{h.c.}.
\end{equation}
Here $G_F$ is the Fermi constant extracted from muon decay, $G_F = 1.1663785(6) \times 10^{-5}$ GeV$^{-2}$ \cite{ParticleDataGroup:2024cfk}, $V_{ud}$ is the $ud$ element of the CKM matrix.
$g_V$ is the single nucleon vector coupling. $g_V$ is equal to 1 at leading order in $\alpha$, but it receives electromagnetic corrections which we will discuss shortly.
$g^{in}_{A}$ and $g^{fn}_{A}$ denote the Gamow-Teller matrix elements between initial and final states and the intermediate states. At leading order in chiral EFT, they are given by the matrix elements of the single nucleon axial current
\begin{align}\label{eq:axial}
g^{in}_{A} g^{fn}_{A} = \frac{1}{3} \lambda^2 M_{GT}(n)\, \qquad M_{GT}(n) = \langle f| \sum_k  \vec \sigma_k\tau_k^+ |n\rangle \cdot \langle n | \sum_l \vec \sigma_l \tau_l^+ | i\rangle\,,
\end{align}
with $\lambda =g_A/g_V$, the sum over $k,l$ over individual nucleons, $\tau^+$ the isospin raising operator, and where $| i \rangle$, $| n \rangle$, $ | f \rangle$ refer to initial, intermediate and final state.
We find it convenient to factor out a factor of $g_V$ in Eq. \eqref{eq:weak} so that the coupling $g_A/g_V$ in Eq. \eqref{eq:axial} is renormalization-scale independent and can be directly related to the nucleon axial coupling as extracted from neutron decay \cite{ParticleDataGroup:2024cfk}
\begin{equation}
    \lambda = 1.2753 \pm 0.0013, 
\end{equation}
where the experimental number 
captures a set of $\mathcal O(\alpha)$ corrections \cite{Cirigliano:2022hob,Cirigliano:2024nfi}.
The axial couplings $g^{in}_{A}$ and $g^{fn}_{A}$ have an expansion in the chiral EFT expansion parameter and in $\alpha$. Higher-order corrections include, for example, the contributions of two-body weak currents, or energy-independent nuclear-structure dependent electromagnetic corrections, similar to $\delta_{\rm NS}$ for Gamow-Teller decays \cite{Cirigliano:2024msg,Cirigliano:2024rfk,Combes:2024pvm}.

At higher order in the multipole expansion, there arise corrections from weak magnetism \cite{Morabit:2024sms}, which, in the nuclear-level EFT we can express as 
\begin{equation}\label{eq:weak1}
    \mathcal L^{(1)}_{\rm weak} = - \frac{2 G_F}{\sqrt{2}} V_{ud} \, g_V  \sum_n  \left\{ \frac{g_{M}^{i n}}{m_{A_i}}\, \bar{\mathcal  B}^j_n \mathcal A_i  + \frac{g_{M}^{ fn}}{m_{A_f}} \bar{\mathcal A}_f \mathcal B^j_n \right\} \varepsilon^{jk\ell }  \partial_k \left(\bar e \gamma^\ell P_L\nu\right) + \textrm{h.c.}.
\end{equation}
The coefficients $g^{in}_M$ and $g^{fn}_M$ can be extracted from matching to the nucleon-level EFT used in Ref.~\cite{Morabit:2024sms}, to obtain
\begin{equation}
    \frac{g^{in}_M g_A^{fn}}{m_{A_i}} + \frac{g^{in}_A g_M^{fn}}{m_{A_f}} = \frac{g_A g_M}{6 m_N} M_{GT}(n),    
\end{equation}
with $g_M = 4.7$.
Photon exchange also leads to energy-dependent corrections 
that scale as the electron energy times the radius and are enhanced by the nuclear charge \cite{Wang:2016rqh,Hayen:2017pwg,Cirigliano:2024msg}. These can be also captured in an Effective Lagrangian formalism. For example, one could write down the operators   
\begin{equation}\label{eq:LagNLO}
 \delta   \mathcal L^{(1)}_{\text{weak}} = - \frac{2 G_F}{\sqrt{2}} V_{ud} g_V \sum_n \frac{1}{m_{A_i}}\bar{\mathcal{B}}^j_n \mathcal{A}_i \, \left\{ g^{in}_1 D_0 \left(\bar e_L\gamma^j \nu_L \right) 
    + g^{in}_2 \bar e_L \overleftrightarrow D_0 \gamma^j \nu_L
  + g^{in}_3 m_e \bar e_R \sigma^{0j} \nu_L 
    \right\} + \textrm{h.c.} + \ldots,
\end{equation}
where $\overleftrightarrow D_0 = \overrightarrow D_0 - \overleftarrow D_0$. The Lagrangian in Eq. \eqref{eq:LagNLO} only includes a few possible terms, as denoted by $\ldots$.
The coefficients $g_i$ once again need to be fixed by matching onto a nucleon-level EFT. As in the case of superallowed $\beta$ decays, we expect their typical scale to be approximately determined by the charge radius \cite{Cirigliano:2024msg}, and that some of these coefficients will be enhanced by powers of $Z$. 
$\mathcal O(\alpha Z R E_e)$ corrections are usually approximated by the solution of the Dirac equation in a realistic nuclear charge distribution \cite{Hayen:2017pwg}. Developing an accurate mapping between the corrections to the phase space factor and those included in Eq. \eqref{eq:LagNLO} or those arising from the charge radii in Eq.~\eqref{eq:HEFT_em}
will be important to achieve a rigorous accounting of all electromagnetic corrections to $2\nu\beta\beta$.

Finally, Ref.~\cite{Morabit:2024sms} pointed out that there can arise non-factorizable contributions to the $2\nu\beta\beta$ decay rate and spectrum. These can for example be induced by pion exchanges between the two-weak currents, or by shorter-range effects. In the nuclear level EFTs, these effects are captured by local two-nucleon four-lepton operators  
\begin{equation}\label{eq:L2beta}
    \mathcal L_{2\beta} = - 2 G_F^2 V_{ud}^2 \left[ C_{2\beta}\, \bar{\mathcal A}_f \mathcal A_i\, \left(\bar e_L \gamma^j \nu_L\right) \left(\bar e_L \gamma^j \nu_L\right) + C^v_{2\beta}\, \bar{\mathcal A}_f \mathcal A_i\, \left(\bar e_L \gamma^0\, \nu_L\right) \left(\bar e_L \gamma^0\, \nu_L\right)\right] + \textrm{h.c.}\,,
\end{equation}
We will also encounter these operators in Sec~\ref{sec:largeomega} when we integrate out intermediate nuclear excited states.

\subsection{The vector coupling $g_V$}

To get the complete radiative corrections to $2\nu\beta\beta$ at $\mathcal O(\alpha)$, we need to consider the corrections to the vector coupling $g_V$. These arise from the exchange of virtual photons from the electroweak scale all the way down to the scale of QCD, and they contain both a perturbative and a nonperturbative component, the latter encoded in the $W\gamma$ box \cite{Sirlin:1977sv}.
We follow here the formalism developed in Ref. \cite{Cirigliano:2023fnz}, to which we refer for more details. At a matching scale $\mu_\chi \sim m_N \sim 1$ GeV, the vector coupling is given by
\begin{align}
g_V (\mu_\chi) &= 
 \left[ 1 +  \overline \Box^V_\text{had} (\mu_0) -  \frac{\alpha (\mu_\chi)}{2 \pi}  \, 
\kappa \left( \mu_0,\mu_\chi \right)   \right]
  \bar  C^r_\beta (\mu_\chi)\,.
\label{eq:gVM}
\end{align}
Here $\mu_\chi$ denotes 
the renormalization scale in the $\overline{\text{MS}}_\chi$ scheme commonly used in chiral perturbation theory \cite{Gasser:1983yg}.
This can be converted to the standard  $\overline{\text{MS}}$ scheme used in the rest of the paper by setting
\begin{equation}
     \mu^2 = \mu_\chi^2  e^{-1}.
\end{equation}
$\bar C^r_\beta (\mu_\chi)$ is
the matching coefficient between the Standard Model and the Low-Energy EFT, defined in such a way to be scheme independent. 
$\bar C^r_\beta (\mu_\chi)$ encodes terms of $\mathcal O(\alpha)$,
with resummation of QCD and QED leading logarithms. Recently, next-to-leading logarithmic corrections have become available \cite{Moretti:2025ghw,Cirigliano:2026ios}, but they are not included in our numerical values.  
$\overline{\Box}_\text{had}^V (\mu_0)$ captures non-perturbative contributions, and is related to the standard $\gamma W$ box 
$\Box_{\gamma W}^V$ of Refs.~\cite{Seng:2018qru,Seng:2018yzq},
while $\kappa \left( \mu_0,{\mu_\chi} \right)$ is a perturbative correction to connect perturbative QCD, the hadronic box, and  HB$\chi$PT.
Using the perturbative input from Refs. \cite{Sirlin:1981ie,Hill:2019xqk,Cirigliano:2023fnz}
and the nonperturbative input from Refs. \cite{Seng:2018qru,Seng:2018yzq,Czarnecki:2019mwq,Seng:2020wjq,Shiells:2020fqp,Hayen:2020cxh}
\begin{equation}
    g_V(m_N) -1= 1.153(12) \cdot 10^{-2}. 
\end{equation}
The evolution of $g_V$ from $\mu_\chi \sim m_N$ to low-energy is governed by the anomalous dimension
\begin{align}\label{gVanom}
    \mu_\chi \frac{\mathrm{d} g_V\left( \mu_\chi \right)}{\mathrm{d} \mu_\chi} &=  \gamma (\alpha) \  g_V \left( \mu_\chi \right), \qquad               \gamma (\alpha) = \tilde{\gamma}_0 \, \frac{\alpha}{\pi} + \tilde{\gamma}_1 \left( \frac{\alpha}{\pi} \right)^2 \  +  \   \cdots
\end{align}
with \cite{Cirigliano:2023fnz} 
    \begin{align}
    \tilde{\gamma}_0 &= -\frac{3}{4},  \qquad
    \tilde{\gamma}_1 =  \frac{5 \tilde{n}}{24} + \frac{5}{32} - \frac{\pi^2}{6},
\end{align}
and $\tilde n$ is the effective number of active charged leptons 
\begin{equation}
    \tilde n(\mu_\chi) = \sum_\ell Q_\ell^2 n_\ell \theta(\mu_\chi - m_\ell).
\end{equation}
We provide numerical values at $\mu = 2E_0$ in the $\overline{\text{MS}}$ scheme for various isotopes in Table~\ref{tab:gV} (second row), which are used in the results section. The first row gives the corresponding values in the $\overline{\text{MS}}_\chi$ scheme at the same numerical values of $\mu_\chi$ which are not used in this paper but kept for completeness.

\begin{table}[t]
\centering
\begin{tabular}{|c|c|c|c|c|c|}
\hline
$\mu_\chi$ (MeV) & $m_e$ & 6.12 & 7.0 & 7.1 & 8.1  \\
\hline
$g_V-1$ (\%) & 2.499 & 2.054 & 2.031 & 2.028 & 2.003  \\
\hline
$\mu$ (MeV) & $m_e$ & 6.12 & 7.0 & 7.1 & 8.1  \\
\hline
$g_V-1$ (\%) & 2.404 & 1.964 & 1.941 & 1.938 & 1.914  \\
\hline
\end{tabular}
\caption{Representative values of the vector coupling $g_V$ as a function of the renormalization scale. The third to sixth columns correspond to the choice $\mu_\chi = 2E_0$, using the $\mathcal Q$ values of $^{76}$Ge, $^{136}$Xe, $^{130}$Te, and $^{100}$Mo, respectively.}
\label{tab:gV}
\end{table}

\section{Tree-level analysis}
\label{sec:treelevel}

\begin{figure}[t]
\includegraphics[width=\textwidth]{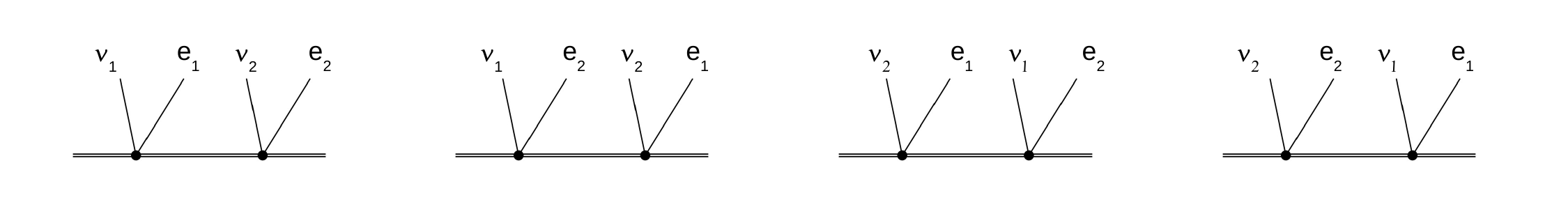}
\caption{Tree-level diagrams for $2\nu\beta\beta$ decay. The double line denotes the initial ($i$), intermediate ($n$) and final ($f$) nuclei. The four diagrams correspond to the different assignments of the leptons.}\label{fig:Tree}
\end{figure}

Before turning to the radiative corrections, we briefly derive the tree-level amplitude and verify that it reproduces the standard result of Ref.~\cite{Simkovic:2018rdz}. The four tree-level diagrams are shown in Fig.~\ref{fig:Tree}. In the first diagram, in which $e_1$ and $\nu_1$ are emitted at the first vertex, the propagator of the intermediate nucleus gives
\begin{equation}
\frac{i}{E_i - E_n - E_{e_1}-E_{\nu_1}} = \frac{-i}{\omega_n+\epsilon_K} \equiv \frac{-i}{\Omega_n^{+K}}\,,
\end{equation}
where $\omega_n$ is defined in Eq.~\eqref{eq:omegadef}, and
\begin{equation}
\epsilon_K = \frac{E_{e_1} + E_{\nu_1} - E_{e_2} - E_{\nu_2}}{2}\,,\qquad
\epsilon_L = \frac{E_{e_1} + E_{\nu_2} - E_{e_2} - E_{\nu_1}}{2}\,,\qquad
\Omega_n^{\pm K,L} = \omega_n \pm \epsilon_{K,L}\,.
\end{equation}
Here we used energy conservation, $E_i - E_f = E_{e_1}+E_{e_2}+E_{\nu_1}+E_{\nu_2}$, so that $E_n - E_i + E_{e_1} + E_{\nu_1} = \omega_n + \epsilon_K$. In the second diagram, $e_2$ and $\nu_1$ are emitted at the first vertex, and the propagator gives $-i/\Omega_n^{-L}$, 
while the third and fourth diagrams, with $\nu_1\leftrightarrow\nu_2$, give $-i/\Omega_n^{+L}$ and $-i/\Omega_n^{-K}$, respectively. The first and fourth diagrams, and the second and third, combine pairwise into
\begin{equation}
\frac{1}{\Omega_n^{+K,L}}+\frac{1}{\Omega_n^{-K,L}} = \frac{2\,\omega_n}{\omega_n^2-\epsilon_{K,L}^2}\,.
\end{equation}
Summing the four diagrams, and accounting for the relative signs from exchanging external fermions, we obtain
\begin{equation}
    i\mathcal{M}_0 = i\, \frac{8\left(G_{F} V_{ud} \lambda \right)^2 }{3 m_e}  \Bigg( \mathcal{L}_{11;22}[1,1]\, M_{GT}^{K}- \mathcal{L}_{12;21}[1,1]\,M_{GT}^{L}\Bigg) \;,
\end{equation}
where we defined
\begin{equation}\label{eq:GTKL}
M_{GT}^{K,L}=m_e \sum_n \frac{M_{GT}(n) \,\omega_n}{\omega_n^2-\epsilon_{K,L}^2}\,.
\end{equation}
In addition, we introduced a shorthand notation for the product of lepton bilinears 
\begin{equation}
\mathcal{L}_{ij;nm}[V,W] \equiv (\bar{e}_i V \gamma^k P_L \nu_j)\, (\bar{e}_n W \gamma_k P_L \nu_m)\,.
\end{equation}

The leading-order $2\nu\beta\beta$ decay rate is obtained by integrating the squared tree-level amplitude over the phase space, which, after using momentum conservation to eliminate the momentum of the final nucleus, reads
\begin{equation}\label{GammaBorn}
  \text{PS}_{\text{Born}} = \prod_{j} \int \frac{\mathrm{d}^3 p_j}{(2\pi)^3\, 2 E_{j}}\; (2\pi)\, \delta\!\left( E_i - E_f - E_{e_1} - E_{e_2} - E_{\nu_1} - E_{\nu_2}\right),
\end{equation}
with $j \in \{ e_{1}, e_{2}, \nu_1, \nu_2 \}$. We introduce the variables
\begin{equation}
\epsilon = E_{e_1} + E_{e_2} - 2 m_e\,, \qquad \Delta = \frac{E_{e_1} - E_{e_2}}{2}\,, \qquad y_{12} = \hat p_{e_1} \cdot \hat p_{e_2}\,,
\end{equation}
with $\hat p = \vec p/|\vec p\,|$, so that $E_{e_{1,2}} = m_e \pm \Delta + \epsilon/2$, and recall that $\mathcal Q = E_i - E_f - 2 m_e$.

Even at leading order, the expression of $\mathcal M_0$ shows that 
there is no complete factorization into NMEs and leptonic functions, because $M^{K,L}_{GT}$ depend on the lepton energies. Observables such as the summed and single-electron energy distributions therefore depend on nuclear matrix elements. This is very different from the case of allowed $\beta$ decays, where, at leading order in the multipole expansion, the electron spectrum is independent of nuclear matrix elements, which only appear at higher orders \cite{Holstein:1974zf,Hayen:2017pwg, Behrens:1971rbq}. 

Ref.~\cite{Simkovic:2018rdz} noticed that 
the matrix elements $M_{GT}^{K,L}$ in Eq.~\eqref{eq:GTKL} can be expanded in powers of $\epsilon_{K,L}/\omega_n$. The expansion converges rapidly when the relevant excitation energies are large compared with the lepton energies, see Ref.~\cite{Morabit:2024sms} for an explicit check in the nuclear shell model and Sec.~\ref{sec:results} for a detailed discussion. A direct calculation without this expansion, summing explicitly over the intermediate $1^+$ states, has recently been performed for $^{82}$Se and $^{136}$Xe~\cite{Ghinescu:2026mpj} and it confirms that the expansion describes the summed-energy spectrum of $^{136}$Xe well, while larger deviations appear for $^{82}$Se.
Upon expanding in $\epsilon_{K,L}/\omega_n$, 
at LO in $\alpha$ the decay rate depends on nuclear matrix elements through
\begin{align}\label{eq:MGTm}
M_{GT}^{(-2m-1)} =m_e (2m_e)^{2m} \sum_n \frac{M_{GT}(n)}{\omega_n^{2m+1}}\,,
\end{align}
in which the Gamow--Teller strengths $M_{GT}(n)$ of the transitions through the $n$th intermediate state are weighted by increasing powers of $1/\omega_n$. Matrix elements with larger $m$ are therefore more sensitive to the lowest intermediate states, and are expected to be more reliably calculable with a variety of nuclear-structure methods~\cite{Simkovic:2018rdz}. Their ratios
\begin{equation}
    \xi_{m1} = \frac{M^{(-m)}_{GT}}{M^{(-1)}_{GT}}
\end{equation}
characterize the shape of the spectrum. 
Upon expanding in $\epsilon_{K,L}$, and integrating over $E_{\nu_2}$, the differential decay rate can be expressed as
\begin{eqnarray}\label{totddecay}
\frac{\dd\Gamma}{\dd\epsilon\, \dd \Delta\, \dd y_{12}} = \frac{1}{2} \lambda^4 \left(M_{GT}^{(-1)} \right)^2   \left[\frac{\dd G^{2\nu}_0}{\dd\epsilon\, \dd\Delta\, \dd y_{12}  }+ \frac{\dd G^{2\nu}_2}{\dd\epsilon\, \dd\Delta\, \dd y_{12} }\xi_{31}  +\frac{\dd G^{2\nu}_4}{\dd\epsilon\, \dd\Delta\, \dd y_{12} } \left(\xi_{51} + \frac{1}{3}\xi_{31}^2\right) + \frac{\dd G^{2\nu}_{22}}{\dd\epsilon\, \dd\Delta\, \dd y_{12} }\frac{1}{3}\xi_{31}^2  + \ldots \right]\,,
\end{eqnarray}
where
\begin{align}
\label{dGamma}
\frac{\mathrm{d} G^{2\nu}_{n}}{\mathrm{d}\epsilon\, \mathrm{d}\Delta\, \mathrm{d} y_{12} } = \frac{1}{\ln{\left(2\right)}}\frac{\left(G_{F} V_{ud}\right)^4}{8 \pi^7 m_e^2}  \int^{\mathcal{Q}-\epsilon}_{0} \mathrm{d} E_{\nu_1}  E^2_{e_1}  E^2_{e_2}  E_{\nu_1}^2 E_{\nu_2}^2  \beta_1 \beta_2 F(Z_f,E_{e_1}) F(Z_f,E_{e_2})  \mathcal A_n\,,
\end{align}
with $\vec{\beta}_{1(2)} \equiv \vec{p}_{e_{1(2)}}/E_{e_{1(2)}}$ and energy conservation fixes $E_{\nu_2} = \mathcal Q - \epsilon - E_{\nu_1}$.  The functions $\mathcal A_n$ are
\begin{align}
\label{eq:A02}
\mathcal A_0  &= 1 - \beta_1 \beta_2 y_{12} \,, \qquad
\mathcal A_2  = (1 - \beta_1 \beta_2 y_{12} )\, \frac{\epsilon_K^2 + \epsilon_L^2}{(2m_e)^2} \,,
\qquad
\mathcal A_4  = (1 - \beta_1 \beta_2 y_{12} )\,\frac{\epsilon_K^4 + \epsilon_L^4}{(2m_e)^4} \,,\nonumber\\
\mathcal A_{22} & = \frac{\epsilon_K^2  \epsilon_L^2}{ (2m_e)^4} + \beta_{1} \beta_2\, \frac{y_{12}}{3}\,  \frac{\epsilon_K^4 - 5 \epsilon^2_K \epsilon_L^2 + \epsilon_L^4 }{(2m_e)^4}\,.
\end{align}
For $\mathcal A_0$, the integral over $E_{\nu_1}$ gives
\begin{equation}\label{dGamma0}
\frac{\mathrm{d} G^{2\nu}_{0}}{\mathrm{d}\epsilon\, \mathrm{d}\Delta\, \mathrm{d} y_{12} } = \frac{1}{\ln 2}\,\frac{(G_F V_{ud})^4}{8 \pi^7 m_e^2}\,  \frac{1}{30}\,  E^2_{e_1}  E^2_{e_2}\,    \beta_1 \beta_2\, (\mathcal{Q}-\epsilon)^5\,  F(Z_f,E_{e_1})\,  F(Z_f,E_{e_2})\, \mathcal{A}_0\,,
\end{equation}
while the corresponding expressions for $\mathcal A_2$, $\mathcal A_4$, and $\mathcal A_{22}$ are given in Eqs.~\eqref{eq:G2G4} and~\eqref{eq:G22} of Appendix~\ref{app:tree}.

The electrostatic interaction between the outgoing electrons and the daughter nucleus, with charge $Z_f = Z+2$, is included through the Fermi functions $F(Z_f, E_{e_{1,2}})$. Their expansion starts as
\begin{equation}\label{Fermi_expand}
F(Z, E_e) = 1 + \frac{\pi\alpha Z}{\beta_e} + \mathcal{O}(\alpha^2 Z^2)\,,
\end{equation}
with $\beta_e = \sqrt{1-m_e^2/E_e^2}$. This static Coulomb contribution must not be double counted when computing the $\mathcal{O}(\alpha)$ radiative corrections, as discussed in Sec.~\ref{sec:RC}. Strictly, the term proportional to $\beta_1\beta_2 y_{12}$ in $\mathcal A_n$ involves a different combination of Coulomb-distorted electron wave functions than $F$~\cite{Nitescu:2021pdq,Deppisch:2020mxv}. We use $F$ for both terms, a common approximation whose effect on the normalized spectra is small.  The Fermi functions used in this work are defined in App.~\ref{app:sirlin}

The differential rate in Eq.~\eqref{totddecay} provides the templates that are fitted to data to extract $\xi_{31}$ and $\xi_{51}$, which can then be compared with nuclear-structure calculations~\cite{Simkovic:2018rdz}. It can be extended to higher order in chiral EFT, which introduces weak-magnetism and two-nucleon corrections~\cite{Morabit:2024sms}, and it is modified by the radiative corrections computed in the remainder of this paper. We first compute these corrections for arbitrary $\omega_n$. In Sec.~\ref{sec:largeomega} we then expand them in $\epsilon_{K,L}/\omega_n$ as well, which yields a radiative correction for each term in Eq.~\eqref{totddecay}, and allows us to quantify how the radiative corrections affect the extraction of $\xi_{31}$.

\section{Radiative corrections}\label{sec:RC}
We now turn to the computation of electromagnetic radiative corrections to $2\nu\beta\beta$ at $\mathcal{O}(\alpha)$.
Working within the EFT framework of Sec.~\ref{sec:EFT}, the corrections arise from virtual photon exchange among the two outgoing electrons and the three nuclear states as well as from real photon emission from all charged lines.
We use dimensional regularization throughout, with $d = 4 - 2\varepsilon$, and work in the $\overline{\mathrm{MS}}$ scheme.

\subsection{Diagram classification}
\label{sec:diagrams_classification}

\begin{figure}
    \centering
    \includegraphics[width=\linewidth]{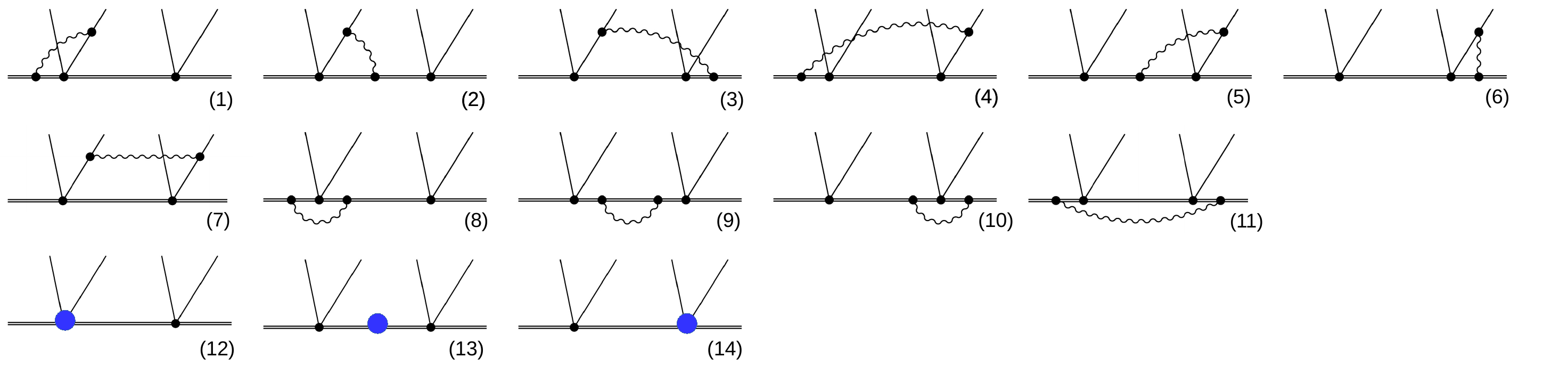}
    \caption{One-loop diagrams for $2\nu\beta\beta$ decay, grouped by class: photon exchange between an electron and the nucleus (1--6), between the two electrons (7), nuclear self-energy and vertex corrections (8--11), and counterterms (12--14). Double lines denote nuclear states, plain and wiggly lines denote leptons and photons, respectively. Black dots denote vertices from the leading-order Lagrangians, blue dots insertions of the counterterm Lagrangian. Only one lepton ordering is shown.}
    \label{fig:virtual}
\end{figure}

The $\mathcal{O}(\alpha)$ virtual corrections arise from the exchange of a single photon among the charged particles in the process.
The relevant diagrams are shown in Figs.~\ref{fig:virtual} and fall into four topologically distinct classes:
\begin{itemize}
    \item \textbf{Diagrams~1--6 (electron--nucleus triangles).}
    A photon is exchanged between one of the outgoing electrons and one of the three nuclear lines ($\mathcal{A}_i$, $\mathcal{B}_n$, or $\mathcal{A}_f$).
    These diagrams have direct analogues in single-$\beta$ decay. They contain the Coulomb interaction with the nucleus, which is enhanced by the nuclear charge and is accounted for by the Fermi function (Sec.~\ref{sec:Zdrop}). However, because the photon can couple to the intermediate nucleus, the diagrams obtain a dependence on $\omega_n$ that has no analogue in single-$\beta$ decay.
    \item \textbf{Diagram~7 (electron--electron box).}
    A photon is exchanged between the two outgoing electrons.
    This topology has no analogue in single-$\beta$ decay.
    It introduces dependence on the two-electron invariant mass $s = (p_{e_1} + p_{e_2})^2$, and is an important source of the differences from the sum of two single-$\beta$ Sirlin functions.

    \item \textbf{Diagrams~8--11 (nuclear self-energies and vertex corrections).}
    A photon connects two nuclear lines, either as a self-energy insertion on one of the nuclear propagators or as a vertex correction between two nuclear lines.
    These diagrams are purely hadronic and do not depend on the lepton momenta directly.

    \item \textbf{Diagrams~12--14 (counterterms).}
    The ultraviolet divergences of the virtual diagrams are cancelled by field-renormalization and coupling counterterms, given in Eqs.~\eqref{eq:CTterms} and~\eqref{eq:CTmatrixelement} of Appendix~\ref{app:virtual}.
\end{itemize}
In addition to the virtual corrections, real photon emission ($2\nu\beta\beta + \gamma$) contributes at the same order in $\alpha$. The relevant diagrams are shown in Fig.~\ref{fig:real}.
The infrared divergence of the real emission diagrams cancels against the infrared poles of the virtual diagrams, yielding a finite inclusive correction.
The real-emission diagrams are discussed in Sec.~\ref{sec:real}.

\subsection{Cancellation of $Z$-dependent terms}
\label{sec:Zdrop}

With nuclear charges $Z \simeq 30$--$60$, terms of $\mathcal O(\alpha Z)$ or $\mathcal O(\alpha Z^2)$ would dominate the radiative corrections. We show here that they cancel in the sum of the diagrams of each class, with one exception. The static Coulomb field of the daughter nucleus of $\mathcal O(\alpha Z/\beta)$ survives but is already accounted for by the Fermi functions. emitted electrons. We consider the first lepton ordering, with $(e_1,\nu_1)$ emitted at the first vertex and the intermediate propagator $-i/\Omega_n^{+K}$, the other three orderings follow in the same way.  We drop the common prefactors, $\alpha (G_F V_{ud}\lambda)^2 M_{GT}(n)/3$ and the sum over $n$. Diagram 7 does not depend on $Z$ and is not included in this section. The $Z$ dependence of the counterterms is given in Eqs.~\eqref{eq:CTterms} and~\eqref{eq:CTmatrixelement}.

\subsubsection{Nuclear self-energies and vertex corrections (diagrams~8--11)}
\label{sec:Zdrop_nuclear}

A photon connecting two nuclear lines carries the product of their charges. Writing $\Omega \equiv \Omega_n^{+K}$ for brevity, the sum of diagrams~8--11 is proportional to
\begin{equation}
\label{eq:M8to11_Z}
    \mathcal{L}_{11;22}[1,1]\int [\mathrm{d}k]\frac{1}{k^2}\frac{1}{v\cdot k - \Omega}\left[\frac{Z(Z+1)}{\left(v\cdot k\right)\left(-\Omega\right)}+\frac{(Z+1)^2}{\Omega^2}+\frac{(Z+2)(Z+1)}{\left(-\Omega\right)(v\cdot k)}+\frac{Z(Z+2)}{\left(v\cdot k\right)^2}\right],
\end{equation}
with the integration measure $[\dd k]$ defined in Eq.~\eqref{eq:integrationconvention}.
With $Z+(Z+2) = 2(Z+1)$ and $Z(Z+2) = (Z+1)^2-1$, the terms proportional to $(Z+1)^2$ combine into
\begin{equation}
(Z+1)^2\int [\mathrm{d}k]\frac{1}{k^2}\frac{1}{v\cdot k - \Omega}\left(\frac{1}{v\cdot k}-\frac{1}{\Omega}\right)^2
= (Z+1)^2\int [\mathrm{d}k]\frac{1}{k^2}\,\frac{v\cdot k-\Omega}{\Omega^2\,(v\cdot k)^2} = 0\,,
\end{equation}
since the remaining integrals are scaleless and vanish in dimensional regularization. Only the $Z$-independent term, from the $-1$ in $Z(Z+2)$, survives.

\subsubsection{Electron--nucleus triangles (diagrams~1--6)}
\label{sec:Zdrop_triangles}

Consider first diagrams~1--3, in which the photon connects electron~1 with the initial, intermediate, or final nucleus, respectively. In diagram~1, the photon is emitted by the initial nucleus before the first vertex and absorbed by the electron, while in diagrams~2 and~3 it is emitted by the electron and absorbed by a nucleus after the first vertex. This determines the signs of the photon momentum in the propagators. Writing the electron propagator as
\begin{equation}
S_\pm(k) \equiv \frac{\mathcal L_{11;22}[\gamma^0(\slashed p_{e_1} + m_e \pm \slashed k),1]}{k^2 \pm 2 k\cdot p_{e_1}}\,,
\end{equation}
the three diagrams are proportional to
\begin{equation}\label{eq:M13_Z}
\int[\mathrm{d}k]\frac{1}{k^2}\left[\frac{Z\, S_+(k)}{(-\Omega)\,(v\cdot k)} + \frac{(Z+1)\, S_-(k)}{(v\cdot k-\Omega)(-\Omega)} + \frac{(Z+2)\, S_-(k)}{(v\cdot k)(v\cdot k-\Omega)}\right].
\end{equation}
To isolate the $Z$ dependence, we write the charges as $Z = (Z+2) - 2$ and $Z+1 = (Z+2)-1$. The terms proportional to $Z+2$ in the second and third diagrams combine, since
\begin{equation}
\frac{1}{v\cdot k-\Omega}\left(\frac{1}{-\Omega}+\frac{1}{v\cdot k}\right) = \frac{1}{(-\Omega)\, (v\cdot k)}\,,
\end{equation}
which removes the intermediate-state propagator. Changing the integration variable $k\to -k$ in this combined term, so that it contains $S_+(k)$ like diagram~1, the terms proportional to $Z+2$ add up to
\begin{equation}\label{eq:M16_Z}
\frac{Z+2}{-\Omega}\int[\mathrm{d}k]\frac{1}{k^2}\left(\frac{1}{v\cdot k}+\frac{1}{-v\cdot k}\right) S_+(k)\,.
\end{equation}
Diagrams~4--6, with electron~2, give the analogous expression with $S_+(k)$ replaced by $\mathcal L_{11;22}[1,\gamma^0(\slashed p_{e_2}+m_e+\slashed k)]/(k^2+2k\cdot p_{e_2})$. The remaining terms, proportional to $-2$ and $-1$, are not enhanced by $Z$ and are part of the radiative corrections computed below. 

The two terms in the bracket of Eq.~\eqref{eq:M16_Z} would cancel if they had the same $i\eta$ prescription. However, each heavy-nucleus propagator carries $+i\eta$ with respect to its own energy variable, so that after the change of variables the two terms read $1/(v\cdot k+i\eta)$ and $1/(-v\cdot k+i\eta)$, and
\begin{equation}
    \frac{1}{v\cdot k +i\eta} + \frac{1}{-v\cdot k +i\eta} = - 2\pi i\, \delta(v\cdot k)\,.
\end{equation}
The surviving contribution comes from static photons, $v\cdot k = 0$, i.e.\ from the Coulomb field of the daughter nucleus with charge $Z+2$. Evaluating the remaining integral we find
\begin{equation}
\begin{split}
    i \mathcal{M}^{Z}_{1-6}&= -2\pi i\,\frac{\alpha (Z+2)}{4\pi}\left[i\mathcal{M}_0^{+K}\left(\left(\frac{1}{\beta_1}+\frac{1}{\beta_2}\right)\left(\frac{1}{\varepsilon_{\rm IR}}+1+i\pi \right)+\frac{1}{\beta_{1}}\ln{\frac{\mu^2}{4\vec{p}_{e_1}^{\,2}}}+\frac{1}{\beta_{2}}\ln{\frac{\mu^2}{4\vec{p}_{e_2}^{\,2}}}\right)\right.\\
    &\left.\qquad-\frac{m_e}{E_{e_1}\beta_1}\,i\mathcal{M}_0^{+K}[\gamma^0,1]-\frac{m_e}{E_{e_2}\beta_2}\,i\mathcal{M}_0^{+K}[1,\gamma^0]\right] +\ldots \;,
\end{split}
\end{equation}
where $\mathcal{M}_0^{+K}$ is the contribution of this lepton ordering to the tree-level amplitude, and $\mathcal{M}_0^{+K}[V,W]$ the same with $\mathcal{L}_{11;22}[1,1]$ replaced by $\mathcal{L}_{11;22}[V,W]$.
All terms except the one proportional to $i\pi$ are purely imaginary relative to $\mathcal M_0^{+K}$. They form the Coulomb phase~\cite{Weinberg:1965nx,Weinberg:1995mt}, including its infrared divergence, and do not contribute to the decay rate at $\mathcal O(\alpha)$.  The real part has a $\pi$-enhancement \cite{Hill:2023acw} and multiplies the rate by $1+\pi\alpha(Z+2)(\beta_1^{-1}+\beta_2^{-1})$, which is precisely the $\mathcal O(\alpha)$ term of the product of the Fermi functions, see Eq.~\eqref{Fermi_expand}. We therefore omit this term from the double-weak Sirlin function, to avoid double counting it in the differential decay rate.

\subsection{Virtual diagrams}
\label{sec:virtual}
Having shown that no $Z$-enhanced terms survive apart from the Coulomb contribution contained in the Fermi functions, we now compute the remaining, $Z$-independent parts of the virtual diagrams. For one intermediate state and the first lepton ordering, they are
\begin{align}
    i\mathcal{V}_{1-6} &= i \frac{\alpha}{\pi}\left(G_{F} V_{ud} \lambda\right)^2 \sum_n \frac{M_{GT}(n)}{3} \frac{1}{\Omega_n^{+K}}\int [\mathrm{d}k]\frac{1}{k^2}\left(\frac{\mathcal{L}_{11;22}[\gamma^0\left(\slashed{p}_{e_1}+m_e+\slashed{k}\right),1]}{k^2+2k\cdot p_{e_1}}\right. \nonumber \\
    &\times\left.\left(\frac{2}{v\cdot k}+\frac{1}{-v\cdot k-\Omega_n^{+K}}\right)+\frac{\mathcal{L}_{11;22}[1,\gamma^0\left(\slashed{p}_{e_2}+m_e+\slashed{k}\right)]}{k^2+2k\cdot p_{e_2}}\left(\frac{2}{v\cdot k}+\frac{1}{v\cdot k-\Omega_n^{+K}}\right)\right)+\ldots\;, \label{eq:ampV16} \\
    i \mathcal V_{7} &= i\frac{\alpha}{\pi}\left(G_{F} V_{ud} \lambda\right)^2\sum_n\frac{M_{GT}(n)}{3} 
    \int \left[\mathrm{d}k\right]
    \frac{\mathcal{L}_{11;22}\left[\gamma^{\mu} \left(\slashed{p}_{e_1}+m_e-\slashed{k}\right), \gamma_\mu \left(\slashed{p}_{e_2}+m_e+\slashed{k}\right)\right]}
    {k^2 \left(v\cdot k -\Omega_n^{+K}\right)\left(k^2 - 2k\cdot p_{e_1}\right)\left(k^2 + 2k\cdot p_{e_2}\right)}
    + \ldots \;, \label{eq:ampV7} \\
    i \mathcal V_{8-11} &= i\frac{\alpha }{\pi} \left(G_{F} V_{ud} \lambda\right)^2 \sum_n \frac{M_{GT}(n)}{3} \frac{\mathcal{L}_{11;22}}{\left(\Omega_n^{+K}\right)^2}\int [\mathrm{d}k]\frac{1}{k^2}\frac{1}{v\cdot k-\Omega_n^{+K}}+\ldots\;,
    \label{eq:ampV811}
\end{align}
where the dots denote the other three lepton orderings, and $[\mathrm{d}k]$ includes the $\overline{\rm MS}$ factor $\mu^{2\varepsilon}$, with $\mu$ the renormalization scale, see Eq.~\eqref{eq:integrationconvention}.
In every diagram, one propagator is that of a heavy nucleus and is linear in the loop momentum,
\begin{equation}
  \frac{1}{v\cdot k-\Delta+i\eta}\,,\qquad
  \Delta\in\{0,\Omega_n^{\pm K},\Omega_n^{\pm L}\}\,,
\end{equation}
where the residual energy $\Delta$ is set by the excitation energy of the intermediate state and by the lepton energies. The diagrams therefore involve tensor integrals with one heavy-particle propagator
\begin{align}
   \bar{B}^{\mu_1 \ldots \mu_n} &= \int [\mathrm{d}k] \frac{k^{\mu_1} \ldots k^{\mu_n}}{v\cdot k - \Delta} \frac{1}{k^2 - m_0^2}, \\
   \bar{C}^{\mu_1 \ldots \mu_n} &= \int [\mathrm{d}k] \frac{k^{\mu_1} \ldots k^{\mu_n}}{v\cdot k - \Delta} \frac{1}{k^2 - m_0^2}
    \frac{1}{(k + p_1)^2 - m_1^2}, \\
   \bar{D}^{\mu_1 \ldots \mu_n} &= \int [\mathrm{d}k] \frac{k^{\mu_1} \ldots k^{\mu_n}}{v\cdot k - \Delta} \frac{1}{k^2 - m_0^2}
    \frac{1}{(k + p_1)^2 - m_1^2} \frac{1}{(k + p_2)^2 - m_2^2},
\end{align}
with $n \leq 1$ for triangle integrals and $n \leq 2$ for box integrals.

By generalizing tensor reduction techniques for relativistic integrals \cite{Denner:2002ii} to the case of one linear propagator, all diagrams can be expressed in terms of scalar tadpole, bubble, triangle, and box nonrelativistic integrals \cite{Zupan:2002je}:
\begin{equation}
\begin{split}
    \bar{A}_{0}(\Delta) &= \int \left[\mathrm{d}k\right]\frac{1}{v \cdot k - \Delta}\;, \\
    \bar{B}_0(\Delta,m_0)&= \int \left[\mathrm{d}k\right]\frac{1}{v \cdot k - \Delta} \frac{1}{k^2 - m_0^2}\;, \\
    \bar{C}_{0}(v,p_1;\Delta,m_0,m_1) &= \int \left[\mathrm{d}k\right] \frac{1}{v \cdot k - \Delta} \frac{1}{k^2 - m_0^2} \frac{1}{(k + p_1)^2 - m_1^2}\;, \\
    \bar{D}_{0}(v,p_1,p_2;\Delta,m_0,m_1,m_2) &= \int \left[\mathrm{d}k\right] \frac{1}{v \cdot k - \Delta} \frac{1}{k^2 - m_0^2} \frac{1}{(k + p_{1})^2 - m_1^2} \frac{1}{(k + p_{2})^2 - m_2^2}\;,
\end{split}
\label{eq:pave_basis}
\end{equation}
together with standard relativistic scalar integrals.

The tadpole and bubble integrals are known for general $\Delta$. The bubble is given in Eq.~\eqref{eq:scalarpavebubble}. General solutions for the scalar triangle and box integrals were derived in Ref.~\cite{Zupan:2002je}, but in an implicit form and without infrared divergences regulated dimensionally. We obtain closed-form expressions for all of them with the following method~\cite{Saad}. The heavy-particle triangles and boxes are ultraviolet finite and can be obtained from relativistic integrals with an auxiliary propagator~\cite{liu2022automatic}, characterized by a parameter $L$ with dimension of inverse energy
\begin{equation}
    \frac{1}{v\cdot k-\Delta+i\eta} = \lim_{L\to 0^+}\frac{1}{L}\frac{1}{\left(k+p_L\right)^2-m_L^2+i\eta'}\;,
\end{equation}
with $2Lp_L^\mu=v^\mu$, $L\eta'=\eta$, and $4L^2m_L^2=1+4L\Delta$. The expression is evaluated using traditional methods that employ quadratic propagators, and the $L\to 0^+$ limit is taken at the end to remove its dependency. We have verified the results against a direct numerical evaluation of the integrals after isolating the divergent terms and, when possible, against analytical solutions using standard techniques. 

With the bubble integral, diagrams~8--11 give
\begin{equation}
\label{eq:M8to11_final}
    i\mathcal{V}_{8-11} = i\mathcal{M}_0^{+K}\,\frac{\alpha}{4\pi}\left(-\frac{2}{\varepsilon_{\rm UV}}-2L_{\mu}-2L_{\Omega_n^{+K}}-4\right)+\ldots\;,
\end{equation}
with $L_\mu = \ln(\mu^2/m_e^2)$ and $L_{\Delta} = \ln[m_e^2/(4\Delta^2)]$.

Diagrams 1--6 contain a more complicated $\Delta$ dependence. They involve the integrals  $\bar{C}_0(v,\pm p_e;0,0,m_e)$ which are infrared divergent for on-shell electrons, $\bar{C}_0(v,\pm p_e;\Omega_n^{+K},0,m_e)$ which are infrared finite but complex (Eqs.~\eqref{eq:CVpos} and \eqref{eq:CVneg}), and standard relativistic bubble integrals (Eq.~\eqref{eq:noneeftpave}). At leading order in $1/\Omega_n^{+K}$ we obtain
\begin{equation}
\begin{split}
    i \mathcal{V}_{1-6}&= \frac{\alpha}{4\pi}\left[i \mathcal{M}_0^{+K}\left(\frac{2}{\varepsilon_{\rm UV}}+2L_{\mu}-2L_{\Omega_n^{+K}}+8+4\frac{w_1}{\beta_1}\left(\frac{1}{\varepsilon_{\rm IR}}+L_{\mu}+1-w_1+\frac{\textrm{Li}_2\left(\frac{2\beta_1}{1+\beta_1}\right)}{w_1}\right)\right.\right.\\
    &\qquad\left.+4\frac{w_2}{\beta_2}\left(\frac{1}{\varepsilon_{\rm IR}}+L_{\mu}+1-w_2+\frac{\textrm{Li}_2\left(\frac{2\beta_2}{1+\beta_2}\right)}{w_2}\right)\right)\\
    &\qquad\left.- 4\frac{w_1}{\beta_1}\frac{m_e}{E_{e_1}}\,i\mathcal{M}_0^{+K}[\gamma^0,1]- 4\frac{w_2}{\beta_2}\frac{m_e}{E_{e_2}}\,i\mathcal{M}_0^{+K}[1,\gamma^0]\right]+\mathcal O\!\left(\frac{E_e}{\Omega_n^{+K}}\right)+\ldots\;,
\end{split}
\end{equation}
with $w_i = \tanh^{-1}\beta_i$. 
The most interesting of these master integrals is $\bar{C}_0(v,\pm p_e;0,0,m_e)$ which also appears in single-$\beta$ decay. Besides the $\pi$ enhancement discussed in Sec.~\ref{sec:Zdrop_triangles}, it contains the finite terms
\begin{equation}\label{Sirlinpart}
\frac{1}{\beta_e} {\rm Li}_2\left(\frac{2\beta_e}{1+\beta_e}\right) + \frac{w_e^2}{\beta_e}\,,
\end{equation}
which are characteristic of the single-$\beta$ Sirlin function, Eq.~\eqref{eq:sirlin1beta}. The full expressions for diagrams~1--6 are given in Eqs.~\eqref{eq:diag13full} and~\eqref{eq:diag46full}.

The most difficult diagram is the electron box, which, after tensor reduction, depends on the infrared-divergent scalar box $\bar{D}_0(v,-p_{e_1},p_{e_2};\Delta,0,m_e,m_e)$, given in Eq.~\eqref{eq:diag7eftbox}, the infrared-convergent $\bar{C}_0(v,p_{e_2}+p_{e_1};\Delta-E_{e_1},m_e,m_e)$, given in  Eq.~\eqref{eq:diag7efttriangle}, and an infrared-divergent relativistic triangle $\bar{C}_0$, given in Eq.~\eqref{eq:noneeftpave}. These integrals are functions of the electron invariant mass $s$ and, due to the presence of two $k$ terms in the numerator, the full Passarino--Veltman decompositions is too lengthy to be presented here. However, in Eqs.~\eqref{eq:scalarmasterlargedelta1} and \eqref{eq:scalarmasterlargedelta2}, we show that the master integral can be expanded in the large-$\omega_n$ limit, substantially simplifying its expression. We give diagram~7 at leading order in this expansion in Eq.~\eqref{eq:M7largeomegaexpansion}.

The counterterm diagrams are proportional to the tree-level amplitude.
Including them, and multiplying by the square root of the residue 
of the propagators of the initial- and final-state particles, we obtain for the divergent part of the virtual diagrams
\begin{align}\label{eq:div1}
    i \left[\mathcal V\right]_{\text{div}} &= i \mathcal M_0 \left[ \frac{\alpha}{4\pi} \times   
     \frac{4}{\varepsilon}
\left(  \frac{w_1}{\beta_1} +  \frac{w_2}{\beta_2}
   - \frac{1 + \beta_{12}^2}{2}   \frac{w_{12}}{\beta_{12}}  
\right) + \delta g^{in}_{A} + \delta g^{fn}_{A} - \delta Z_{\mathcal B_n} +  \frac{1}{2}\left( 2 R_e + R_{\mathcal A_i} + R_{\mathcal A_f}\right) \right] \nonumber  \\
& =  i \mathcal M_0 \frac{\alpha}{4\pi} \times  \frac{4}{\varepsilon} \left[ -\frac{3}{2} +  \frac{w_1}{\beta_1} +  \frac{w_2}{\beta_2}
   - \frac{1 + \beta_{12}^2}{2}   \frac{w_{12}}{\beta_{12}}   \right]\;,
\end{align}
with the counterterms and residues of Eqs.~\eqref{eq:CTterms} and~\eqref{eq:CTmatrixelement}. Here $\beta_{12} = \sqrt{1-4m_e^2/s}$ and $w_{12} = \tanh^{-1}\beta_{12}$.  The remaining $1/\varepsilon$ poles are of infrared origin and cancel exactly against those of the real emission, Eq.~\eqref{eq:realdiv}.

After summing over all lepton orderings, we split the finite part into a term proportional to the tree-level amplitude and a remainder
\begin{equation}
    i \mathcal V_{\text{fin}} = i \mathcal V_{\text{Born}} + i \delta \mathcal V.
\end{equation}
The real part of $\mathcal V_{\text{Born}}$ is given by
\begin{align}
   i \mathcal V_{\rm Born} &= i \mathcal   M_0 \frac{\alpha}{4\pi} \times 4 \left[   - \frac{3}{4} L_\mu  
   + \left( \frac{w_1}{\beta_1} + \frac{w_2}{\beta_2} - \frac{1 + \beta_{12}^2}{2} \frac{w_{12}}{\beta_{12}} \right) L_\mu
   + \frac{w_1}{\beta_1}  
    + \frac{w_2}{\beta_2}
- \frac{w^2_1}{\beta_1}  -  \frac{w_2^2}{\beta_2}
+ \frac{1 +\beta^2_{12}}{2} \frac{w_{12}^2}{\beta_{12}} \right. \nonumber \\ & \left. 
-  \frac{1}{\beta_1}   {\rm Li}_2\left(\frac{2\beta_1}{1+\beta_1}\right) - \frac{1}{\beta_2} {\rm Li}_2\left(\frac{2\beta_2}{1+\beta_2}\right)    + \frac{1+\beta_{12}^2}{2\beta_{12}} \left(   {\rm Li}_2\left(\frac{2\beta_{12}}{1+\beta_{12}}\right) - \frac{\pi^2}{2} \right)
\right]\,.
\end{align}
The remainder $\delta \mathcal V$ depends non-trivially on $\Omega_{n}^{\pm K}$ and $\Omega_{n}^{\pm L}$ and is not proportional to the tree-level amplitude. 
The expressions for the interference of the virtual amplitude with the tree level are too lengthy to be reproduced in full.
We provide partial expressions for the diagrams in Appendix~\ref{app:Expressions}.
As we will discuss in Sec~\ref{sec:decay},
in the limit $\omega_n \gg \epsilon_{K,L}$ it is possible to obtain a complete factorization between nuclear and leptonic input to the radiative corrections.

\subsection{Real emission diagrams}\label{sec:real}

\begin{figure}
    \centering
    \includegraphics[width=\linewidth]{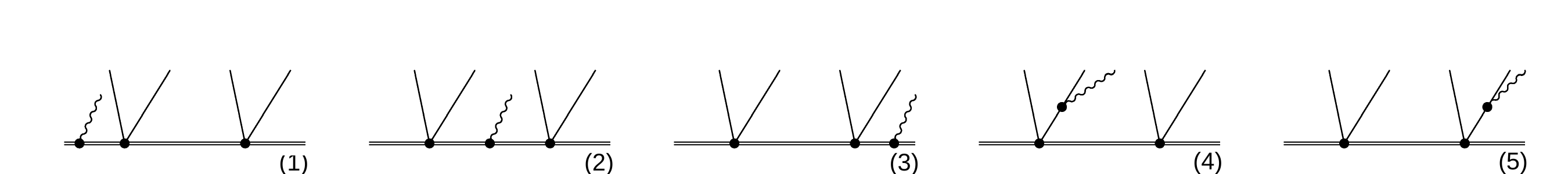}
    \caption{Real emission diagrams. The notation is as in Fig.~\ref{fig:virtual}}
    \label{fig:real}
\end{figure}

The real emission diagrams are shown in Fig. \ref{fig:real}. For the first lepton ordering, the amplitude is
\begin{align}
    i \mathcal R &=   i\, 8 (G_F V_{ud}\lambda)^2\, e\, \varepsilon^*_{\lambda} \sum_n \frac{M_{GT}(n)}{3}\left\{
   \frac{v^\lambda}{\Omega_n^{+ K} } \left[ \frac{1}{ v\cdot k + 2 \Omega_{n}^{+ K} } + \frac{1}{ v\cdot k - 2 \Omega_{n}^{+ K}  } - \frac{2}{v\cdot k}\right]  \mathcal L_{11;22}[1,1] \right. \nonumber \\
   & \left.
+ \mathcal L_{11;22}[\gamma^\lambda (\slashed{p}_{e_1} + \slashed{k} + m_e),1]\,
\frac{1}{p_{e_1} \cdot k}\, \frac{1}{2\Omega_{n}^{+ K} + v\cdot k} + \mathcal L_{11;22}[1,\gamma^\lambda (\slashed{p}_{e_2} + \slashed{k} + m_e)]\, \frac{1}{ p_{e_2} \cdot k}\, \frac{1}{2\Omega_{n}^{+K} - v\cdot k}        \right\}\,,
\end{align}
where $k$ and $\varepsilon_\lambda$ are the photon momentum and polarization. The factors $2\Omega_n^{+K}\pm v\cdot k$ arise because the photon carries away part of the energy. Also for the real emission, $\mathcal O(Z)$ terms cancel at the amplitude level, so that the total correction will be $\mathcal O(\alpha)$ rather than $\mathcal O(\alpha Z^2)$ or $\mathcal O(\alpha Z)$. 

In the soft limit, we can easily sum the lepton orderings and the amplitude reduces to
\begin{equation}\label{Rsoft}
    i \mathcal R_{\text{soft}} = i\mathcal{M}_0\times \varepsilon^*_{\lambda} e   \left[ -\frac{2 v^\lambda}{v\cdot k} + \frac{p_{e_1}^\lambda}{p_{e_1} \cdot k} +  \frac{p_{e_2}^\lambda}{p_{e_2} \cdot k} \right]\,.
\end{equation}
 The dependence on $\Omega_{n}^{\pm K}$ and $\Omega_{n}^{\pm L}$ is the same as that of the tree-level amplitude, which makes the cancellation of the infrared divergences between virtual and real corrections much more transparent.

The emission of an extra photon of course changes the phase space from Eq.~\eqref{GammaBorn} into
\begin{equation}
 \text{PS}_{\text{real}} = \prod_{j}  \int \frac{\mathrm{d}^3 p_j}{(2\pi)^3\, 2 E_{j}} \int \frac{\mathrm{d}^3 k}{(2\pi)^3\, 2 |\vec k|}\, (2\pi)\, \delta\!\left( E_i - E_f - E_{e_1} - E_{e_2} - E_{\nu_1} - E_{\nu_2} - |\vec k|\right)\,,
\end{equation}
with $j \in \{e_1,e_2,\nu_1,\nu_2\}$. We parametrize the photon energy as $|\vec k| = \mathcal Q\xi$, with $\xi \in [0,1]$, and treat the infrared region in dimensional regularization with subtraction techniques~\cite{Frixione:1995ms,Alioli:2010xd}. The photon phase space becomes
\begin{equation}
  \mu^{2\varepsilon}\int  \frac{\mathrm{d}^{d-1} k}{(2\pi)^{d-1}\, 2 |\vec k|}\, f(k) =  \frac{2^{2-2\varepsilon}}{(4\pi)^{2}} \frac{\Gamma(1-\varepsilon)}{\Gamma(1-2\varepsilon)} \left(\frac{\mu^2 e^{\gamma_E} } {\mathcal Q^2}\right)^{\varepsilon} \int_0^1 \mathrm{d} \xi\,  \frac{1}{\xi^{1+2 \varepsilon}}\,  \frac{1}{2}\int_{-1}^1  \mathrm{d} y\, (1-y^2)^{-\varepsilon} \, \frac{1}{\pi} \int_0^\pi \mathrm{d} \varphi\,  (\sin \varphi)^{-2 \varepsilon} \left( \vec{k}^2 f(k)\right)\,,
\end{equation}
where $y$ and $\varphi$ are the polar and azimuthal angles of the photon, and we expand
\begin{equation}
    \frac{1}{\xi^{1+2\varepsilon}} = - \frac{1}{2\varepsilon}\,  \delta(\xi)  + \left[\frac{1}{\xi}\right]_{+} + \mathcal O(\varepsilon)\,,
\qquad
    \int_0^1 \mathrm{d}\xi\, \left[\frac{1}{\xi}\right]_{+} g(\xi) = \int_0^1 \mathrm{d}\xi\, \frac{g(\xi) - g(0)}{\xi}\,.
\end{equation}
The $\delta(\xi)$ piece has the same phase space as the tree-level process and thus can be grouped with virtual diagrams.
This term generates infrared divergences and finite terms proportional to the tree-level amplitude. The plus distribution describes the emission of photons with nonzero energy, with the soft region subtracted. The resulting decay rate is therefore inclusive. That is, it contains both $2\nu\beta\beta$ events and $2\nu\beta\beta+\gamma$ events with photons of any energy. How this inclusive rate maps onto measured spectra is discussed in Secs.~\ref{sec:etot} and~\ref{sec:results}.

Squaring the soft amplitude and summing over photon polarizations, the infrared-divergent part is
\begin{align}
 |\mathcal R|_{\text{div}}^2 = |\mathcal M_0|^2  \frac{\alpha}{4\pi} \frac{1}{\varepsilon} \frac{\int d\Omega}{4 \pi} \left[ 8 + \frac{2m_e^2 |\vec k|^2}{(p_{e_1} \cdot k)^2} + 
  \frac{2m_e^2 |\vec k|^2}{(p_{e_2} \cdot k)^2}
 - \frac{8 E_{e_1} |\vec k|}{p_{e_1}\cdot k}
  - \frac{8 E_{e_2} |\vec k|}{p_{e_2}\cdot k} + \left(1 - \frac{2m_e^2}{s}\right)\frac{2s\, |\vec k|^2}{p_{e_1}\cdot k\, p_{e_2}\cdot k}\right]\,.
  \label{eq:r2div}
\end{align}
Performing the angular integrals with the master integrals of Eq.~\eqref{eq:photonanglemasters} (see also Ref.~\cite{Alioli:2010xd}), we obtain
\begin{align}\label{eq:realdiv}
 |\mathcal R|^2_{\text{div}} = |\mathcal M_0|^2  \frac{\alpha}{2\pi} \frac{4}{\varepsilon}  \left[ \frac{3}{2}  -  \frac{w_1}{\beta_1}  
 - \frac{w_2}{\beta_2} 
 +  \frac{1+\beta_{12}^2}{2} \frac{w_{12}}{\beta_{12}} \right].
\end{align}
The divergences in Eq. \eqref{eq:realdiv}
cancel the corresponding divergences in the virtual emission in Eq.~\eqref{eq:div1} after computing the interference with the tree level amplitude. 

By expanding the photon phase space and the amplitude at $\mathcal O(\varepsilon)$, it is also possible to obtained closed-form expressions for the soft part of the real emissions
\begin{align}
  &  |\mathcal R|^2_{\text{soft}} = |\mathcal M_0|^2 \frac{\alpha}{2\pi} \times 4 \Bigg\{ \left[ \frac{3}{2}  - \frac{w_1}{\beta_1}  - \frac{w_2}{\beta_2}
 +   \frac{1+\beta_{12}^2}{2} \frac{w_{12}}{\beta_{12}} \right] \left(L_\mu + \ln \left(\frac{m_e^2}{4 \mathcal Q^2}\right)\right)  +   2    +  \frac{w_1}{2\beta_1}   +  \frac{w_2}{2\beta_2}   -   \frac{w_1^2}{\beta_1} - \frac{w_2^2}{\beta_2} 
\nonumber \\
& - \frac{1}{\beta_1} {\rm Li}_2\left(\frac{2\beta_1}{1+\beta_1} \right) 
 -   \frac{1}{\beta_2} {\rm Li}_2\left(\frac{2\beta_2}{1+\beta_2} \right)   +  \frac{1}{4} I_\varepsilon(p_{e_1}, p_{e_2})
\Bigg\},
\end{align}
where the function $I_\varepsilon(p_{e_1}, p_{e_2})$ arises from the $\mathcal O(\varepsilon)$ terms of the angular integrals. It can be reconstructed from Eqs.~(A42)--(A50) of Ref.~\cite{Alioli:2010xd} and is given in Appendix~\ref{app:Expressions}.

Combining with the part of the virtual corrections proportional to the tree-level amplitude, we find
\begin{align}
  &  2\,{\rm Re}\left(\mathcal M^*_0\, \mathcal V_{\text{Born}}\right) + |\mathcal R|^2_{\text{soft}}
    = | \mathcal M_0|^2\, \frac{\alpha}{2\pi}  \left\{ 3 L_\mu  +
    2\left[ 3  - 2\frac{w_1}{\beta_1}  - 2\frac{w_2}{\beta_2}
 +   (1+\beta_{12}^2) \frac{w_{12}}{\beta_{12}} \right] \ln \frac{m_e^2}{4 \mathcal Q^2}  + 8 + 6 \frac{w_1}{\beta_1} + 6 \frac{w_2}{\beta_2} - 8 \frac{w_1^2}{\beta_1}\right. \nonumber\\  & \left.
  - 8 \frac{w_2^2}{\beta_2}   -  \frac{8}{\beta_1} {\rm Li}_2\left(\frac{2\beta_1}{1+\beta_1} \right)
 -  \frac{8}{\beta_2} {\rm Li}_2\left(\frac{2\beta_2}{1+\beta_2} \right)
 + 2\frac{1+\beta_{12}^2}{\beta_{12}} \left( w_{12}^2 +  {\rm Li}_2\left(\frac{2\beta_{12}}{1+\beta_{12}} \right) - \frac{\pi^2}{2} \right) +  I_{\varepsilon}(p_{e_1}, p_{e_2}) \right\}.\label{eq:Bornlike}
\end{align}
The only dependence on $\mu$ in Eq.~\eqref{eq:Bornlike}, the term $3L_\mu$, is compensated by that of the vector coupling. This can be seen explicitly by noting the dependence of the tree-level amplitude on $g_V^2$ resulting in an extra $\mu$-dependent term $|\mathcal M_0|^2 (4\tilde \gamma_0 \alpha/\pi) \ln \mu $ where $\tilde \gamma_0=-3/4$ is the $\mathcal O(\alpha)$ anomalous dimension of $g_V$ given in Eq.~\eqref{gVanom}.

The finite part of the real emission, which arises from the plus distributions, depends on the intermediate states in a nontrivial way, and we give it only in the large-$\omega_n$ limit, in Eq.~\eqref{eq:r2largeomega}. As we will see in the next section, Eq.~\eqref{eq:Bornlike} provides most of the terms that survive in this limit.

\section{Decay rate and the large-$\omega_n$ limit}\label{sec:decay}

The good convergence of the expansion of the tree-level rate in $\epsilon_{K,L}/\omega_n$ discussed in Sec.~\ref{sec:treelevel} raises the question of whether a similar expansion can be applied to the radiative corrections.
We therefore expand the full results described the previous section
to second order $\epsilon_{K,L}/\omega_n$. The main advantage of this expansion is that the radiative corrections to the decay rate can then be given in closed form.
At leading order in $1/\omega_n$, we find that radiative corrections modify Eq.~\eqref{totddecay} by shifting
\begin{align}
\label{eq:dGalpha}
\frac{\mathrm{d} G^{2\nu}_0}{\mathrm{d}\epsilon\, \mathrm{d}\Delta\, \mathrm{d} y_{12}  }  \mapsto  g_V^4 \frac{\mathrm{d} G^{2\nu}_0}{\mathrm{d} \epsilon\, \mathrm{d}\Delta\, \mathrm{d} y_{12}  } + \frac{\alpha}{2\pi} \frac{\mathrm{d} G^{2\nu}_\alpha}{\mathrm{d} \epsilon\, \mathrm{d} \Delta\, \mathrm{d} y_{12}  }\,, 
\end{align}
where $\mathrm{d} G_\alpha^{2\nu}$ contains contributions of ultrasoft photons, while hard photons cause the vector coupling $g_V$ to deviate from 1 (Table~\ref{tab:gV}).
The expression for $\dd G^{2\nu}_{\alpha}$ was reported in Ref.~\cite{deVries:2026udq} and is given by
 \begin{align}
\label{eq:largew}
&\frac{\mathrm{d} G^{2\nu}_\alpha}{\mathrm{d}\epsilon\, \mathrm{d}\Delta\, \mathrm{d} y_{12}} = \frac{\mathrm{d} G^{2\nu}_{0}}{\mathrm{d}\epsilon\, \mathrm{d}\Delta\, \mathrm{d} y_{12}} f_{2\nu}(\beta_1, \beta_2, \beta_{12}) \nonumber \\ 
&f_{2\nu} = \Bigg\{ 6\frac{w_1}{\beta_1}+6\frac{w_2}{\beta_2}-8\frac{w_1^2}{\beta_1}-8\frac{w_2^2}{\beta_2}-\frac{8}{\beta_1}\text{Li}_2{\left(\frac{2\beta_{1}}{1+\beta_{1}}\right)}-\frac{8}{\beta_2}\text{Li}_2{\left(\frac{2\beta_{2}}{1+\beta_{2}}\right)}+2\frac{1+\beta_{12}^2}{\beta_{12}}\left(w_{12}^2+\text{Li}_2{\left(\frac{2\beta_{12}}{1+\beta_{12}}\right)}-\frac{\pi^2}{2}\right)\nonumber \\
&\quad+2\left(\ln{\left(\frac{m_e^2}{4(\mathcal{Q}-\epsilon)^2}\right)}+\frac{137}{30}\right)\left(3-2 \left(\frac{w_1}{\beta_1}+\frac{w_2}{\beta_2}\right)+\frac{1+\beta_{12}^2}{\beta_{12}}w_{12}\right) -\frac{4m_e^2}{\mathcal{A}_0 E_{e_1}E_{e_2}}\left(\beta_{12}w_{12}+\frac{w_1 E_{e_2}}{\beta_{1}E_{e_1}}+\frac{w_2 E_{e_1}}{\beta_{2}E_{e_2}}\right)\nonumber \\
&\quad+I_{\epsilon} +3 L_{\mu} -\frac{\mathcal{Q}-\epsilon}{\mathcal{A}_0}\Bigg[\frac{1-\frac{\beta_2}{\beta_1}y_{12}}{E_{e_1}}+\frac{1-\frac{\beta_1}{\beta_2}y_{12}}{E_{e_2}} +\frac{\mathcal{Q}-\epsilon}{21 E_{e_1}E_{e_2}}\left(2-\frac{E_{e_1}E_{e_2}}{m_e^2}\left(\frac{\beta_1}{\beta_2}+\frac{\beta_2}{\beta_1} -2 \beta_1\beta_2\right)y_{12}\right) \nonumber \\
&\quad-\frac{w_1}{3\beta_1 E_{e_1}}\left(2+\frac{m_e^2}{E_{e_1}E_{e_2}}-\left(3-\beta_{1}^2\right)\frac{\beta_2}{\beta_1}y_{12}+\frac{\mathcal{Q}-\epsilon}{7E_{e_1}}\left(1-y_{12}\frac{\beta_2}{\beta_1}\right)\right) \nonumber \\
&\quad-\frac{w_2}{3\beta_2 E_{e_2}}\left(2+\frac{m_e^2}{E_{e_1}E_{e_2}}-\left(3-\beta_{2}^2\right)\frac{\beta_1}{\beta_2}y_{12}+\frac{\mathcal{Q}-\epsilon}{7E_{e_2}}\left(1-y_{12}\frac{\beta_1}{\beta_2} \right)\right)\Bigg]\Bigg\}\,.
\end{align}
The first line of Eq.~\eqref{eq:largew}
and the dependence on $L_\mu$ and $I_\varepsilon$
arise from  $\mathcal V_{\text{Born}}$
and from the soft part of the real emission diagrams in Eq.~\eqref{eq:Bornlike}.
In the second line, we see that real emission convert the $\ln m_e^2/\mathcal Q^2$ in Eq.~\eqref{eq:Bornlike}
into a $\ln m_e^2/(\mathcal Q-\epsilon)$. The mass-dependent terms in the second line arise from $\delta \mathcal V$. Finally, the terms proportional to polynomials of $\mathcal Q - \epsilon$ come from the real emission diagrams. The $\mu$ dependence of $3L_\mu$ cancels against that of $g_V^4$, so that the combination in Eq.~\eqref{eq:dGalpha} is independent of $\mu$.

At second order in $1/\omega_n$ we obtain further shifts 
\begin{align}
\label{eq:dGalpha2}
\xi_{31} \frac{\mathrm{d} G^{2\nu}_2}{\mathrm{d}\epsilon\, \mathrm{d}\Delta\, \mathrm{d} y_{12}  }  \mapsto \xi_{31} \left[ g_V^4 \frac{\mathrm{d} G^{2\nu}_2}{\mathrm{d} \epsilon\, \mathrm{d}\Delta\, \mathrm{d} y_{12}  } + \frac{\alpha}{2\pi} \left( \frac{\mathrm{d} G^{2\nu}_\alpha}{\mathrm{d} \epsilon\, \mathrm{d} \Delta\, \mathrm{d} y_{12}  } \frac{(\mathcal Q-\epsilon)^2 + 28 \Delta^2}{56 m_e^2} + 
\frac{\mathrm{d} G^{2\nu}_{2\alpha}}{\mathrm{d} \epsilon\, \mathrm{d} \Delta\, \mathrm{d} y_{12}  }
\right) \right] + \zeta_{31} \frac{\alpha}{2\pi} \frac{\mathrm{d} G^{2\nu}_{L\alpha}}{\mathrm{d} \epsilon\, \mathrm{d} \Delta\, \mathrm{d} y_{12}  }  \,, 
\end{align}
where the new ratio of matrix elements $\zeta_{31}$ is given by
\begin{align}\label{eq:LMGT} 
LM_{GT}^{(-3)} =m_e (2m_e)^{2} \sum_n \frac{M_{GT}(n)}{\omega_n^{3}} L_{\omega_n}  \,, \qquad \zeta_{31} = \frac{LM_{GT}^{(-3)}}{M_{GT}^{(-1)}} \,, \qquad
L_{\omega_n} = \ln\frac{m_e^2}{4\omega_n^2}\,. 
\end{align}
We give the expressions for $G^{2\nu}_{2\alpha}$ and $G^{2\nu}_{L\alpha}$ in Appendix~\ref{app:NLO}. 

The first two terms in Eq.~\eqref{eq:dGalpha2} correspond
to multiplying the tree-level
decay rate, including the term proportional to $\xi_{31}$, by the double-weak Sirlin function of Eq.~\eqref{eq:largew}. The loop and real-emission diagrams, however, induce an additional contribution proportional to $\xi_{31}$, $\mathrm{d}G^{2\nu}_{2\alpha}$, and a term proportional to a new ratio of matrix elements, $\zeta_{31}$, which cannot be absorbed into $\xi_{31}$ or $\xi_{51}$. An analysis at this order that includes radiative corrections
therefore requires, in principle, one more nuclear input than a tree-level analysis.
In practice this is not an obstacle because $\zeta_{31}$ is built from the
same $M_{GT}(n)$ with an additional logarithmic weight, so that any
calculation able to provide $\xi_{31}$ can provide
$\zeta_{31}$ at no extra cost. Furthermore its contribution to the spectrum turns out to be negligible (Sec.~\ref{sec:dS}). The values used in this work are collected in
Table~\ref{tab:NMEratios}. For $^{100}$Mo, we also give the values for single-state dominance through the $1^+$ ground state of $^{100}$Tc, for which $\xi_{31} = (2m_e/\omega_1)^2$, $\xi_{51} = \xi_{31}^2$, and $\zeta_{31} = \xi_{31}L_{\omega_1}$, with $\omega_1 \simeq 1.69$ MeV.

\begin{table}[t]
\centering
\begin{tabular}{lccc|cc|ccc}
\hline\hline
 & \multicolumn{3}{c|}{NSM} & \multicolumn{2}{c|}{QRPA} & \multicolumn{3}{c}{SSD}\\
isotope & $\xi_{31}$ & $\xi_{51}$ & $\zeta_{31}$ & $\xi_{31}$ & $\xi_{51}$ & $\xi_{31}$ & $\xi_{51}$ & $\zeta_{31}$\\
\hline
$^{76}$Ge  & 0.13 & 0.022 & $-0.58$ & 0.104 & 0.017 & --    & --    & --      \\
$^{100}$Mo & 0.33   & 0.11    & --      & 0.565 & 0.208 & 0.368 & 0.135 & $-1.38$ \\
$^{130}$Te & 0.12   &  0.028  & --      & 0.202 & 0.042 & --    & --    & --      \\
$^{136}$Xe & 0.17 & 0.046 & $-0.72$ & 0.387 & 0.121 & --    & --    & --      \\
\hline\hline
\end{tabular}
\caption{
Ratios of nuclear matrix elements used in this work: nuclear shell model (NSM)~\cite{Dekens:2024hlz,Coraggio:2018tuo,Coraggio:2022vgy,LuigiCoraggioPrivate,CUORE:2025xue}, quasiparticle random-phase approximation with $g_A^{\rm eff}=1.269$ (QRPA)~\cite{Simkovic:2018rdz}, and single-state dominance (SSD)~\cite{Simkovic:2018rdz}. $\zeta_{31}$, defined in Eq.~\eqref{eq:LMGT}, requires the individual transitions which we only have for the NSM calculations in $^{76}$Ge and $^{136}$Xe. For SSD it follows from $\zeta_{31} = \xi_{31}L_{\omega_1}$.
}
\label{tab:NMEratios}
\end{table}

\subsection{Distributions in $\epsilon_{\text{tot}}$}\label{sec:etot}

The distributions of Eq.~\eqref{eq:dGalpha} refer to the summed kinetic energy of the electrons, $\epsilon$, and are appropriate if the energy of the photon is not measured. If the photon is absorbed in the detector, as in calorimetric experiments, the measured quantity is instead the total energy of electrons and photon. We therefore define $\epsilon_{\text{tot}}$ as the energy available to electrons and photons,
\begin{equation}
    \epsilon_{\text{tot}} = \mathcal Q - E_{\nu_{1}} - E_{\nu_{2}}.
\end{equation}
In the Born kinematics $\epsilon_{\text{tot}} = \epsilon$,
while in the real emission kinematic $\epsilon_{\text{tot}} = \epsilon + E_\gamma$.
The virtual correction and the soft part of the real emissions are thus the same as for the $\epsilon$ distributions, giving
\begin{align}
\frac{\mathrm{d} G^{2\nu}_V}{\mathrm{d}  \epsilon_{\text{tot}}\, \mathrm{d}\Delta \, \mathrm{d}  y_{12}} &=  \frac{\alpha}{2\pi} \frac{\mathrm{d} G^{2\nu}_0}{\mathrm{d}  \epsilon_{\text{tot}}\, \mathrm{d}  \Delta \, \mathrm{d}  y_{12}}  f_t^{2\nu}(\beta_1, \beta_2, \beta_{12}) 
    \\ f_t^{2\nu}(\beta_1, \beta_2, \beta_{12}) &=   \Bigg\{ 
     6\frac{w_1}{\beta_1}+6\frac{w_2}{\beta_2}-8\frac{w_1^2}{\beta_1}-8\frac{w_2^2}{\beta_2}-\frac{8}{\beta_1}\text{Li}_2{\left(\frac{2\beta_{1}}{1+\beta_{1}}\right)}-\frac{8}{\beta_2}\text{Li}_2{\left(\frac{2\beta_{2}}{1+\beta_{2}}\right)}
    \nonumber \\ & \quad+2\frac{1+\beta_{12}^2}{\beta_{12}}\left(w_{12}^2+\text{Li}_2{\left(\frac{2\beta_{12}}{1+\beta_{12}}\right)}-\frac{\pi^2}{2}\right)
+2 \ln{\left(\frac{m_e^2}{4\mathcal Q^2}\right)}\left(3-2 \left(\frac{w_1}{\beta_1}+\frac{w_2}{\beta_2}\right)+\frac{1+\beta_{12}^2}{\beta_{12}}w_{12}\right) \nonumber \\
&\quad -\frac{4m_e^2}{\mathcal{A}_0 E_{e_1}E_{e_2}}\left(\beta_{12}w_{12}+\frac{w_1 E_{e_2}}{\beta_{1}E_{e_1}}+\frac{w_2 E_{e_1}}{\beta_{2}E_{e_2}}\right)+I_{\epsilon} +3 L_{\mu}.
    \Bigg\}. \label{eq:dGammaVtot}
\end{align}
The function $f^{2\nu}_t$ differs from $f_{2\nu}$ in two ways. The soft-photon logarithm involves the maximal photon energy $\mathcal Q$ rather than $\mathcal Q-\epsilon$ and the constant and polynomial terms from the hard real emission are absent. Both are replaced by the real-emission contribution at fixed $\epsilon_{\rm tot}$. For fixed $\epsilon_{\text{tot}}$,  the electron energies depend on the photon energy and we cannot carry out the integration over $\xi$ analytically. Instead we write
\begin{align}
    \frac{\mathrm{d} G^{2\nu}_R}{\mathrm{d} \epsilon_{\text{tot}}\, \mathrm{d} \Delta\,  \mathrm{d} y_{12}} &=  
  \frac{\alpha}{2\pi}   \frac{\mathrm{d} G^{2\nu}_0}{\mathrm{d}  \epsilon_{\text{tot}}\, \mathrm{d}  \Delta \, \mathrm{d}  y_{12}}  \frac{1}{\mathcal A_0}  \nonumber \\ 
&\times  \Bigg[    \int_0^{\xi_{\text{max}}}  \frac{d\xi}{\xi} \Bigg( 
\frac{\bar{E}_{e_1}^2 \bar{E}_{e_2}^2 \bar{\beta}_1 \bar{\beta}_2
F(\bar{E}_{e_1},Z_f)  F(\bar{E}_{e_2},Z_f) }{{E}_{e_1}^2 {E}_{e_2}^2 {\beta}_1 {\beta}_2 F({E}_{e_1},Z_f)  F({E}_{e_2},Z_f)}
g( \bar E_{e_1}, \bar E_{e_2}, y_{12}, \xi) -  g(E_{e_1},E_{e_2}, y_{12},0) \Bigg) \nonumber \\ & + 
 g( E_{e_1}, E_{e_2}, y_{12} , 0) \log \xi_{\text{max}} \Bigg],
\label{eq:dGammaRtot}
\end{align}
where
\begin{equation}
    \xi_{\text{max}} = \frac{\epsilon_{\text{tot}}- 2 |\Delta|}{\mathcal Q},
\end{equation}
and $|\Delta | < \epsilon_{\text{tot}}/2$.
Barred variables are evaluated in the real-emission kinematics, with $\epsilon = \epsilon_{\text{tot}} - \mathcal Q\xi$, while unbarred quantities use the Born kinematics $\epsilon = \epsilon_{\text{tot}}$.
The function $g$, which describes the photon spectrum, is
\begin{equation}
    g(\bar E_{e_1}, \bar E_{e_2}, y_{12}, \xi) = g_{-1}(\bar E_{e_1}, \bar E_{e_2}, y_{12}) + \mathcal Q \xi\, g_0(\bar E_{e_1}, \bar E_{e_2}, y_{12}) + 
(\mathcal Q \xi)^2\,  g_1(\bar E_{e_1}, \bar E_{e_2}, y_{12}),
    \end{equation}
where
\begin{align}
    g_{-1} &= - 4 (1 - \beta_{1} \beta_2 y_{12}) \left[ 3  - 2\left(\frac{w_1}{\beta_1} 
 + \frac{w_2}{\beta_2} \right)
  +  (1+\beta_{12}^2)\frac{w_{12}}{\beta_{12}} \right]  \label{eq:gm1}\\
    g_0 &=
    -6 \Bigg[ \frac{1-\frac{\beta_2}{\beta_1}y_{12}}{E_{e_1}}+\frac{1-\frac{\beta_1}{\beta_2}y_{12}}{E_{e_2}} -\frac{w_1}{3\beta_1 E_{e_1}}\left(2+\frac{m_e^2}{E_{e_1}E_{e_2}}-\left(3-\beta_{1}^2\right)\frac{\beta_2}{\beta_1}y_{12}\right) \nonumber \\ & \quad -\frac{w_2}{3\beta_2 E_{e_2}}\left(2+\frac{m_e^2}{E_{e_1}E_{e_2}}-\left(3-\beta_{2}^2\right)\frac{\beta_1}{\beta_2}y_{12}\right)\Bigg] \label{eq:g0} \\
g_1 &=    - 2 \Bigg[\frac{1}{E_{e_1}E_{e_2}}\left(2-\frac{E_{e_1}E_{e_2}}{m_e^2}\left(\frac{\beta_1}{\beta_2}+\frac{\beta_2}{\beta_1} -2 \beta_1\beta_2\right)y_{12}\right)-\frac{w_1}{\beta_1 E^2_{e_1}}\left(1-y_{12}\frac{\beta_2}{\beta_1}\right)-\frac{w_2}{\beta_2 E^2_{e_2}}\left(1-y_{12}\frac{\beta_1}{\beta_2} \right) \Bigg]. \label{eq:g1}
\end{align}

In the case of single $\beta$-decay, Sirlin pointed out that as consequence of the Kinoshita--Lee--Nauenberg theorem~\cite{Kinoshita:1962ur,Lee:1964is}, the neutrino spectrum has a  
finite limit when $m_e \rightarrow 0$ \cite{Sirlin:2011wg},
in contrast to the $\beta$ spectrum
\footnote{We thank R.~Plestid for pointing this out to us.}.
The same applies for the differential distribution with respect to $\epsilon_{\text{tot}}$. 
In the massless limit,
and setting the Fermi function to 1,
the $\xi$ and $\Delta$ integrals in Eqs.~\eqref{eq:dGammaVtot}
and \eqref{eq:dGammaRtot}
can  be performed analytically, and we indeed find that  $\mathrm{d}\Gamma/ \mathrm{d} \epsilon_{\text{tot}} \mathrm{d} y_{12}$ does 
 not depend logarithmically 
on $m_e$. Since these logarithms arise from a combination of the virtual corrections, including the photon exchange between the two electrons, and the collinear region of the real emission, their cancellation provides a nontrivial check of the calculation. For concreteness, we give the decay rate integrated with respect to the lepton angle, which is
\begin{equation}
\lim_{m_e \rightarrow 0}    \frac{\mathrm{d}\Gamma}{\mathrm{d} \epsilon_{\text{tot}}}= 
\lambda^4 \left(M_{GT}^{(-1)} \right)^2  
\frac{1}{\ln{\left(2\right)}}\frac{(G_F V_{ud})^4}{8 \pi^7 m_e^2}  \frac{1}{900}   (\mathcal{Q}-\epsilon_{\text{tot}})^5  \epsilon_{\textrm{tot}}^5
\left[ g_V^4 + 
\frac{\alpha}{2\pi}        \left( 6 \log\frac{\mu}{2 \epsilon_{\text{tot}}} + \frac{227}{10} - \frac{16}{3} \pi^2\right) \right],
\end{equation}
at leading order in $1/\omega_n$.

The distributions in $\epsilon$ and $\epsilon_{\text{tot}}$ represent the two extremes in which the measured energy is only the energy of the two electrons or 
the total energy of the electron and photon systems. For more 
general functions of the electron and photon energies, specified by the observable $\mathcal O$, we can write the real emission diagrams as 
\begin{equation}
 G^{2\nu}_R(\mathcal O) =  
\int \dd\epsilon\, \dd \Delta\, \dd y_{12}\,  \dd \xi \, \frac{\dd G^{2\nu}_R}{ \dd\epsilon\, \dd \Delta\, \dd y_{12} \, \dd \xi } \mathcal O(\epsilon,\xi),
 \end{equation}
with 
\begin{align}
  \frac{\mathrm{d} G^{2\nu}_R}{ \mathrm{d}\epsilon\, \mathrm{d} \Delta\, \mathrm{d} y_{12}\, \mathrm{d} \xi }  &=   \frac{1}{\ln 2}\frac{(G_F V_{ud})^4}{8 \pi^7 m_e^2} \,  \frac{\alpha}{2\pi}   \, \frac{1}{30}\, \, \left[\frac{1}{\xi}\right]_{+}  E^2_{e_1}  E^2_{e_2}\,    \beta_1 \beta_2\,  F(Z_f,E_{e_1})\, F(Z_f,E_{e_2}) \nonumber \\ & \times \Big\{ \left(\mathcal{Q}(1-\xi) - \epsilon \right)^5\,  \theta(\mathcal{Q}(1-\xi) - \epsilon )\, g(E_{e_1},E_{e_2}, y_{12}, \xi)\Big\},
\end{align}
where the plus distribution, defined on $\xi\in[0,1]$, acts on the full $\xi$ dependence, including that of the observable $\mathcal O(\epsilon,\xi)$.  
The distribution of the photon direction is given in Sec.~\ref{sec:2nugamma}.

\subsection{The radiative mode $2\nu\beta\beta+\gamma$}
\label{sec:2nugamma}

The real-emission contribution can also be read as the rate of the radiative decay mode $2\nu\beta\beta+\gamma$. At leading order in $1/\omega_n$, its photon energy spectrum is
\begin{equation}\label{eq:radchannels}
\begin{split}
\frac{\mathrm{d} \Gamma(2\nu\beta\beta+\gamma)}{\mathrm{d} E_\gamma} &= \frac{1}{2} \lambda^4 \left(M_{GT}^{(-1)} \right)^2  \frac{1}{\ln 2}\frac{(G_F V_{ud})^4}{8 \pi^7 m_e^2} \,  \frac{\alpha}{2\pi} \\
&\times\int_0^{\mathcal Q - E_\gamma} \mathrm{d} \epsilon \int^{\epsilon/2}_{-\epsilon/2} \mathrm{d} \Delta \int_{-1}^1 \mathrm{d} y_{12}    \, \frac{1}{30} E^2_{e_1}  E^2_{e_2}    \beta_1 \beta_2  (\mathcal{Q} - \epsilon - E_\gamma)^5  F(Z_f,E_{e_1})\, F(Z_f,E_{e_2})\,\mathcal A_\gamma\,.
\end{split}
\end{equation}
with the photon spectrum
\begin{align}\label{eq:Agamma}
\begin{split}
\mathcal A_\gamma &=\frac{4}{E_\gamma} \Bigg\{ -3 \mathcal{A}_0 - \frac{3}{2} \frac{E_\gamma}{E_{e_1}} \left(1- y_{12} \frac{\beta_2}{\beta_1}\right) - \frac{3}{2} \frac{E_\gamma}{E_{e_2}} \left(1- y_{12} \frac{\beta_1}{\beta_2}\right) - \frac{E_\gamma^2}{E_{e_1} E_{e_2}} \left(1 - \frac{y_{12}}{2} \left(\frac{E_{e_1} \beta_1}{E_{e_2}\beta_2} + \frac{E_{e_2} \beta_2}{E_{e_1}\beta_1} \right)\right) \\
& + 2\left[ \mathcal{A}_0 + \frac{E_\gamma}{2 E_{e_1}} \left(1 + \frac{m_e^2}{2 E_{e_1} E_{e_2}} - y_{12}\frac{\beta_2}{\beta_1} \left(1+\frac{m_e^2}{2 E_{e_1}^2} \right)\right) + \frac{E_\gamma^2}{4 E_{e_1}^2} \left(1 - y_{12} \frac{\beta_2}{\beta_1}\right) \right] \frac{w_{1}}{\beta_{1}}  \\ 
& + 2\left[ \mathcal{A}_0 + \frac{E_\gamma}{2 E_{e_2}} \left(1 + \frac{m_e^2}{2 E_{e_1} E_{e_2}} - y_{12}\frac{\beta_1}{\beta_2} \left(1+\frac{m_e^2}{2 E_{e_2}^2} \right)\right) + \frac{E_\gamma^2}{4 E_{e_2}^2} \left(1 - y_{12} \frac{\beta_1}{\beta_2}\right) \right] \frac{w_{2}}{\beta_{2}}  - \frac{4 E_{e_1} E_{e_2}}{s} \mathcal{A}_0^2 \frac{w_{12}}{\beta_{12}} \Bigg\}\,\\
&= \frac{1}{E_\gamma} \Big(g_{-1}(E_{e_1},E_{e_2},y_{12}) + E_\gamma\, g_0(E_{e_1},E_{e_2},y_{12}) + E^2_\gamma\, g_1(E_{e_1},E_{e_2},y_{12}) \Big),
\end{split}
\end{align}
where the functions $g_{-1}$, $g_0$, and $g_1$ are given in Eqs.~\eqref{eq:gm1},~\eqref{eq:g0}, and~\eqref{eq:g1}. The spectrum is infrared divergent and scales as $1/E_\gamma$. Physical quantities are rates with a photon above a certain threshold, or the inclusive rates discussed above for which the soft region is combined with the virtual corrections and the infrared divergences cancel.

\subsubsection{Photon angular distribution}
\label{sec:photonangle}

For applications in detector simulations, the direction of the photon might be needed as well.
Squaring the real-emission amplitude, summing over photon polarizations, and integrating over
the directions of the neutrinos, we find, at leading order in $1/\omega_n$ 
\begin{align}\label{eq:R2full}
  \frac{\sum_{\rm pol}|\mathcal R|^2}{|\mathcal M_0|^2}
  = e^2\Bigg[&-J^2
  +\frac{m_e^2}{E_{e_1}E_{e_2}\mathcal A_0}\left(\frac{1}{p_{e_1}\!\cdot k}+\frac{1}{p_{e_2}\!\cdot k}
   -\frac{p_{e_1}\!\cdot k}{(p_{e_2}\!\cdot k)^2}-\frac{p_{e_2}\!\cdot k}{(p_{e_1}\!\cdot k)^2}\right)
  \\
  &+\frac{2}{E_{e_1}E_{e_2}\mathcal A_0\;v\!\cdot k}\left(\frac{E_{e_2}\,p_{e_1}\!\cdot k}{p_{e_2}\!\cdot k}
   +\frac{E_{e_1}\,p_{e_2}\!\cdot k}{p_{e_1}\!\cdot k}-E_{e_1}-E_{e_2}\right)
  +\frac{1}{E_{e_1}E_{e_2}\mathcal A_0}\left(\frac{p_{e_1}\!\cdot k}{p_{e_2}\!\cdot k}
   +\frac{p_{e_2}\!\cdot k}{p_{e_1}\!\cdot k}-2\right)\Bigg]\,,\nonumber
\end{align}
where $\mathcal A_0=1-\beta_1\beta_2\,y_{12}$, and
\begin{equation}
  J^\mu=-\frac{2v^\mu}{v\cdot k}+\frac{p_{e_1}^\mu}{p_{e_1}\cdot k}+\frac{p_{e_2}^\mu}{p_{e_2}\cdot k}
\end{equation}
is the current describing soft emission from the nucleus, see Eq.~\eqref{Rsoft}. The second and third terms arise from the
interference between the soft and non-soft amplitudes, and the last term from the
non-soft amplitude squared.

In the rest frame of the initial nucleus we define $k^\mu=E_\gamma(1,\hat n)$ and
$p_{e_i}^\mu=E_{e_i}(1,\beta_i\hat p_i)$ such that
$p_{e_i}\cdot k=E_\gamma E_{e_i}a_i$ where 
\begin{equation}
  a_i = 1-\beta_i\,\hat n\cdot\hat p_i\,.
\end{equation}
Including the photon phase space, the real-emission rate differential in the photon direction becomes
\begin{equation}\label{eq:radangle}
  \frac{\dd\Gamma(2\nu\beta\beta+\gamma)}{\dd\epsilon\,\dd\Delta\,\dd y_{12}\,\dd E_\gamma\,\dd\Omega_\gamma}
  = \mathcal N(\epsilon,\Delta,E_\gamma)\,\frac{\alpha}{2\pi}\,\frac{1}{E_\gamma}\,\frac{g(\hat n)}{4\pi}\,,
  \qquad
  g(\hat n)=2\left[\mathcal A_0\,F(\hat n)+\frac{E_\gamma\,G(\hat n)}{E_{e_1}E_{e_2}}
  +\frac{E_\gamma^2\,H(\hat n)}{E_{e_1}^2E_{e_2}^2}\right],
\end{equation}
where $\mathcal N$ is the normalization of Eq.~\eqref{eq:radchannels}, i.e.\ the phase space,
the Fermi functions, and the factor $(\mathcal Q-\epsilon-E_\gamma)^5$. The three terms
correspond to those of Eq.~\eqref{eq:R2full}
\begin{align}
  F(\hat n)&=\frac{4}{a_1}+\frac{4}{a_2}-4-\frac{m_e^2}{E_{e_1}^2a_1^2}
  -\frac{m_e^2}{E_{e_2}^2a_2^2}-\frac{2\,p_{e_1}\cdot p_{e_2}}{E_{e_1}E_{e_2}\,a_1a_2}\,,\\
  G(\hat n)&=m_e^2\left(\frac{1}{E_{e_1}a_1}+\frac{1}{E_{e_2}a_2}-\frac{E_{e_1}a_1}{E_{e_2}^2a_2^2}
  -\frac{E_{e_2}a_2}{E_{e_1}^2a_1^2}\right)
  +2\left(\frac{E_{e_1}a_1}{a_2}+\frac{E_{e_2}a_2}{a_1}-E_{e_1}-E_{e_2}\right),\\
  H(\hat n)&=\frac{(E_{e_1}a_1-E_{e_2}a_2)^2}{a_1a_2}\,,
\end{align}
with $p_{e_1}\cdot p_{e_2}=E_{e_1}E_{e_2} \mathcal A_0$. We checked that the angular averages reproduce the coefficients of
the photon spectrum in Eq.~\eqref{eq:radchannels}
\begin{equation}
  \int\frac{\dd\Omega_\gamma}{4\pi}\,g(\hat n)=g_{-1}+E_\gamma\,g_0+E_\gamma^2\,g_1\,,
  \end{equation}
so that Eq.~\eqref{eq:radchannels} is reproduced once we integrate over the photon direction.
A convenient frame has the $z$ axis along $\hat p_1$ and $\hat p_2$ in the $xz$ plane, in which
\begin{equation}
  a_1=1-\beta_1\cos\theta_\gamma\,,\qquad
  a_2=1-\beta_2\left(y_{12}\cos\theta_\gamma+\sqrt{1-y_{12}^2}\,\sin\theta_\gamma\cos\phi_\gamma\right),
\end{equation}
with $\dd\Omega_\gamma=\dd\cos\theta_\gamma\,\dd\phi_\gamma$.

\section{Large-$\omega_n$ effective field theory}\label{sec:largeomega}
The calculations of the previous sections are lengthy because of the presence of heavy-nucleus loop integrals depending on several scales. As shown above, the analytic expressions simplify greatly in an expansion of leptons energies over $\omega_n$. In this approximation, it is possible to obtain the same result as Eq.~\eqref{eq:dGalpha} by explicitly integrating out the intermediate nuclear states. While it is not immediately obvious that this is a sensible approximation because intermediate nuclear excited $1^+$ states in, for example, ${}^{76}$Ge or ${}^{136}$Xe can have excitation energies of a few hundred keV \cite{Dekens:2024hlz}, so that $\omega_n \gtrsim E_0/2$ is comparable to the lepton energies, the calculation provides an important check of the full calculation. Furthermore, we show in Sec.~\ref{sec:validity} that the large-$\omega_n$ approximation nevertheless works well numerically when the relevant excitation energies are spread over several states. 

In the large-$\omega_n$ limit, the calculation of $2\nu\beta\beta$ radiative corrections is split into two parts. The first part consists of a matching calculation of the EFT described by Eqs.~\eqref{eq:HEFT_em} and \eqref{eq:weak} onto a lower-energy EFT containing only the ground-state nuclei, photons, and leptons. In a second step, the low-energy EFT, which we will call large-$
\omega_n$ EFT, is used to compute radiative corrections. This computation is now simplified because loop integrals no longer depend on the scale $\omega_n$.At leading order in the expansion in $E_e/\omega_n$, the large-$\omega_n$ EFT is described by the kinetic terms of the ground-state nuclei and leptons and by the two-nucleon four-lepton operators of Eq.~\eqref{eq:L2beta}. At tree level, their Wilson coefficients are
\begin{equation}
C^{(0)}_{2\beta} =
\sum_n \frac{g^{in}_A\,g^{fn}_A}{\omega_n}\,,\qquad C^{v,(0)}_{2\beta} =0\,,
\end{equation}
where the $(0)$ superscript denotes that this is the leading-order coefficient.

The one-loop matching coefficients are obtained by expanding the loop integrals from the previous sections in $E_e/\omega_n$, before actually performing the loop integrations. For example, at leading order in this expansion, the sum of diagrams 1-6 in Fig.~\ref{fig:virtual} depends on the integral
\begin{align}
 \int [\dd k]
     \left(\frac{1}{k^2}\right)^2 \frac{k^\mu}{v \cdot k + \omega_n} \,.
\end{align}
These integrals are easily evaluated using the master integrals \cite{Broadhurst:1991fz}
\begin{eqnarray} I_n(\Delta) \equiv \int [dk] \frac{1}{(k^2)^n}\frac{1}{v\cdot k-\Delta}\,, \qquad I_n(\Delta) = -2(\mu^2 e^{\gamma_E})^\varepsilon\frac{\Gamma(2n+1-d)\Gamma(d/2-n)}{(-1)^n\Gamma(n)(2\Delta)^{2n+1-d}}\,,
\end{eqnarray}
which are much simpler than the integrals from the previous sections. The explicit matching at a scale $\mu$ in dimensional regularization gives 
\begin{equation}\label{eq:Coneloop}
C^{(1)}_{2\beta}(\mu) = \sum_n \frac{g^{in}_A g^{fn}_A}{\omega_n}\,\frac{\alpha}{4\pi} \left(\ln \frac{4\omega_n^2}{\mu^2} -\frac{5}{2}\right)\,,\qquad
C^{v,(1)}_{2\beta}(\mu) = \sum_n \frac{g^{in}_A g^{fn}_A}{\omega_n}\,\frac{\alpha}{4\pi} \left(3 \ln \frac{4\omega_n^2}{\mu^2}-\frac{3}{2}\right)\,.
\end{equation}
At leading order in the large-$ \omega_n$ EFT this is the only non-analytic dependence on $\omega_n$ because all other contributions arise from the low-energy EFT without intermediate states which therefore cannot depend on $\omega_n$. The $\mathcal O(1/\omega_n)$ contribution to the differential decay rate depends on the interference
\begin{equation}
3 \mathrm{Re}( C^{(0)}_{2\beta} C^{(1)*}_{2\beta}) - \mathrm{Re}( C^{(0)}_{2\beta} C^{v,(1)*}_{2\beta})\,
\end{equation}
for which the $\ln \omega_n$ dependence cancels leaving only a constant piece. This result explains the universality of the double-weak Sirlin function. Since the one-loop matching does not introduce any new dependence on the intermediate states into the rate, the radiative corrections at this order depend on the nuclear structure only through $C^{(0)}_{2\beta}$ and thus through $M^{(-1)}_{GT}$. At next-to-leading order, the logarithms of $\omega_n$ no longer cancel. They appear weighted by $M_{GT}(n)/\omega_n^3$, which is precisely the structure of $LM^{(-3)}_{GT}$ in Eq.~\eqref{eq:LMGT}, and give rise to $\zeta_{31}$ (note that we have not checked whether the large-$\omega_n$ EFT reproduces the coefficient of the $\zeta_{31}$ term).

\begin{figure}
    \centering
    \includegraphics[width=\linewidth]{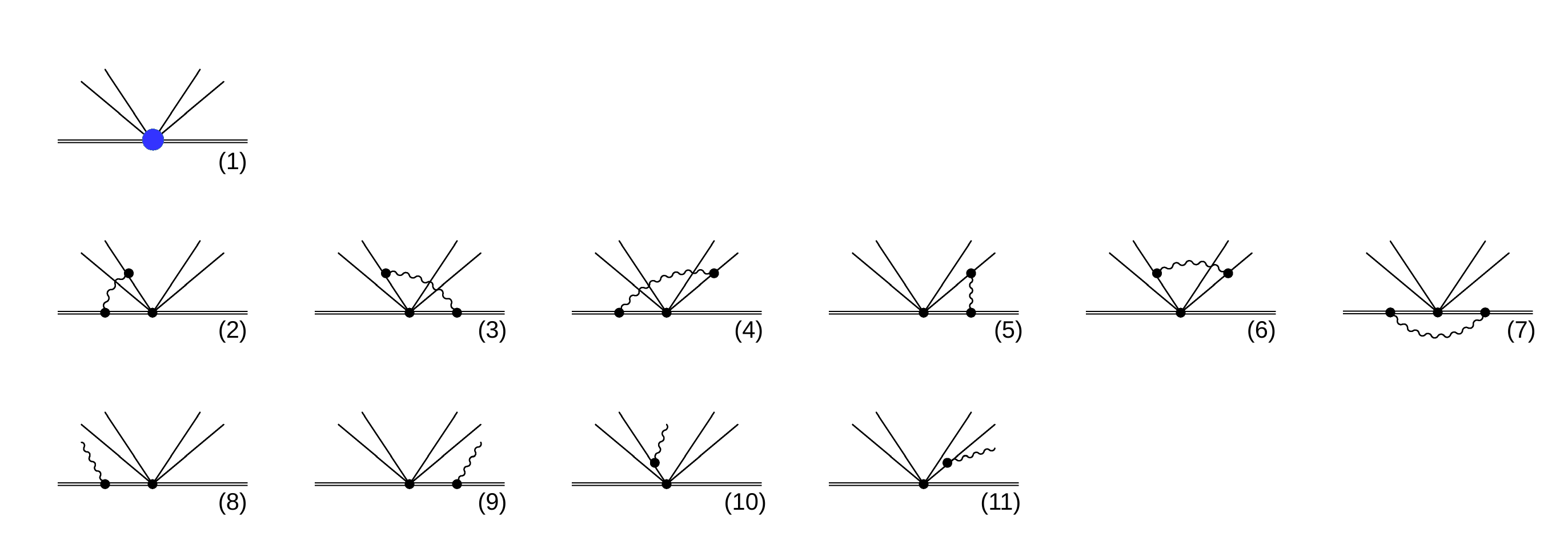}
    \caption{Tree level, virtual and real emission diagrams in the large-$\omega_n$ EFT. Double, plain and wavy lines denote the nuclear ground states, lepton and photons, respectively. A blue dot denotes an insertion of the matching coefficients $\bar{C}_{2\beta}$ and $\bar C^v_{2\beta}$ at $\mathcal O(\alpha)$. Black dots denote insertions from the LO lepton and ground-state Lagrangians, Eqs.~\eqref{eq:QED} and~\eqref{eq:HEFT_em}, and from Eq.~\eqref{eq:L2beta}, with tree-level coefficient $C^{(0)}_{2\beta}$.
    }
    \label{fig:largeomegafigs}
\end{figure}

In the large-$\omega_n$ EFT, the $\mathcal O(\alpha)$ corrections consist of the tree-level contributions of $C^{(1)}_{2\beta}$ and $C^{v,(1)}_{2\beta}$, and of one-loop and real emission diagrams proportional to $C^{(0)}_{2\beta}$.
The diagrams are shown in Fig.~\ref{fig:largeomegafigs}.
Diagram 1 represents an insertion of $C^{(1)}_{2\beta}$ and $C^{v,(1)}_{2\beta}$. 
Diagram 7 in  Fig.~\ref{fig:largeomegafigs}
is  the nuclear vertex correction, analogous to diagrams 8--11 in Fig.~\ref{fig:virtual}. The nuclear vertex correction now involves a scaleless integral and vanishes. Diagrams 2--5 contain a photon exchange between an outgoing electron and an initial- or final-state nucleus. The same diagrams appear in radiative correction to neutron $
\beta$-decay \cite{Ando:2004rk,Cirigliano:2022hob}
and involve the $\bar{C}_0(v,p_e;0,0,m_e)$ integral which contains an IR divergence and the function in Eq.~\eqref{Sirlinpart} that appears in the Sirlin function.  Diagram 6 in Fig.~\ref{fig:largeomegafigs}, associated to the electron-electron repulsion, is the analogue of diagram 7 in Fig.~\ref{fig:virtual} but
no longer contains a heavy-nucleus propagator and is thus a standard relativistic one-loop integral \cite{Ellis:2007qk}. The real emission, diagrams 8--11,  is computed as in the full theory, and the infrared divergences cancel in the same way. We have verified that the sum of all contributions reproduces the large-$\omega_n$ limit of the full calculation, Eq.~\eqref{eq:largew}, which provides a nontrivial check of the calculation.

Because all loop integrals are standard, the large-$\omega_n$ EFT is the natural starting point for extensions of this work, for instance to other observables or to other double-weak processes. 
In the next section we use it to compute radiative corrections to $0\nu\beta\beta$ decay.

\section{Radiative corrections to neutrinoless double beta decays}\label{sec:0nu}

As a further application of the methods developed in this paper, we compute
the $\mathcal{O}(\alpha)$ radiative corrections to neutrinoless double-beta
decay ($0\nu\beta\beta$). While these corrections are less phenomenologically
urgent than those to $2\nu\beta\beta$ because the $0\nu\beta\beta$ signal has not
yet been observed, they can be computed cleanly
within the same EFT framework and, to our knowledge, have not appeared in the
literature. Potentially relevant is that real-photon emission in $0\nu\beta\beta$ events generates a radiative tail below the
$0\nu\beta\beta$ endpoint because photons take away energy from the electrons. This affects the signal efficiency of experiments that search for a peak at the endpoint on top of the $2\nu\beta\beta$ continuum (Sec.~\ref{sec:endpoint}).

$0\nu\beta\beta$ shares the initial and final nuclear states as $2\nu\beta\beta$, and its two outgoing electrons are subject to the same electromagnetic interactions. The computation of ultrasoft radiative corrections therefore proceeds along very similar lines, but with an important simplification. The standard scenario, the exchange of a light Majorana neutrino, involves the exchange of a virtual neutrino momenta of order $|\vec q\,| \sim m_\pi \sim k_F \gg \mathcal{Q}$. The intermediate nuclear states are therefore already integrated out at the scale of the nuclear dynamics, and the process is accurately described by an EFT with only the ground-state nuclei as in Sec.~\ref{sec:largeomega}. This is reflected by the observation that closure corrections arising from ultrasoft neutrino exchange that do depend on intermediate nuclear states are subleading \cite{Cirigliano:2017tvr, Castillo:2024jfj}.

\subsection{Tree-level amplitude and decay rate}
\label{sec:0vbb_tree}

In the ground-state EFT, the $0\nu\beta\beta$ amplitude is generated by a single local operator coupling the initial and final nuclear ground states to two outgoing electrons. At leading order, the tree-level amplitude takes the form
\begin{equation}
\label{eq:0vbb_tree}
    i\mathcal{M}_{0\nu} = i\,c\,\bar u_{e_1}C P_R\bar u_{e_2}^{T}\,,
\end{equation}
where $C$ is the charge-conjugation matrix and $c$ is a matching coefficient that encodes all nuclear and short-distance physics. In the standard light Majorana neutrino exchange mechanism,
\begin{equation}
\label{eq:c_scaling}
    c =\frac{G_F^2 V_{ud}^2\,m_{\beta\beta}\,M_{0\nu}}{R_A}\,,
\end{equation}
where $m_{\beta\beta} = \sum_i U_{ei}^2 m_i$ is the effective neutrino Majorana mass, $M_{0\nu}$ is the $0\nu\beta\beta$ nuclear matrix element containing long- and short-distance neutrino-exchange pieces \cite{Cirigliano:2018hja}, and $R_A$ is the nuclear radius to compensate for a similar factor in the definition of the NME. Non-standard mechanisms induced by higher-dimensional lepton-number-violating operators \cite{Cirigliano:2017djv, Cirigliano:2018yza, Deppisch:2020ztt} can lead to a different Lorentz structure for the outgoing leptons and the calculation of radiative corrections  needs to be revisited. 

Squaring the amplitude and summing over electron spins gives
\begin{equation}
\label{eq:0vbb_Msq}
    |\mathcal{M}_{0\nu}|^2 = 2\,|c|^2 E_{e_1} E_{e_2}(1-\beta_1 \beta_2 y_{12})\,.
\end{equation}
 The tree-level differential decay rate, after integrating over the two-body phase space and including the Fermi functions, is
\begin{align}
\label{eq:0vbb_rate}
    \frac{\dd\Gamma^{(0)}_{0\nu}}{\dd\epsilon\,\dd\Delta\,\dd y_{12}} &=
    \frac{1}{\ln 2}\frac{|c|^2}{8\pi^3}\,
    E_{e_1}^2\,E_{e_2}^2\,\beta_1\,\beta_2\,
    F(Z+2,E_{e_1})\,F(Z+2,E_{e_2})(1-\beta_1 \beta_2 y_{12})
    \,
    \delta\!\left(\epsilon - \mathcal{Q}\right)\,,
\end{align}
where the superscript $(0)$ denotes the $\mathcal O(\alpha^0)$ result, and the $\delta$ function enforces the two-body kinematics, in which the electrons share the full $\mathcal Q$ value.

\subsection{Virtual and real radiative corrections}
\label{sec:0vbb_virtual}

The $\mathcal{O}(\alpha)$ virtual corrections to $0\nu\beta\beta$ arise from photon exchange between the two outgoing electrons and the initial and final nuclear lines. The diagram topology is analogous of that encountered in the large-$\omega_n$ EFT for $2\nu\beta\beta$ in Sec.~\ref{sec:largeomega}. The Feynman diagrams are the same as in Fig.~\ref{fig:largeomegafigs}, with   the only difference that the neutrino lines are absent.
The nuclear vertex correction is again scaleless and vanishes in dimensional regularization, the electron-nucleus triangle diagrams generate a contribution proportional to $Z+2$ already captured by the Fermi function as well as $Z$-independent pieces, while the electron-electron box generates the two-electron correlation term. A slight complication occurs due to the lepton bilinears where the Dirac algebra now involves the charge-conjugation matrix. 

As we found for $2\nu\beta\beta$, the infrared divergence of the virtual diagrams is proportional to the tree-level amplitude and cancels against the soft real emission. Combining virtual and soft real-emission contributions, the radiatively corrected rate is
\begin{align}
\label{eq:0vbb_RC}
    \frac{\dd\Gamma_{0\nu}}{\dd\epsilon\,\dd\Delta\,\dd y_{12}}
    \Bigg|_{\mathcal{O}(\alpha)}
    &=
    \frac{\dd\Gamma^{(0)}_{0\nu}}{\dd\epsilon\,\dd\Delta\,\dd y_{12}}
    \times
    \left(g_V^4 + \frac{\alpha}{2\pi}\,f_{0\nu}(\beta_1,\beta_2,\beta_{12})\right)+ \frac{\dd\Gamma^{(2)}_{0\nu}}{\dd\epsilon\,\dd\Delta\,\dd y_{12}}\,,
\end{align}
where $g_V$ is the vector coupling defined in Sec.~\ref{sec:EFT} and the radiative correction function is
\begin{align}
\label{eq:f0nu}
    f_{0\nu}(\beta_1,\beta_2,\beta_{12}) &= 6 + f_{t}^{2\nu}(\beta_1,\beta_2,\beta_{12})\,,
\end{align}
with $f_t^{2\nu}$ given in Eq.~\eqref{eq:dGammaVtot}, evaluated for $\epsilon = \mathcal Q$.
The real emission diagrams induce further corrections, in which the sum of electron energies is not constrained to the $\mathcal Q$ value. They are
\begin{equation}\label{eq:0vbb_real}
\frac{\mathrm{d}\Gamma^{(2)}_{0\nu}}{\mathrm{d}\epsilon\,\mathrm{d}\Delta\,\mathrm{d} y_{12}} = \frac{1}{\ln 2}\frac{|c|^2}{8\pi^3}\,\frac{\alpha}{2\pi}\,
    E_{e_1}^2\,E_{e_2}^2\,\beta_1\,\beta_2\,
    F(Z_f,E_{e_1})\,F(Z_f,E_{e_2}) \left\{
    \left[\frac{1}{\mathcal Q-\epsilon} \right]_+ g_{-1} + g_0 + (\mathcal Q-\epsilon)\, g_1
    \right\},
\end{equation}
where $g_i$  functions are defined in Eqs.~\eqref{eq:gm1}, \eqref{eq:g0} and \eqref{eq:g1}.
The plus distribution is defined by its action on a test function $\varphi(\epsilon)$,
\begin{equation}
    \int_0^{\mathcal Q} \mathrm{d}\epsilon\, \left[\frac{1}{\mathcal Q - \epsilon}\right]_+ \varphi(\epsilon) \equiv \int_0^{\mathcal Q} \mathrm{d} \epsilon\, \frac{\varphi(\epsilon)-  \varphi(\mathcal Q)}{\mathcal Q-\epsilon}\,,
\end{equation}
where, in the subtraction term, all kinematic variables, including the limits of the $\Delta$ integration, are evaluated at $\epsilon = \mathcal Q$.
The functions $f_{0\nu}$,
and $g_i$ closely resemble those appearing
in Eq.~\eqref{eq:largew} for the large $\omega_n$ expansion of the $2\nu\beta\beta$ rate.
Indeed, 
 $f_{0\nu}$ is very similar to the virtual and real soft pieces of Eq.~\eqref{eq:largew}, 
differing by them only by a constant.
$g_{-1,0,1}$ correspond to the coefficients of $\ln( \mathcal Q - \epsilon)$, $\mathcal Q - \epsilon$ and
$(\mathcal Q - \epsilon)^2$ terms in Eq.~\eqref{eq:largew}, respectively.

\subsection{Distributions in $\epsilon_{\text{tot}}$}

As in the case of $2\nu \beta\beta$, we can consider distributions with respect to the total energy of electrons and photons,
\begin{equation}
\label{eq:0vbb_RCtot}
    \frac{\mathrm{d}\Gamma_{0\nu}}{\mathrm{d}\epsilon_{\text{tot}}\,\mathrm{d}\Delta\,\mathrm{d} y_{12}}
    =
    \frac{\mathrm{d}\Gamma^{(0)}_{0\nu}}{\mathrm{d}\epsilon_{\text{tot}}\,\mathrm{d}\Delta\,\mathrm{d} y_{12}}
    \left(g_V^4 + \frac{\alpha}{2\pi}\,f_{0\nu}+ \frac{\alpha}{2\pi} f^{R}_{0\nu}   \right)\,,
\end{equation}
with the real-emission contribution
\begin{align}
  f^R_{0\nu} &= \frac{1}{\mathcal A_0}\Bigg[    \int_0^{\xi_{\text{max}}}  \frac{\mathrm{d}\xi}{\xi} \Bigg(
\frac{\bar{E}_{e_1}^2 \bar{E}_{e_2}^2 \bar{\beta}_1 \bar{\beta}_2
F(Z_f,\bar{E}_{e_1})  F(Z_f,\bar{E}_{e_2}) }{{E}_{e_1}^2 {E}_{e_2}^2 {\beta}_1 {\beta}_2 F(Z_f,{E}_{e_1})  F(Z_f,{E}_{e_2})}
g( \bar E_{e_1}, \bar E_{e_2}, y_{12}, \xi) -  g(E_{e_1},E_{e_2}, y_{12},0) \Bigg) \nonumber \\ & +
 g( E_{e_1}, E_{e_2}, y_{12} , 0) \ln \xi_{\text{max}} \Bigg]\,,
\end{align}
where $\xi_{\rm max} = (\mathcal Q - 2|\Delta|)/\mathcal Q$. As in Sec.~\ref{sec:etot}, barred quantities are evaluated in the real-emission kinematics, with $\epsilon = \mathcal Q - \mathcal Q \xi$ at fixed $\Delta$, while unbarred quantities have $\epsilon = \mathcal Q$.

\subsection{The radiative mode $0\nu\beta\beta+\gamma$}
\label{sec:0nugamma}
 
The real emission likewise gives the rate of the radiative mode $0\nu\beta\beta+\gamma$. In this case the sum of the electron energies is fixed to $\epsilon = \mathcal Q - E_\gamma$, and the photon energy spectrum is
\begin{equation}\label{eq:0nugamma}
\frac{\mathrm{d} \Gamma(0\nu\beta\beta+\gamma)}{\mathrm{d} E_\gamma} = \frac{1}{\ln 2}\frac{|c|^2}{8\pi^3}\,
 \,  \frac{\alpha}{2\pi}  \int^{(\mathcal Q - E_\gamma)/2}_{-(\mathcal Q-E_\gamma)/2} \mathrm{d} \Delta \int_{-1}^1 \mathrm{d} y_{12}        E_{e_1}^2\,E_{e_2}^2\,\beta_1\,\beta_2\,
    F(Z_f,E_{e_1})\,F(Z_f,E_{e_2})\,   \mathcal A_\gamma\,,
\end{equation}
with the same photon spectrum $\mathcal A_\gamma$ as for $2\nu\beta\beta$, Eq.~\eqref{eq:Agamma}, evaluated for electrons sharing the energy $\mathcal Q-E_\gamma$.
 
The same holds for the distribution in the photon direction. Normalized to the respective Born terms, the squared real-emission amplitudes of $2\nu\beta\beta$ and $0\nu\beta\beta$ decay coincide, so that the photon distribution in $0\nu\beta\beta$ decay is given by Eq.~\eqref{eq:radangle}, with $\mathcal N$ replaced by the tree-level $0\nu\beta\beta$ rate, Eq.~\eqref{eq:0vbb_rate}, differential in $\Delta$ and $y_{12}$ and divided by $\mathcal A_0$, evaluated for electrons sharing the energy $\mathcal Q-E_\gamma$.

\section{Results}\label{sec:results}

We now turn to the numerical impact of the radiative corrections. We first show that the large-$\omega_n$ expansion reproduces the full $\mathcal O(\alpha)$ correction to high accuracy. This leads to a simple expression for the radiatively corrected rate, Eq.~\eqref{totddecay2}, which we use in the remainder of this section. We then discuss the double-weak Sirlin function and the distortions of the energy and angular distributions, the photon emission rates, the bias on $\xi_{31}$ and its dependence on the detection of the photon, and the implications for $0\nu\beta\beta$ searches. Unless stated otherwise, all results include the Fermi functions, use $\mu = 2E_0$ with the corresponding values of $g_V$ in Table~\ref{tab:gV}, and the ratios of nuclear matrix elements of Table~\ref{tab:NMEratios}.

\subsection{The validity of the large-$\omega_n$ limit}
\label{sec:validity}

\begin{figure*}[t]
\centering
\includegraphics[width=0.85\linewidth]{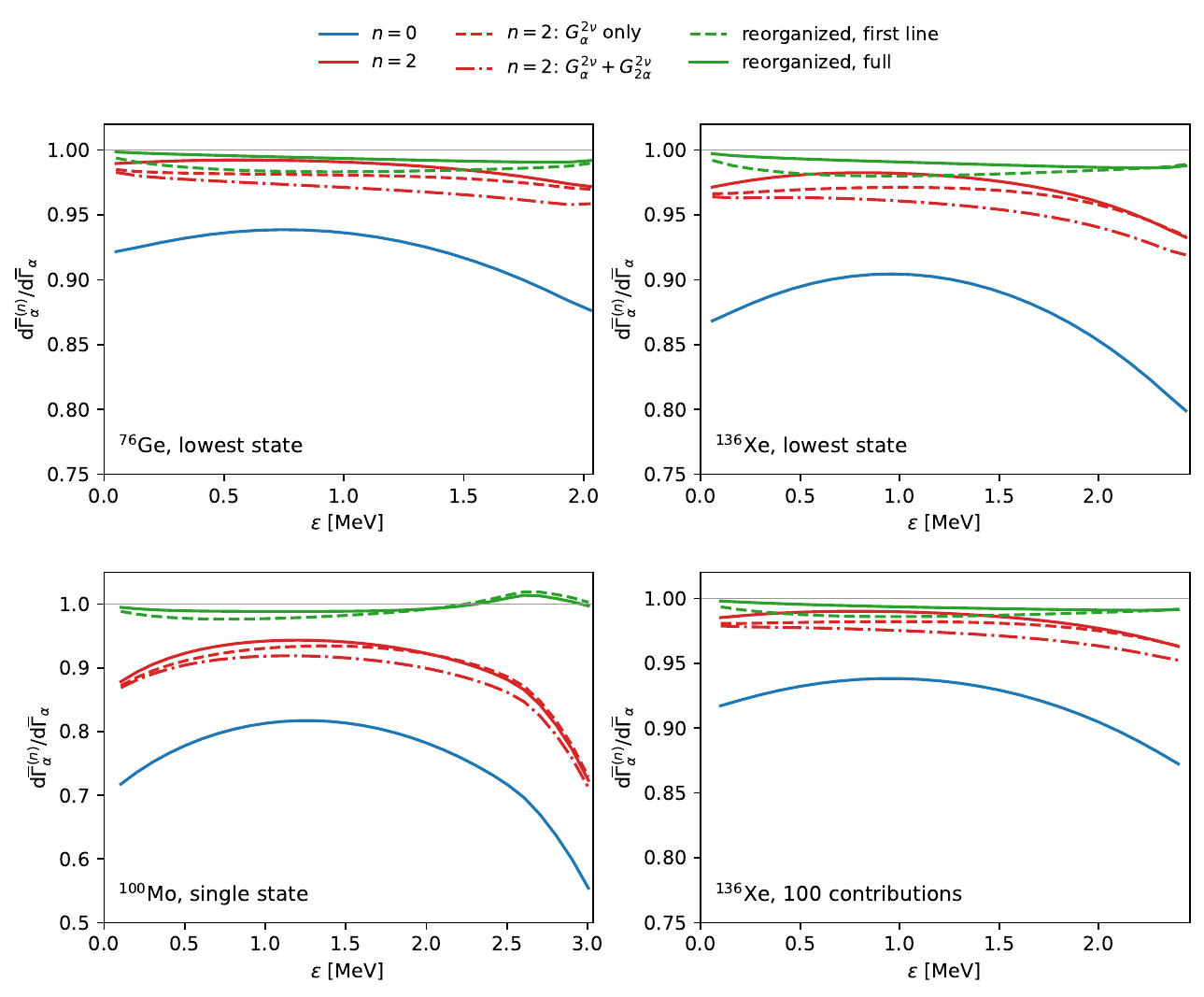}
\caption{Ratio of the $\mathcal O(\alpha)$ correction in the large-$\omega_n$ expansion to the full result, as a function of $\epsilon$, for a single intermediate state in $^{76}$Ge, $^{136}$Xe, and $^{100}$Mo, and for 100 contributions in $^{136}$Xe. Blue and red solid lines: expansion to order $n=0$ and $n=2$, with the tree-level spectrum expanded to the same order; red dashed and dot-dashed lines: partial $n=2$ results including only $G^{2\nu}_\alpha$, and $G^{2\nu}_\alpha+G^{2\nu}_{2\alpha}$. Green lines: the reorganized rate, Eq.~\eqref{totddecay2} (dashed), and including the corrections of Eq.~\eqref{totddecay2b} (solid).}
\label{fig:largeomega}
\end{figure*}

The large-$\omega_n$ expansion simplifies the radiative corrections considerably, but the lowest $1^+$ states of the intermediate nuclei have $\omega_n \gtrsim E_0/2$ of comparable size to the lepton energies and the success of the expansion is not clear. We therefore explicitly check the accuracy of the expansion. We denote by $\mathrm{d}\overline\Gamma_\alpha/\mathrm{d}\epsilon$ the $\mathcal O(\alpha)$ part of the spectrum computed with the full dependence on $\omega_n$, and by $\mathrm{d}\overline\Gamma{}^{(n)}_\alpha/\mathrm{d}\epsilon$ its expansion up to order $n$ in $1/\omega_n$, in which the tree-level spectrum is expanded to the same order. At $n=0$ this is the term proportional to $\mathrm{d}G^{2\nu}_\alpha$ in Eq.~\eqref{eq:dGalpha} while at $n=2$ it includes the terms of Eq.~\eqref{eq:dGalpha2}. Note that these quantities depend on the renormalization scale and are only used for the comparison in this subsection.

To compute the correction for generic $\omega_n$, we need the spectrum of the intermediate states and their Gamow--Teller matrix elements. We use the NSM results of Refs.~\cite{Dekens:2024hlz,Castillo:2025wyr} for $^{76}$Ge and $^{136}$Xe, and the single-state description of $^{100}$Mo of Sec.~\ref{sec:largeomega}, with $\omega_1 \simeq 1.69$ MeV. Fig.~\ref{fig:largeomega} shows the ratio $\mathrm{d}\overline\Gamma{}^{(n)}_\alpha/\mathrm{d}\overline\Gamma_\alpha$ as a function of $\epsilon$.
 To speed up the evaluation of the expressions with full $\omega_n$ dependence,  Fig.~\ref{fig:largeomega} was obtained setting the Fermi function to 1.

We first consider the contribution of the lowest intermediate state, which has the smallest $\omega_n$ and therefore provides the most stringent test. For $^{76}$Ge and $^{136}$Xe, with $\omega_1 = 2.03$ and $1.91$ MeV, the leading-order expansion ($n=0$) reproduces $88$--$94\%$ and $80$--$90\%$ of the full correction, respectively. The largest deviations occur near the endpoint. At $n=2$, the agreement improves to $97$--$99\%$ for $^{76}$Ge and $93$--$98\%$ for $^{136}$Xe.
The dashed and dot-dashed lines include only part of the $n=2$ result, corresponding to the functions $\dd G^{2\nu}_{\alpha}$
and $\dd G^{2\nu}_{2\alpha}$ in Eq.~\eqref{eq:dGalpha2}.

These partial results show that most of the $n=2$ correction is captured by the term proportional to $G^{2\nu}_\alpha$ which amounts to multiplying the $\xi_{31}$ term of the tree-level spectrum by the double-weak Sirlin function. The contributions of $G^{2\nu}_{2\alpha}$ and of the term proportional to $\zeta_{31}$ are of 1--2\% each and partially cancel. 

Of course, realistic spectra contain many intermediate states, with larger values of $\omega_n$ that improve the convergence.
The NSM calculation of Refs.~\cite{Dekens:2024hlz,Castillo:2025wyr} includes about 60 states for both Ge and Xe, with the tree-level matrix elements converging after summing 40 states for $^{136}$Xe. Computing the unexpanded results including the interference of 60$\times$60 states is computationally prohibitive. In the case of $^{136}$Xe, we identify 100 contributions that provide 95\% of the tree-level cross section, and keep only these contributions. Including these 100 contributions for $^{136}$Xe, the agreement is $87$--$94\%$ at $n=0$ and $96$--$99\%$ at $n=2$. 

The most demanding case is $^{100}$Mo, where a single low-lying state dominates and $\xi_{31}$ is large. Here the expansion of the $\mathcal O(\alpha)$ correction converges slowly, reproducing only $55$--$82\%$ of the full correction at $n=0$ and $72$--$94\%$ at $n=2$. The reason is the tree-level spectrum itself. The terms proportional to $\xi_{51}$ and $\xi_{31}^2$ are not small compared with those proportional to $\xi_{31}$, and truncating the tree-level spectrum at $n=2$ misses a sizable part of its shape.

These observations suggest that the radiative corrections largely factorize into the double-weak Sirlin function times the full tree-level spectrum. We therefore propose the reorganization
\begin{eqnarray}\label{totddecay2}
\frac{\mathrm{d}\Gamma_\alpha}{\mathrm{d}\epsilon\, \mathrm{d} \Delta\, \mathrm{d} y_{12}} &=& \frac{1}{2} \lambda^4 \left(M_{GT}^{(-1)} \right)^2   \left[\frac{\mathrm{d} G^{2\nu}_0}{\mathrm{d}\epsilon\, \mathrm{d}\Delta\, \mathrm{d} y_{12}  }+ \frac{\mathrm{d} G^{2\nu}_2}{\mathrm{d}\epsilon\, \mathrm{d}\Delta\, \mathrm{d} y_{12} }\,\xi_{31}  +\frac{\mathrm{d} G^{2\nu}_4}{\mathrm{d}\epsilon\, \mathrm{d}\Delta\, \mathrm{d} y_{12} } \left(\xi_{51} + \frac{1}{3}\xi_{31}^2\right) + \frac{\mathrm{d} G^{2\nu}_{22}}{\mathrm{d}\epsilon\, \mathrm{d}\Delta\, \mathrm{d} y_{12} }\,\frac{1}{3}\xi_{31}^2  + \ldots \right]\nonumber\\
&&\times 
\left( g_V^4 + \frac{\alpha}{2\pi} f_{2\nu} \right)\,.
\end{eqnarray}
This is simply the tree-level rate of Eq.~\eqref{totddecay} multiplied by $g_V^4 + \frac{\alpha}{2\pi} f_{2\nu}$, with $f_{2\nu}$ given in Eq.~\eqref{eq:largew}. The remaining terms of Eq.~\eqref{eq:dGalpha2} give the correction
\begin{equation}\label{totddecay2b}
\frac{\mathrm{d}\,\delta\Gamma_\alpha}{\mathrm{d}\epsilon \, \mathrm{d}\Delta\, \mathrm{d} y_{12} } =  \frac{1}{2} \lambda^4 \left(M_{GT}^{(-1)} \right)^2   \frac{\alpha}{2\pi}  \left[ \xi_{31}
\frac{\mathrm{d} G^{2\nu}_{2\alpha}}{\mathrm{d} \epsilon\, \mathrm{d} \Delta\, \mathrm{d} y_{12}  }
 + \zeta_{31} \frac{\mathrm{d} G^{2\nu}_{L\alpha}}{\mathrm{d} \epsilon\, \mathrm{d} \Delta\, \mathrm{d} y_{12}  }  \right].
\end{equation}
As shown by the green lines in Fig.~\ref{fig:largeomega}, Eq.~\eqref{totddecay2} reproduces the full $\mathcal O(\alpha)$ correction to within $2\%$ in all cases, even for $^{100}$Mo. Including Eq.~\eqref{totddecay2b} improves the agreement to about $1\%$. Eq.~\eqref{totddecay2} is the main result for the phenomenology in the remainder of this section. It is independent of the renormalization scale, requires no nuclear input beyond that of a tree-level analysis, and is straightforward to implement. Take the tree-level spectrum, including the terms proportional to $\xi_{31}$ and, if necessary, $\xi_{51}$, and multiply by  $g_V^4 + (\alpha/(2\pi)) f_{2\nu}$. The correction of Eq.~\eqref{totddecay2b} is very small and we quantify its impact on the spectral shape in Sec.~\ref{sec:dS}. Since the radiative corrections factorize to sufficient accuracy, Eq.~\eqref{totddecay2} can also be applied to tree-level spectra computed without the expansion in the lepton energies, such as those of Ref.~\cite{Ghinescu:2026mpj}, by multiplying them by $g_V^4 + \frac{\alpha}{2\pi}f_{2\nu}$.

To display the effect of the different corrections on the shape of the spectra, we follow Ref.~\cite{Morabit:2024sms} and use normalized distributions. For a variable $x = \epsilon$ or $y_{12}$, we define
\begin{equation}\label{eq:Sdef}
S^{(n)}_i(x) \equiv \frac{1}{\Gamma^{(n)}_i}\frac{\mathrm{d}\Gamma^{(n)}_i}{\mathrm{d}x}\,,\qquad
\delta S^{(n)}_i(x) \equiv \frac{S^{(n)}_i(x)}{S^{(0)}_0(x)} - 1\,,
\end{equation}
where $\Gamma^{(n)}_i$ is the total rate, the subscript $i=0$ denotes the tree-level rate and $i=\alpha$ the radiatively corrected rate of Eq.~\eqref{totddecay2}, and the superscript $n=0$ or $n=2$ indicates whether the terms proportional to $\xi_{31}$ are included. All distortions are thus defined relative to the leading-order tree-level spectrum $S^{(0)}_0$. Where the terms proportional to $\xi_{51}$ and $\xi_{31}^2$ are included, we state so explicitly. If the photon is absorbed, we use the distributions in $\epsilon_{\rm tot}$ of Sec.~\ref{sec:etot}, neglecting the small $\xi_{31}$-dependent radiative term.

\subsection{The double-weak Sirlin function and spectral distortions}\label{sec:dS}

From the success of the large-$\omega_n$ expansion we conclude that radiative corrections are essentially captured by the value of $g_V^4-1$ combined with the $f_{2\nu}$ function. We follow Ref.~\cite{deVries:2026udq} and define the dimensionless function $G_\alpha$ 
\begin{align}
\label{eq:Ga}
\frac{m_e^{10}}{\ln 2}\frac{(G_F V_{ud})^4}{8 \pi^7 m_e^2}    {G}_\alpha(\epsilon) \equiv  (g_V^4 - 1)\frac{\mathrm{d} G^{2\nu}_0}{\mathrm{d}\epsilon} + \frac{\alpha}{2\pi}\frac{\mathrm{d} G^{2\nu}_\alpha}{\mathrm{d}\epsilon}\,.
\end{align}
We compare this to the analogous function $\hat G_\alpha(\epsilon)$ for the approximation used in Ref.~\cite{Nitescu:2024tvj} which uses the sum of two single-$\beta$ decay Sirlin functions. 
$\hat G_\alpha(\epsilon)$ is defined in Eq.~\eqref{eq:sirlin0}.
The factors on the left-hand-side of Eq.~\eqref{eq:Ga} are only chosen to make $G_{\alpha}$ of $\mathcal O(1)$. The interpretation of the dimensionless function $G_\alpha(\epsilon)$, up to the normalization, is that of a ‘double-weak Sirlin function’. It captures the universal radiative-correction factor for double-weak decays at $\mathcal O(\alpha)$ and is renormalization scale independent.

\begin{figure}[t]
\centering
\includegraphics[width=0.6\linewidth]{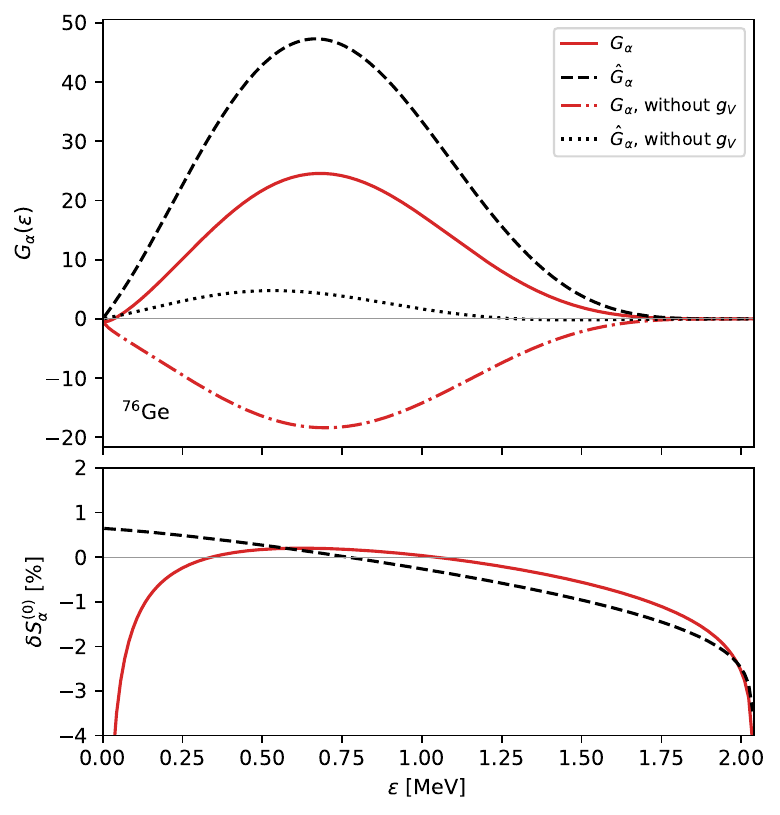}
\caption{Top: The double-weak Sirlin function $G_\alpha(\epsilon)$ for $^{76}$Ge (solid
red), compared with the approximation $\hat G_\alpha$ obtained from the sum of two
single-$\beta$ Sirlin functions (dashed black). The dot-dashed red and dotted black
curves exclude the radiative corrections to $g_V$. All curves are evaluated at $\mu=2E_0$. Bottom: spectral distortion
$\delta S^{(0)}_\alpha$ induced by the full correction (solid red) and by the
two-single-$\beta$ approximation (dashed black).}
\label{fig:Galpha}
\end{figure}

We depict the double-weak Sirlin function in Fig.~\ref{fig:Galpha} which updates Fig.~2 of Ref.~\cite{deVries:2026udq}. It compares $G_\alpha$ with $\hat G_\alpha(\epsilon)$ (the sum of single-$\beta$ Sirlin functions). In Ref.~\cite{deVries:2026udq} we made a typo in a sign in one of the terms in the single-$\beta$ Sirlin function\footnote{To be precise, the term $-2(2+\beta^2)$ in the published version of Ref.~\cite{deVries:2026udq} should read $-2(2-\beta^2)$.} and we corrected it here in Fig.~\ref{fig:Galpha}. The correction only affects the approximation $\hat G_\alpha$ and increases its maximum
by about 10$\%$ while the contribution of the two Sirlin functions
alone (dotted) becomes positive over almost the entire spectrum whereas in Ref.~\cite{deVries:2026udq} it changed sign. That being said, the conclusions are unchanged and the sum of two single-$\beta$ Sirlin
functions overestimates the total radiative correction by almost a factor of two. The spectral distortion has a different shape from that of the full calculation, being too large at low energies and too small in the central part of the spectrum.

\begin{figure*}[t]
\centering
\includegraphics[width=0.9\linewidth]{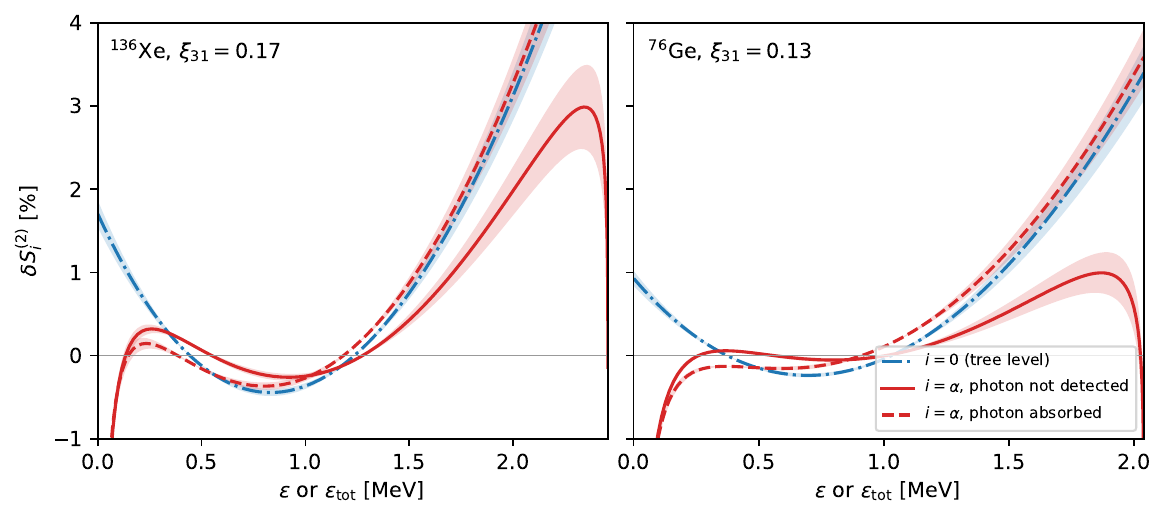}
\caption{Distortion of the normalized summed-energy spectrum with respect to the
leading-order spectrum, for $^{136}$Xe (left) and $^{76}$Ge (right), with NSM values
of $\xi_{31}$. Dot-dashed blue: tree level including $\xi_{31}$. Solid red: including
radiative corrections, photon not detected (spectrum in $\epsilon$). Dashed red:
including radiative corrections, photon absorbed (spectrum in $\epsilon_{\rm tot}$).
The bands correspond to a $\pm10\%$ variation of $\xi_{31}$.}
\label{fig:dS}
\end{figure*}

Fig.~\ref{fig:dS} shows the distortion of the normalized summed-energy spectrum
induced by $\xi_{31}$ and by the radiative corrections calculated here. If the photon is not detected we use Eq.~\eqref{totddecay2} at first order in
$\xi_{31}$. We also show what happens if we assume the photon energy is fully absorbed. If the photon is not detected, the radiative corrections partially cancel the distortion induced by $\xi_{31}$ in the high-energy part of the spectrum. This leads to the bias on $\xi_{31}$ which we discuss in more detail in Sec.~\ref{sec:xi31bias}. If
the photon is absorbed, the radiatively corrected spectrum follows the tree-level more closely and the bias is correspondingly smaller. In both cases a  significant negative distortion remains at the lowest energies. We discuss below how an experimental analysis should implement the radiative corrections in their template fits. 

\begin{figure*}[t]
\centering
\includegraphics[width=\linewidth]{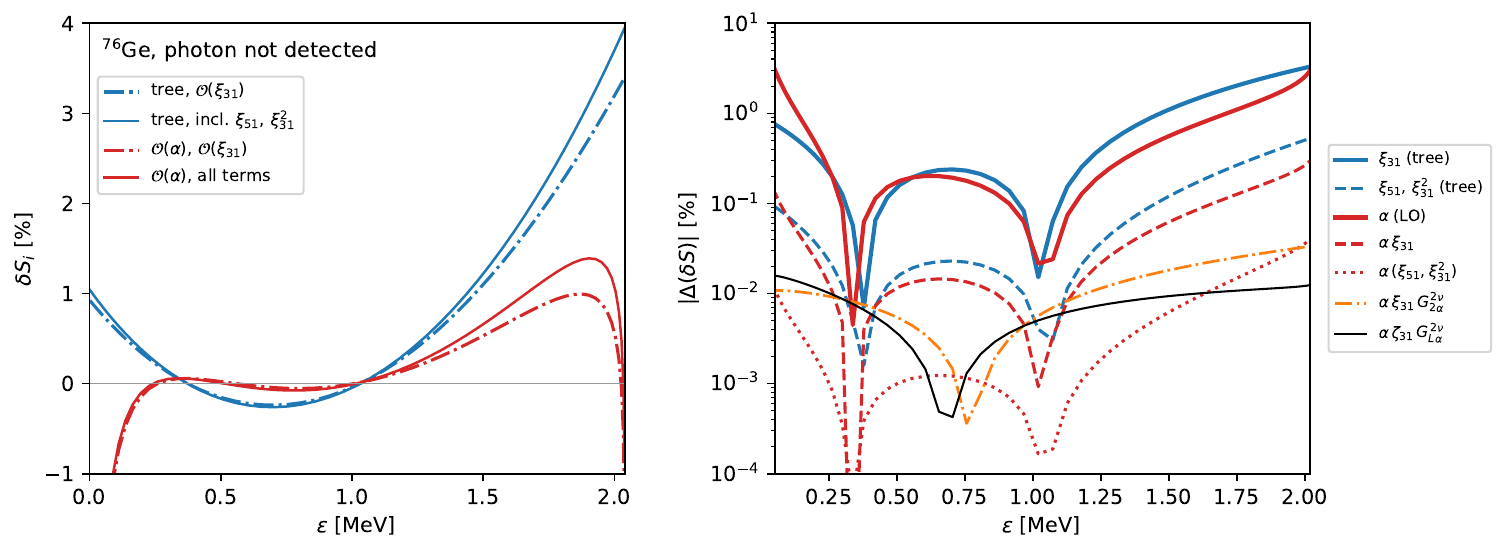}
\caption{Left: distortion of the normalized summed-energy spectrum of $^{76}$Ge at first order in $\xi_{31}$ (dot-dashed, as in Fig.~\ref{fig:dS}) and
including all terms of Eqs.~\eqref{totddecay2} and \eqref{totddecay2b} (solid), for the tree-level spectrum (blue)
and including radiative corrections (red). Right: size of the individual contributions,
shown as the change in $\delta S$ when each term is added. We use $\xi_{31}=0.13$, $\xi_{51}=0.022$ and $\zeta_{31}=-0.58$.}
\label{fig:higherorder}
\end{figure*}

So far we have kept only the terms of first order in $\xi_{31}$. Fig.~\ref{fig:higherorder}
shows, for $^{76}$Ge, the effect of the remaining terms of Eq.~\eqref{totddecay2}. They include the
tree-level contributions proportional to $\xi_{51}$ and $\xi_{31}^2$, the radiative
corrections multiplying them, and the terms proportional to $G^{2\nu}_{2\alpha}$ and to the
new nuclear matrix element $\zeta_{31}$ in Eq.~\eqref{totddecay2b}. 

The left panel shows that the higher-order terms shift the tree-level and the radiatively
corrected spectra by similar amounts, so that the relative effect of the radiative
corrections is essentially unchanged with respect to Fig.~\ref{fig:dS}. The right panel
displays the hierarchy of the individual contributions. The leading effects of $\xi_{31}$
and of the radiative corrections are of comparable size, reaching a few percent towards the endpoint. This confirms that radiative corrections are important to extract the correct value of $\xi_{31}$. Among the subleading terms, the tree-level contributions of $\xi_{51}$ and
$\xi_{31}^2$ are the largest, about $0.4\%$ at $\epsilon=1.9$ MeV. They are followed by the
$\mathcal O(\alpha\,\xi_{31})$ term of Eq.~\eqref{totddecay2} ($0.2\%$), while the
radiative corrections to the $\xi_{51}$ and $\xi_{31}^2$ terms and the
$G^{2\nu}_{2\alpha}$ term contribute at the $10^{-4}$ level. The contribution of
$\zeta_{31}$ remains around $10^{-4}$ over the whole spectrum. We conclude that in practice the radiative corrections are effectively captured by Eq.~\eqref{totddecay2}. This amounts to multiplying the tree-level
spectrum, including the $\xi^2_{31}$ and $\xi_{51}$ corrections, by $g_V^4+\frac{\alpha}{2\pi}f_{2\nu}$, which requires no nuclear input beyond that already used in tree-level analyses and is explicitly independent of the renormalization scale $\mu$. 

\subsection{Angular Distributions}
\label{sec:angular}

\begin{figure*}[t]
\centering
\includegraphics[width=0.9\linewidth]{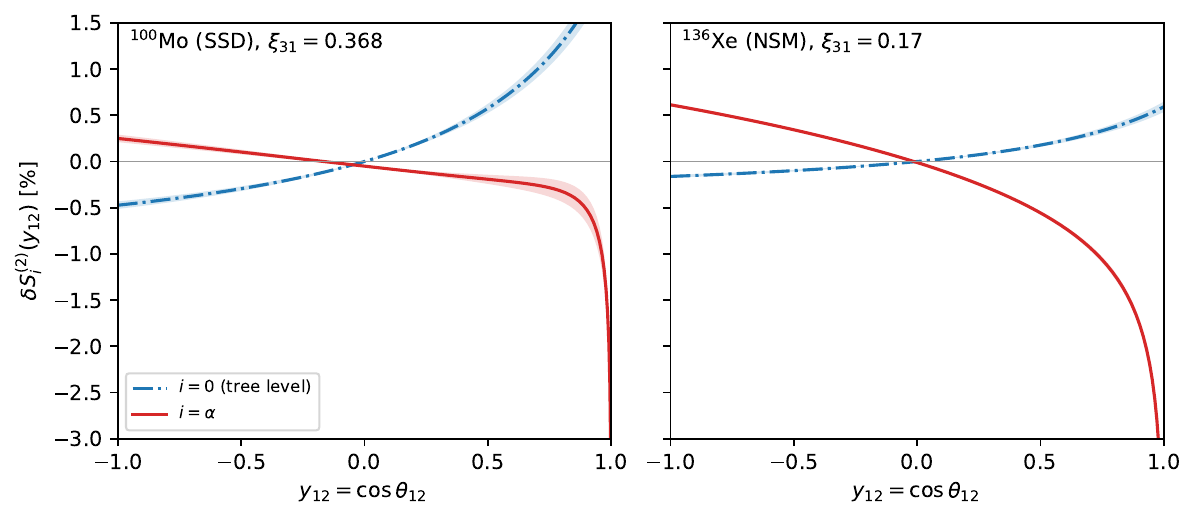}
\caption{Interplay between radiative corrections and $\xi_{31}$ for the angular distribution in $^{100}\mathrm{Mo}$ (left, $\xi_{31}=0.368$) and $^{136}$Xe (right,
$\xi_{31}=0.17$). The bands correspond to a $\pm10\%$ variation of $\xi_{31}$.}
\label{fig:angular}
\end{figure*}

The angular correlation between the two outgoing electrons can be measured by tracking experiments \cite{NEMO-3:2019gwo,NEXT:2025yqw}
and provides information that is complementary to the energy spectra. The possible absorption of the
photon changes the measured energy but not the electron directions. As such, the angular
distribution integrated over all electron energies is the same in both cases and therefore probes the radiative corrections independently of the
detector treatment of the photon. Fig.~\ref{fig:angular} shows the distortion for
$^{100}$Mo and $^{136}$Xe.

The radiative corrections tilt the distribution towards back-to-back electrons in the opposite direction to
the tilt induced by $\xi_{31}$. For $^{100}$Mo, with its large $\xi_{31}$, the two effects
largely compensate. For $^{136}$Xe, where $\xi_{31}$ is
smaller, the radiative tilt dominates. In addition, the
distribution is suppressed for nearly collinear electrons, by $4\%$ at $y_{12}=0.99$. This suppression arises from the Coulomb
interaction between the two electrons in the electron-electron box diagram and is a genuinely
double-weak effect without an analogue in single-$\beta$ decay. Our calculations show that
the angular distribution alone is not a reliable probe of $\xi_{31}$ unless radiative
corrections are included.

\subsection{Photon emission rate}
\label{sec:photonrate}

\begin{figure}[t]
\centering
\includegraphics[width=0.6\linewidth]{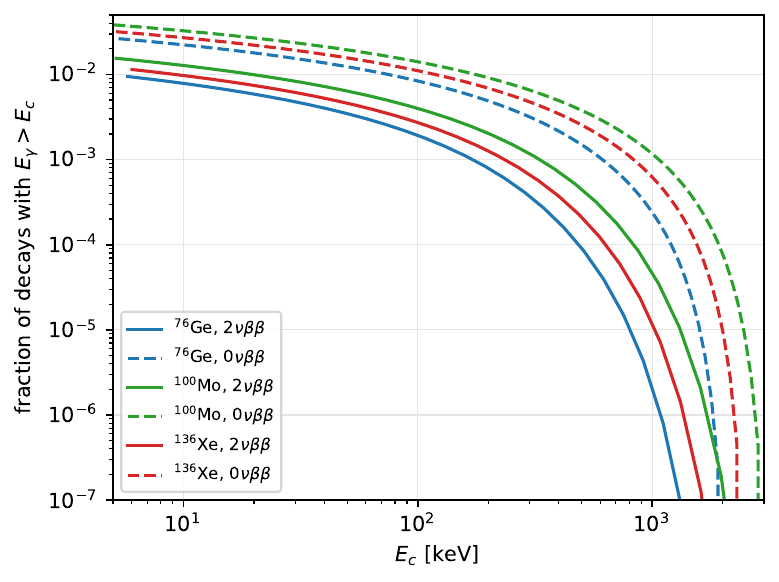}
\caption{Fraction of decays with a real photon of energy above $E_c$, normalized to the
tree-level rate, for $2\nu\beta\beta$ (solid) and $0\nu\beta\beta$ (dashed). For a given
$E_c$, this is the fraction of events removed by cuts rejecting a separate energy deposit
of a photon harder than $E_c$, and the rate of $2\nu\beta\beta+\gamma$
($0\nu\beta\beta+\gamma$) events with an observable photon above $E_c$. We do not show ${}^{130}$Te but the curves almost overlap with those of ${}^{136}$Xe.}
\label{fig:photonBR}
\end{figure}

The real-emission contributions computed in this work also predict the rate of the
radiative modes $2\nu\beta\beta+\gamma$ and $0\nu\beta\beta+\gamma$. The photon spectrum is
infrared divergent and the physically meaningful quantity is the fraction of decays in
which a photon is emitted with energy above a certain threshold $E_c$. We compute
\begin{equation}
  \mathrm{BR}(E_\gamma>E_c) = \frac{1}{\Gamma_{\rm tree}}\int_{E_c}\dd E_\gamma\,
  \frac{\dd\Gamma(\beta\beta+\gamma)}{\dd E_\gamma}\,,
\end{equation}
with the differential rates of Eqs.~\eqref{eq:radchannels} and \eqref{eq:0nugamma}
and normalized to the tree-level rate $\Gamma_{\rm tree}$. Normalizing instead to the
radiatively corrected rate changes these fractions by a relative amount of order $5\%$ which amounts to a higher-order correction in $\alpha$. At leading order in $1/\omega_n$, the fractions are
independent of the nuclear matrix elements for $2\nu\beta\beta$ and $0\nu\beta\beta$ and for the latter they are in addition independent of the actual value of the neutrino Majorana mass $m_{\beta\beta}$.

The results are shown in Fig.~\ref{fig:photonBR} as function of the photon energy in keV. For $E_c=30$, $100$, and $300$ keV, a photon with $E_\gamma> E_c$ is emitted in $0.48\%$,
$0.19\%$, and $0.04\%$ of $^{76}$Ge $2\nu\beta\beta$ decays,  in $0.62\%$, $0.27\%$, and
$0.07\%$ of $^{136}$Xe decays, and $0.84\%$. $0.4\%$, and $0.12\%$ for ${}^{100}$Mo. For $0\nu\beta\beta$ the corresponding fractions are larger,
for instance $0.84\%$ ($^{76}$Ge) and $1.11\%$ ($^{136}$Xe) for $E_c=100$ keV. This happens because the electrons carry the full $\mathcal Q$ value and therefore radiate more than in
$2\nu\beta\beta$. For both modes the fraction
grows with $\mathcal Q$ and decreases steeply once $E_c$ approaches its maximum value.

These fractions are the basic input for the photon-related effects discussed below. For $2\nu\beta\beta$, they give the fraction of events that are
removed if a photon above $E_c$ leads to the rejection of the event, which we use in
Sec.~\ref{sec:xi31bias_photon} to study how the potential bias on $\xi_{31}$ from neglecting radiative corrections depends on the detection of
the photon. For $0\nu\beta\beta$, they determine the loss of signal efficiency discussed in
Sec.~\ref{sec:endpoint}. Finally, they set the rate of
the radiative modes themselves. In detectors with sufficient spatial resolution, photons
above roughly $100$ keV deposit their energy away from the electron tracks, so that
$2\nu\beta\beta+\gamma$ events could in principle be identified as events with an additional
energy deposit. For an experiment recording $10^6$ $2\nu\beta\beta$ decays, this corresponds
to roughly $10^3$ events with a photon above $100$ keV. A more quantitative assessment requires
a dedicated detector simulation, for which the energy and angular distributions given in this work can provide the input. 

\subsection{Bias induced on $\xi_{31}$}
\label{sec:xi31bias}

We now study how the radiative corrections calculated above can affect the determination of
the nuclear structure parameters such as $\xi_{31}$. In this section, for simplicity we only keep
terms linear in $\xi_{31}$ and ignore higher-order terms proportional to $\xi^2_{31}$, $\xi_{51}$, 
and $\zeta_{31}$ but we can straightforwardly extend the method. Experimental collaborations determine $\xi_{31}$ by fitting the measured
summed-electron energy spectrum to a theoretical template whose normalization is
left free \cite{KamLAND-Zen:2019imh, CUPID-Mo:2023lru,CUORE:2025xue, PandaX:2025yly}. To quantify the effect of neglecting the radiative corrections, we fit
the tree-level template to the radiatively corrected spectrum and determine the value
$\xi_{31}^{\rm eff}$ that the fit returns. In this section we assume that the photon escapes undetected, so that the spectrum is that of
$\epsilon = E_{e_1}+E_{e_2}-2m_e$ inclusive over the photon phase space. In the next subsection we discuss the case
in which (part of) the photon energy is deposited in the detector.

We define $T_{0,2}(\epsilon)$ to be the integrals of $\dd G^{2\nu}_{0,2}$ over $\Delta$ and
$y_{12}$, and $R_{0,2}(\epsilon)$  as the same integrals with $\dd G^{2\nu}_{0,2}$ multiplied
by $f_{2\nu}$ of Eq.~\eqref{eq:largew} and we include the Fermi functions everywhere.
To linear order in $\xi_{31}$, the true spectrum is then given by
\begin{equation}
  d(\epsilon) = g_V^4\left[T_0+\xi_{31}T_2\right]
    +\frac{\alpha}{2\pi}\left[R_0+\xi_{31}R_2\right],
\end{equation}
while the tree-level template is $t(\epsilon)=A\,T_0+B\,T_2$, with $\xi_{31}^{\rm eff}=B/A$.

We aim to assess the potential bias on $\xi_{31}$ from ignoring radiative corrections without doing a dedicated experimental analysis. We use an Asimov data set \cite{Cowan:2010js}, that is we fit the expected numbers of events. This is the unbinned, continuum limit of the binned likelihood analysis used in Ref.~\cite{deVries:2025hqa} to assess the sensitivity of $2\nu\beta\beta$ spectra to new physics. For large numbers of events, the Poisson likelihood ratio for Asimov data reduces to the
 $\chi^2$
\begin{equation}
  \chi^2(A,B) = \frac{N}{D}\int_{\epsilon_{\min}}^{\mathcal Q}\dd\epsilon\,
    \frac{\left[d(\epsilon)-A\,T_0(\epsilon)-B\,T_2(\epsilon)\right]^2}{d(\epsilon)}\,.
  \label{eq:asimovchi2}
\end{equation}
Here $D$ is the un-normalized integral $D=\int_{\epsilon_{\min}}^{\mathcal Q}\dd\epsilon\,d(\epsilon)$ and $N$ is the number of events in the fit window defined by the threshold energy $\epsilon_{\min}$. The best fit values of $A$ and $B$ to the Asimov data are independent of $N$. Because the assumed template is linear in $A$ and $B$ the minimum follows from solving
\begin{equation}
  \sum_{j=0,2}M_{ij}\,\theta_j = v_i,\quad
  M_{ij}=\int\dd\epsilon\,\frac{T_iT_j}{d},\quad
  v_i=\int\dd\epsilon\,T_i,
\end{equation}
with $(\theta_0,\theta_2)=(A,B)$ and all integrals are assumed over the fit window.
The covariance matrix of $(A,B)$ is $(D/N)\,M^{-1}$ and the statistical uncertainty on
$\xi_{31}^{\rm eff}$ then follows from
\begin{equation}
  \sigma_\xi=\frac{c}{\sqrt N},\qquad
  c^2 = D\,g^TM^{-1}g,\qquad g=\left(-\frac{B}{A^2},\,\frac{1}{A}\right).
\end{equation}
The bias $\Delta\xi=\xi_{31}^{\rm eff}-\xi_{31}$ equals the statistical uncertainty for
$N^*=(c/\Delta\xi)^2$ events.

\begin{table*}[t]
\centering
\begin{tabular}{llcccccc}
\hline\hline
isotope & $\xi_{31}$ & $\epsilon_{\min}$ [MeV] & $\xi_{31}^{\rm eff}$ & $\Delta\xi_{31}$ & $\Delta\xi/\xi_{31}$ & $c$ & $N^*$\\
\hline
$^{76}$Ge  & 0.13 (NSM)  & 0   & 0.015  & $-0.115$ & $-89\%$  & 42 & $1.3\times10^5$ \\
 & & 0.5 & 0.052 & $-0.078$ & $-60\%$ & 41 & $2.7\times10^5$\\
           &             & 1.0 & 0.060 & $-0.070$ & $-54\%$ & 35 & $2.5\times10^5$\\
           & 0.10 (QRPA) & 0   & $-0.014$ & $-0.114$ & $-114\%$ & 42 & $1.3\times10^5$\\
           &  & 0.5 & 0.023 & $-0.077$ & $-77\%$ & 40 & $2.8\times10^5$\\
           &             & 1.0 & 0.031 & $-0.069$ & $-69\%$ & 35 & $2.5\times10^5$\\
$^{136}$Xe & 0.17 (NSM)  & 0   & 0.082  & $-0.088$ & $-52\%$  & 30 & $1.2\times10^5$\\           
 &   & 0.5 & 0.098 & $-0.072$ & $-42\%$ & 30 & $1.8\times10^5$\\
           &             & 1.0 & 0.107 & $-0.063$ & $-37\%$ & 25 & $1.6\times10^5$\\
           & 0.39 (QRPA) & 0   & 0.288  & $-0.102$ & $-26\%$  & 34 & $1.1\times10^5$\\
           & & 0.5 & 0.307 & $-0.083$ & $-21\%$ & 34 & $1.7\times10^5$\\
           &             & 1.0 & 0.317 & $-0.073$ & $-19\%$ & 29 & $1.6\times10^5$\\
\hline\hline
\end{tabular}
\caption{Bias on $\xi_{31}$ from fitting the tree-level template to the radiatively
corrected spectrum assuming that the photon escapes undetected. The
statistical uncertainty is $\sigma_\xi=c/\sqrt N$ for $N$ events above $\epsilon_{\min}$,
and $N^*$ is the number of events for which $|\Delta\xi_{31}|=\sigma_\xi$.}
\label{tab:xi31bias}
\end{table*}

The results are shown in Table~\ref{tab:xi31bias} for the NSM and QRPA
values of $\xi_{31}$. The radiative corrections shift the fitted value by
$\Delta\xi\simeq-0.06$ to $-0.08$, nearly independently of the true $\xi_{31}$ such that the relative bias is larger for smaller values of $\xi_{31}$. The bias becomes comparable to the
statistical uncertainty for $N\sim 10^5$ events and grows as $\sqrt N$. Because the $3\sigma$
determination of $\xi_{31}\simeq0.17$ requires already $N\gtrsim3\times10^5$, the bias
is relevant as soon as $\xi_{31}$ becomes measurable in the first place. 
If we lower or even remove the energy threshold the
bias grows because the radiative corrections are more important at 
low $\epsilon$. A realistic analysis of course would include backgrounds and detector
systematics  which increases $\sigma_\xi$ and $N^*$.

The idealized analysis of this section shows that the potential bias due to not considering radiative corrections can significantly impact the obtained values of $\xi_{31}$. However, the impact depends on the experimental setup, and in particular on whether the photon energy is (partially) combined with the electron energies as is the case for calorimetric detectors.

\subsubsection{Dependence on photon detection}
\label{sec:xi31bias_photon}

The analysis of the previous subsection assumes that the photon escapes undetected. In calorimetric detectors, such as germanium detectors, bolometers, and
liquid-xenon or liquid-scintillator detectors, the photon is usually deposited in the detector  and the effective measured quantity is
$\epsilon_{\text{tot}}=\epsilon+E_\gamma$. In the following we refer to this case as the photon being absorbed, meaning that its full energy is deposited, through whatever cascade of interactions, and added to that of the electrons. 
We  repeat the analysis of
Sec.~\ref{sec:xi31bias} with the $\epsilon_{\text{tot}}$ distributions of
 Eqs.~\eqref{eq:dGammaVtot} and \eqref{eq:dGammaRtot}, for two
scenarios 
\begin{itemize}
\item \emph{photon absorbed}: the photon energy is always included in the measured energy.
\item \emph{rejection above $E_c$}: photons with $E_\gamma<E_c$ are absorbed together
  with the electrons, while events with harder photons are removed. This is a simple model for detectors with good spatial resolution, in which a sufficiently hard photon deposits its energy far enough from the electrons to be reconstructed as a separate site. Selections based on the site multiplicity are a primary tool to distinguish $\beta\beta$ decays from $\gamma$-ray backgrounds, so that events with such a photon are rejected. Photons from the bremsstrahlung of the electrons in the detector material, which affect the site multiplicity in the same way, are already included in standard simulation tools such as Geant4, but the internal bremsstrahlung of the decay itself, computed here, is not. The fraction of events with a photon above $E_c$ is shown in Fig.~\ref{fig:photonBR}. In the calculation, this scenario is implemented by replacing $\xi_{\text{max}}$ with
  $\min(\xi_{\text{max}},E_c/\mathcal Q)$ in Eq.~\eqref{eq:dGammaRtot}, both in the upper limit of the integral and in the logarithm. A realistic description requires the detector simulation of each experiment, as discussed at the end of this section.

\end{itemize}

The results are shown in Table~\ref{tab:xi31bias_photon}. Since the $\epsilon_{\rm tot}$ distributions are given at leading order in $1/\omega_n$, we neglect the small $\xi_{31}$-dependent radiative term which changes $\Delta\xi$ by about $0.01$ with respect to Table~\ref{tab:xi31bias}. When the photon is absorbed, the
radiative distortion of the spectrum above $0.5$ MeV is small and the resulting bias is tiny
$|\Delta\xi|\lesssim0.01$.  If we remove the energy threshold, a sizable bias remains even in the absorbed case, $\Delta\xi=-0.05$ ($^{76}$Ge)
and $-0.03$ ($^{136}$Xe) which is driven by the radiative distortion of the spectrum below
$\epsilon_{\text{tot}}\simeq0.3$ MeV which does not depend on the photon energy.

\begin{table*}[t]
\centering
\begin{tabular}{lccc@{\hspace{2em}}ccccc}
\hline\hline
 & \multicolumn{3}{c}{$^{76}$Ge ($\xi_{31}=0.13$)} & \multicolumn{3}{c}{$^{136}$Xe ($\xi_{31}=0.17$)}\\
$\epsilon_{\min}$ [MeV] & 0 & 0.5 & 1.0 & 0 & 0.5 & 1.0 \\
\hline
photon not detected     & $-0.107$ & $-0.070$ & $-0.063$ & $-0.077$ & $-0.061$ & $-0.053$ \\
rejection, $E_c=30$ keV  & $-0.087$ & $-0.060$ & $-0.040$   & $-0.062$ & $-0.056$ & $-0.039$ \\
rejection, $E_c=100$ keV & $-0.073$ & $-0.031$ & $-0.024$   & $-0.052$ & $-0.032$ & $-0.024$ \\
rejection, $E_c=300$ keV & $-0.059$ & $-0.005$ & $-0.009$   & $-0.041$ & $-0.011$ & $-0.010$ \\
photon absorbed          & $-0.049$ & $+0.010$ & $+0.002$   & $-0.029$ & $+0.006$ & $+0.002$ \\
\hline\hline
\end{tabular}
\caption{Bias $\Delta\xi=\xi_{31}^{\rm eff}-\xi_{31}$ for different assumptions on the
detection of the real photon, using NSM values of $\xi_{31}$. }
\label{tab:xi31bias_photon}
\end{table*}

The rejection scenario interpolates between these results and shows that the bias is
generated mainly by photons with energies of tens to hundreds of keV. If we reject events
with a photon above $30$ keV, only a small fraction, we are left with about $90\%$ of the
bias of the photon-not-detected case, while for $E_c=300$ keV the bias is reduced to the level of
the absorbed case. The relevant property of an experiment is therefore the photon energy
above which an event is rejected or loses part of its energy. A quantitative statement for a given experiment requires including the
fully differential rate of Eq.~\eqref{eq:dGammaRtot} including the photon direction and pass all events through the full detector simulation and event selection. 

To end this section, we briefly discuss existing determinations of $\xi_{31}$. The analyses of
KamLAND-Zen~\cite{KamLAND-Zen:2019imh} ($^{136}$Xe), CUPID-Mo~\cite{CUPID-Mo:2023lru} ($^{100}$Mo), and
PandaX-4T~\cite{PandaX:2025yly} ($^{136}$Xe) fit the measured spectra with tree-level
templates. All three detectors are calorimetric, so that the photon is largely absorbed,
but the event selections (single-crystal events in CUPID-Mo, single-site events in
PandaX-4T) remove part of the events in which the photon deposits its energy elsewhere in
the detector. For PandaX-4T where the fit extends to small energies (down to 20 keV), this means that the bias on $\xi_{31}$ is in the range $\{-0.03,\,-0.06\}$ with a more precise statement requiring a detector simulation. The measurement reported a value 
$\xi_{31} = 0.59^{+0.41}_{-0.38}$ from $5.5\cdot 10^4$ $2\nu\beta\beta$ events and the bias is still well below the current uncertainty, but considering that the bias is a systematic effect and the statistical uncertainty will decrease in the future, it should be considered in future analyses. 

We also briefly comment on the CUORE analysis of $^{130}$Te~\cite{CUORE:2025xue} which includes radiative corrections in the form of the two single-$\beta$ Sirlin functions of
Ref.~\cite{Nitescu:2024tvj} obtaining $\xi_{31}=0.01_{-0.01}^{+0.31}$. Besides missing the genuinely double-weak contributions discussed
in this work, this correction describes the spectrum of the electrons alone and thus assumes
that the photon escapes undetected. In a bolometric detector most photons are instead absorbed in which case the radiative distortion largely
cancels. Applying the single-$\beta$ correction then introduces a distortion of the template
that is not present in the data and shifts the extracted $\xi_{31}$ upwards. Precise numbers again require a dedicated analysis.

\subsection{The endpoint region and $0\nu\beta\beta$ searches}
\label{sec:endpoint}

We first investigate the role of radiative corrections on the total inclusive $0\nu\beta\beta$ rate by integrating Eq.~\eqref{eq:0vbb_RC}. We divide the contributions into 3 parts arising from modifications of $g_V$ due to radiative corrections, from the virtual and soft corrections obtained in $f_{0\nu}$, and from the real emission of photons. The separate pieces depend on the renormalization scale $\mu$ but the sum is physical. We give the relative changes with respect to the tree-level decay rate in Table.~\ref{tab:0nutotal} for $\mu = 2E_0$. For both ${}^{76}$Ge and ${}^{136}$Xe the total decay rate grows by $5.2\%$. While this is minor in light of the overall uncertainties of the nuclear matrix elements, we stress that the increase is independent of them and on the effective Majorana neutrino mass $m_{\beta\beta}$ and the effect can be easily included in the phase-space factor.

\begin{figure}[t]
   \includegraphics[width=0.59\linewidth]{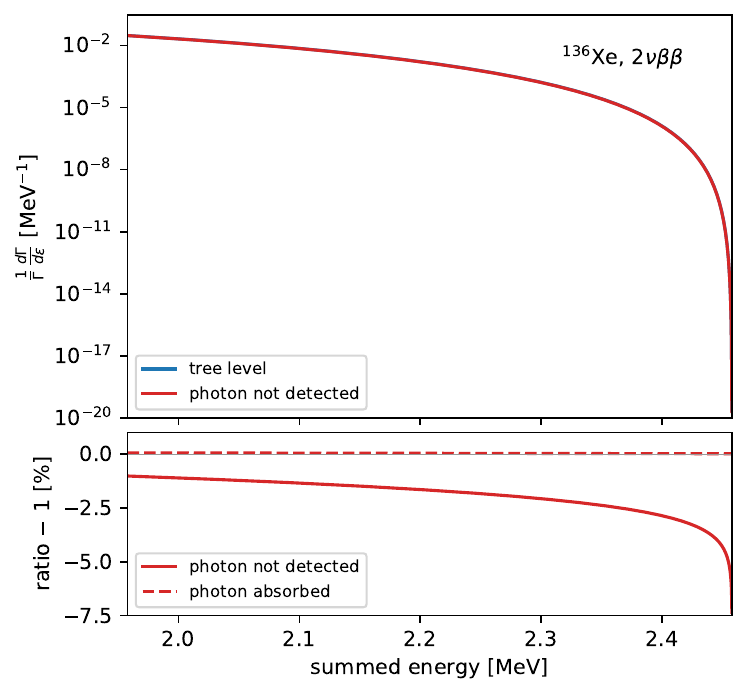}
   \caption{Top: normalized $2\nu\beta\beta$ spectrum of $^{136}$Xe near the endpoint at tree level and including radiative corrections. Bottom: relative change with respect to tree level, for a photon that escapes (solid) and that is absorbed (dashed). The horizontal axis is $\epsilon$ and $\epsilon_{\rm tot}$, respectively.}
   \label{fig:endpoint}
 \end{figure}

There are other potential impacts of the radiative corrections of this work on $0\nu\beta\beta$ experiments. 
Searches for $0\nu\beta\beta$ look for a peak at $\mathcal Q$ on top of the tail of the
$2\nu\beta\beta$ spectrum. Fig.~\ref{fig:endpoint} shows this tail for $^{136}$Xe, together
with the relative change of the normalized spectrum induced by the radiative corrections.
If the photon escapes, the correction grows logarithmically towards the endpoint,
reaching $-3\%$ at $50$ keV and $-7\%$ at $0.1$ keV below $\mathcal Q$. If the
photon is absorbed, the spectrum changes only by about $0.1\%$. 

\begin{table}[t]
\centering
\begin{tabular}{lcc}
\hline\hline
 & $^{76}$Ge & $^{136}$Xe\\
\hline
$\frac{\alpha}{2\pi}\langle f_{0\nu}\rangle$ & $-1.7\%$ & $-1.5\%$\\
real emission                                & $-1.2\%$ & $-1.3\%$\\
$g_V^4-1$                                                     & $+8.1\%$ & $+8.0\%$\\
\hline
Total change                           & $+5.2\%$ & $+5.2\%$\\
\hline\hline
\end{tabular}
\caption{Radiative corrections to the total $0\nu\beta\beta$ rate, inclusive over the
photon, relative to the tree-level rate, for the light-neutrino exchange mechanism.
The individual contributions are evaluated at $\mu=2E_0$ with the corresponding value of
$g_V$ and only their sum is independent of $\mu$.}
\label{tab:0nutotal}
\end{table}

Whether these effects matter depends on the energy resolution, on whether the photon energy is measured, and on the event selection. For high-resolution detectors, the $2\nu\beta\beta$ contribution to the background in the region of interest (ROI) is very small in absolute terms, so that a change of a few percent has no practical impact. For detectors with a worse resolution, in which the $2\nu\beta\beta$ tail is a significant background near $\mathcal Q$, the photon is typically absorbed and the change is small but for a different reason. In either case, the radiatively corrected $2\nu\beta\beta$ background is obtained most reliably with the procedure described at the end of Sec.~\ref{sec:xi31bias_photon}, that is from events generated with the photon as an explicit final-state particle and passed through the detector simulation.

Another effect is the emission of a real photon in
$0\nu\beta\beta$ itself. In that case the electrons share $\mathcal Q-E_\gamma$, and the
event can be lost from the signal sample in two ways. First, if the photon escapes undetected
and $E_\gamma$ is larger than the energy resolution, the event is reconstructed below the
ROI. Second, if the photon is absorbed but produces a separate energy deposit, the event can be
removed by multi-site or veto cuts. 
The resulting signal losses are given in Table~\ref{tab:0nuloss}. They range from a few
tenths of a percent to about $5\%$, are larger for isotopes with larger $\mathcal Q$, and
vanish if the photon is absorbed and the event accepted. These losses then effectively reduce the overall $5.2\%$ increase reported above. The effect is small, but it is one sided and straightforward to include, by generating signal events with the photon spectrum of Sec.~\ref{sec:0nugamma} and passing them through the detector simulation, as for $2\nu\beta\beta$.

\begin{table}[t]
\centering
\begin{tabular}{lccc|ccc}
\hline\hline
 & \multicolumn{3}{c|}{rejection above $E_c$} & \multicolumn{3}{c}{photon not detected}\\
isotope & 30 keV & 100 keV & 300 keV & $\sigma=1$ keV & $3$ keV & $20$ keV\\
\hline
$^{76}$Ge  & 1.5\% & 0.8\% & 0.3\% & 3.3\% & 2.6\% & 1.4\%\\
$^{100}$Mo & 2.3\% & 1.4\% & 0.6\% & 4.7\% & 3.8\% & 2.2\%\\
$^{130}$Te & 2.0\% & 1.1\% & 0.5\% & 4.0\% & 3.2\% & 1.8\%\\
$^{136}$Xe & 1.9\% & 1.1\% & 0.5\% & 3.9\% & 3.1\% & 1.8\%\\
\hline\hline
\end{tabular}
\caption{Fraction of $0\nu\beta\beta$ events lost due to the emission of a real photon.
Left: events with a photon of energy above $E_c$, which are removed by multi-site or
veto cuts. Softer photons are assumed to be absorbed and the loss does not depend on the energy resolution. Right: events reconstructed outside the Region of Interest (ROI) $\mathcal Q\pm2\sigma$ when the
photon escapes undetected, for a Gaussian energy resolution $\sigma$.}
\label{tab:0nuloss}
\end{table}

\section{Combination with other spectral corrections}\label{sec:othercorr}
\begin{figure*}[t]
\centering
\includegraphics[width=\linewidth]{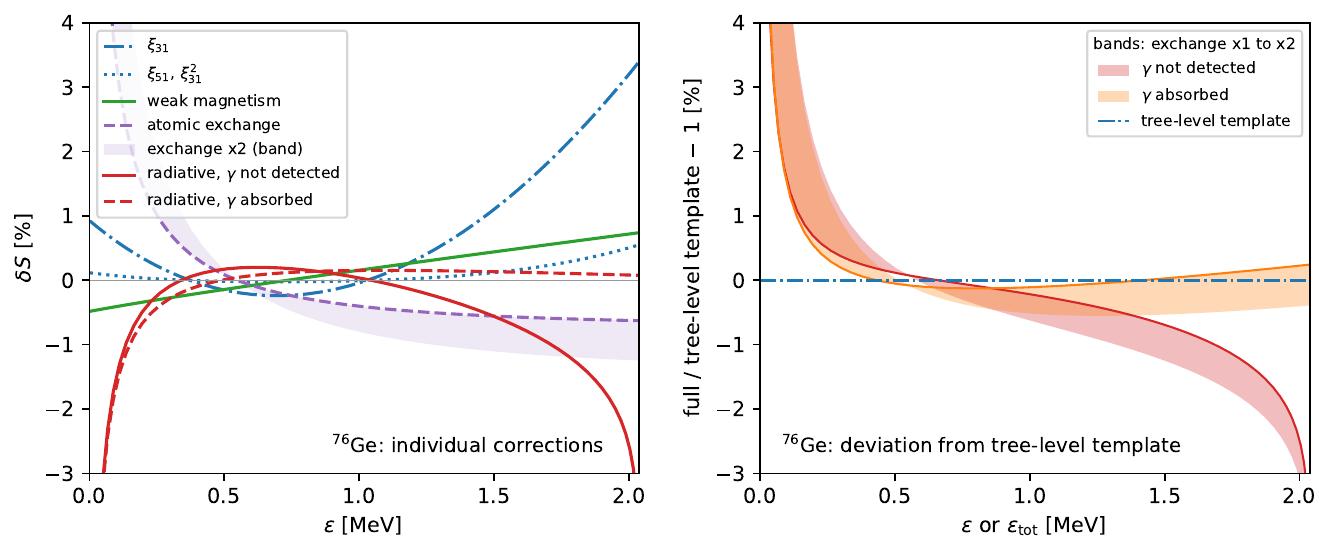}
\caption{Left: distortion of the normalized summed-energy spectrum of $^{76}$Ge with respect
to the leading-order spectrum, induced by the individual corrections discussed in the text:
$\xi_{31}$ and the second-order terms $\xi_{51}$, $\xi_{31}^2$ (NSM values), weak magnetism,
atomic exchange (estimated with the single-$\beta$ parametrization of
Ref.~\cite{Nitescu:2022tlh},  the band shows the effect of doubling $\eta^T$ as suggested by Ref.~\cite{Nitescu:2024tvj}), and the radiative
corrections of this work with the photon not detected and absorbed. Right: deviation of the
spectrum including all corrections from a tree-level template (including $\xi_{31}^2$ and $\xi_{51}$) with the same $\xi_{31}$ and
$\xi_{51}$, for the photon escaping (red) and absorbed (orange). The bands correspond to
the single-$\beta$ estimate of the exchange correction and to twice that value.}
\label{fig:allcorrections}
\end{figure*}

The double-weak radiative corrections derived in this work are not the only significant distortion of the electron spectrum with respect to the tree-level predictions parametrized by $\xi_{31}$ and $\xi_{51}$. We briefly summarize the corrections here that are of size relevant for present and expected measurements.
\begin{itemize}
\item \textbf{Lepton-energy dependence of the nuclear matrix elements.} Expanding the energy
  denominators of the nuclear matrix elements in the lepton energies introduces the ratios
  $\xi_{31}$ and $\xi_{51}$~\cite{Simkovic:2018rdz}, Eq.~\eqref{totddecay2}. The terms
  proportional to $\xi_{51}$ and $\xi_{31}^2$ are the largest second-order effects
  (see Fig.~\ref{fig:higherorder}). These terms are included in the templates of recent experimental analyses~\cite{KamLAND-Zen:2019imh,CUPID-Mo:2023lru,CUORE:2025xue,PandaX:2025yly}.
\item \textbf{Radiative corrections.} The double-weak corrections of this work,
  Eq.~\eqref{totddecay2}. How much this affects the spectra depends on whether the energy of emitted real photons is (partially) measured or not.
  \item \textbf{Weak magnetism.} At next-to-leading order in chiral EFT, the weak-magnetism coupling
  $g_M=4.7$ of the nucleon current adds to the leading lepton function the term~\cite{Morabit:2024sms}
  \begin{equation}
\mathcal{A}_M = \frac{2g_M}{3m_Ng_A}\,\frac{(E_{e_1}+E_{e_2})(2E_{e_1}E_{e_2}-m_e^2)}{E_{e_1}E_{e_2}}\,,
  \end{equation}
  up to an energy-independent constant that only affects the normalization. This contributions grows with the
  electron energies and tilts the spectrum by about $\pm0.5\%$. To our knowledge, it is not included in current experimental templates, but it is straightforward to include. 
  \item \textbf{Pion-exchange two-nucleon currents.} 
  Pions exchanged between nucleons inside the nuclei can undergo beta-decay in flight. Such corrections can be systematically included using chiral EFT in a controlled expansion. The analysis of Ref.~\cite{Morabit:2024sms} shows that they are effectively captured by replacing  $\xi_{31}$ and $\xi_{51}$ by
  $\xi_{31}\Delta_2/\Delta_0$ and $\xi_{51}\Delta_2/\Delta_0$ where $\Delta_2/\Delta_0$ involves pion-exchange nuclear matrix elements. In fits with free $\xi_{31}$ and $\xi_{51}$ this effect is therefore effectively absorbed into
  the extracted values, which should therefore be compared with $\Delta_2/\Delta_0$ times the
  predictions of nuclear-structure calculations. We do not show this effect separately.
  \item \textbf{Atomic exchange.} An electron emitted in  $2\nu\beta\beta$ can be created in a bound state
  of the final atom, while an atomic electron is simultaneously ejected. The analysis of Refs.~\cite{Nitescu:2022tlh,Nitescu:2024tvj} shows that his effect is captured by multiplying the product of Fermi functions by $[1+\eta^T(E_{e_1})][1+\eta^T(E_{e_2})]$, where $\eta^T$
  depends on the overlaps between bound and continuum wave functions.
  A dedicated $2\nu\beta\beta$ calculation is available only for $^{100}$Mo~\cite{Nitescu:2024tvj}.
  Therefore here we use an analytic parametrization obtained for single-$\beta$ decay,
  \begin{equation}
      \eta^T(x)=(a+bx^c)\exp(-dx^e)
      \end{equation}
      with $x=E_e-m_e$ in keV and the parameters listed in
  Table~\ref{tab:exchange}. In
  $2\nu\beta\beta$ the nuclear charge changes by two units, the exchange effect is larger than
  in single-$\beta$ decay \cite{Nitescu:2024tvj} and the detailed calculation for $^{100}$Mo shows that this factor is about a factor $2$ at the relevant electron energies. We indicate this factor 2 by the band in the left and right panels of Fig.~\ref{fig:allcorrections}.
  \end{itemize}

The left panel of Fig.~\ref{fig:allcorrections} shows that the above corrections are of comparable
size, but have different energy dependencies. Weak magnetism tilts the spectrum by about
$\pm0.5\%$. The atomic exchange correction enhances the low-energy part of the spectrum and lowers the spectrum at higher energies. The radiative corrections suppress the spectrum at low energies and cause large distortions near the endpoint if the photon escapes which, as discussed above, counteracts the distortion caused by $\xi_{31}$. Below $\sim0.3$ MeV the radiative and exchange corrections, both of which vary rapidly with energy, dominate.

The right panel shows the deviation of the
corrected spectrum from a tree-level template (including the $\xi_{31}^2$ and $\xi_{51}$ corrections) with the same values of $\xi_{31}$ and $\xi_{51}$ including the radiative, weak magnetism, and atomic exchange corrections in the left panel.
If the photon is not detected, the deviation reaches about $-2\%$ near the end point.
If the photon is fully absorbed, radiative corrections at the end point are small and the weak magnetism and exchange corrections partially cancel (but the amount of cancellation depends on the size of the exchange correction which is unclear for ${}^{76}$Ge). Realistic detectors lie between the two photon limits
(Sec.~\ref{sec:xi31bias_photon}) and we conclude that  a tree-level template is not adequate for any of
these cases. A precise
determination of $\xi_{31}$, and of possible deviations from the Standard Model in the
$2\nu\beta\beta$ spectrum, requires templates that include all of these effects.

Doing so is definitely possible. The expressions derived in this work, including the fully
differential emission of the real photon with energy and direction, can be used to generate a $2\nu\beta\beta$ spectrum generator. Including the weak magnetism and atomic exchange corrections is straightforward once the calculations of Ref.~\cite{Nitescu:2024tvj} are extended to the relevant isotopes. Such a spectrum generator would produce events fully differential in the 
momenta of the two electrons and of the photon, which can then be passed directly through the
detector simulation and event selection of a given experiment. In this way, the absorption,
escape, or separate detection of the photon is treated consistently for each detector, and
the resulting templates can be used to extract $\xi_{31}$, $\xi_{51}$, and possible constraints on BSM
physics. A generator implementing these corrections is in preparation.

  \begin{table}[t]
\centering
\begin{tabular}{lcccccc}
\hline\hline
isotope &  $a$ & $b$ & $c$ & $d$ & $e$\\
\hline
$^{76}$Ge   & 1.7034 & 2.5015 & 0.5692 & 3.1979 & 0.2324\\
$^{100}$Mo & 1.4287 & 1.0662 & 0.6727 & 2.7055 & 0.2482\\
$^{130}$Te  & 1.1568 & 0.6395 & 0.6250 & 2.3902 & 0.2457\\
$^{136}$Xe & 1.1859 & 0.6723 & 0.6192 & 2.4260 & 0.2415\\
\hline\hline
\end{tabular}
\caption{Parameters of the single-$\beta$ parametrization of the atomic exchange correction,
$\eta^T(x)=(a+bx^c)\exp(-dx^e)$ with $x=E_e-m_e$ in keV, taken from Table~III of
Ref.~\cite{Nitescu:2022tlh}.}
\label{tab:exchange}
\end{table}

\section{Summary and conclusions}
\label{sec:conclusions}

We have presented a comprehensive study of the $\mathcal{O}(\alpha)$ radiative corrections to double-$\beta$ decay. We summarize here our main findings:
\begin{enumerate}
\item \textbf{An effective field theory for radiative corrections to double-weak processes.}
  We have developed a heavy-nucleus EFT and used it to compute the ultrasoft radiative corrections to $2\nu\beta\beta$ and $0\nu\beta\beta$ decay. The calculation is organized consistently with the hard corrections contained in $g_V$. We demonstrated  that the corrections enhanced by the nuclear charge cancel, except for the Coulomb interaction with the daughter nucleus that is contained in the Fermi functions.

\item \textbf{Radiative corrections for arbitrary intermediate states.} We have computed the
  virtual and real-emission corrections for intermediate states of arbitrary excitation energy $\omega_n$, including the genuinely double-weak contributions from photons coupling to the intermediate nucleus and from photon exchange between the two electrons. The required one-loop integrals with a heavy-nucleus propagator were obtained in closed form using an auxiliary-mass representation of the heavy propagator~\cite{Saad}. As far as we know, these expressions were not known in closed form and can be used directly in other calculations.

\item \textbf{The large-$\omega_n$ limit and the double-weak Sirlin function.} For excitation energies large compared to the lepton energies, the corrections factorize at leading order into the tree-level rate times a universal double-weak Sirlin function $f_{2\nu}$. We derived this result independently by integrating out the intermediate states before doing the loop integrals. Comparing with the full calculation, we found that multiplying the full tree-level spectrum, including the terms proportional to $\xi_{31}$ and $\xi_{51}$, by $g_V^4 + \frac{\alpha}{2\pi}f_{2\nu}$, Eq.~\eqref{totddecay2}, reproduces the $\mathcal O(\alpha)$ correction to within $2\%$ for all cases considered. Somewhat surprisingly this included the case of $^{100}$Mo, where a single low-lying intermediate state dominates. Eq.~\eqref{totddecay2} is independent of the renormalization scale and requires no nuclear input beyond that of a tree-level analysis. Additional nuclear input, in the form of $\zeta_{31}$ is required at higher orders but is, at least for now, it is irrelevant for phenomenology. 

\item \textbf{Phenomenology.}
  \begin{itemize}
  \item \emph{Double-weak versus single-$\beta$ Sirlin functions.} The genuinely double-weak contributions are large, so that the sum of two single-$\beta$ Sirlin functions, used so far in the literature, overestimates the radiative correction by almost a factor of two and does not reproduce the shape of the spectral distortion \cite{deVries:2026udq}.
  \item \emph{The role of  real photons.} The effect of the radiative corrections on measured spectra depends on whether the energy of the emitted photon is measured. If the photon escapes, the correction to the summed electron spectrum contains logarithms of $m_e$ which become larges at the end point and mimics the distortion induced by $\xi_{31}$. If the photon energy is deposited in the detector these logarithms cancel, as expected from the Kinoshita--Lee--Nauenberg theorem, and the distortion above $\sim 0.5$ MeV is much smaller. We have provided the distribution in the total visible energy, and the rate of the radiative modes $2\nu\beta\beta+\gamma$ and $0\nu\beta\beta+\gamma$, fully differential in the photon energy and direction.
  \item \emph{Angular correlations.} The distribution in the angle between the electrons is independent of the detection of the photon. The radiative corrections affect it significantly and suppress nearly collinear electrons because of the Coulomb repulsion between them.
  \item \emph{$0\nu\beta\beta$ decay.} The radiative corrections increase the inclusive $0\nu\beta\beta$ rate by $5.2\%$, independently of the nuclear matrix element and of $m_{\beta\beta}$. Real-photon emission removes a small fraction of the signal from the region of interest.
  \end{itemize}
\end{enumerate}

 Several corrections remain to be computed or refined. At higher order in $\alpha$, the $\mathcal O(\alpha^2 Z)$ corrections, which have recently been computed for superallowed $\beta$ decays~\cite{Cao:2025zxs,Crosas:2025xyv}, are enhanced by the nuclear charge and may be relevant at the level of precision discussed here. Nuclear-structure-dependent radiative corrections, analogous to $\delta_{\rm NS}$ in superallowed $\beta$ decays~\cite{Cirigliano:2024msg}, are energy independent at leading order, but their energy-dependent part, $\delta^E_{\rm NS}$, can distort the spectrum and should be computed for double-$\beta$ decay. Relatedly, the $\mathcal O(\alpha Z R E_e)$ corrections are currently included through Fermi functions computed with realistic nuclear charge distributions, and a consistent matching onto the operators of the nuclear EFT, Eq.~\eqref{eq:LagNLO}, remains to be established.  The atomic exchange correction \cite{Nitescu:2024tvj} requires dedicated $2\nu\beta\beta$ calculations for all relevant isotopes. 

 The templates used in current analyses of the $2\nu\beta\beta$ spectrum miss physics that affects their interpretation at the level of precision that experiments are now reaching. This is most important for the extraction of $\xi_{31}$ and of possible deviations from the Standard Model in the $2\nu\beta\beta$ spectrum, where the radiative corrections, weak magnetism, and atomic exchange each shift the extracted values by amounts comparable to the uncertainties of the most precise measurements. To a lesser extent, it also affects $0\nu\beta\beta$ searches, through the phase-space factor and the signal efficiency. The templates, however, can be improved with the expressions of this work. If the photon always escapes, the radiative corrections amount to multiplying the tree-level spectrum by $g_V^4 + \frac{\alpha}{2\pi}f_{2\nu}$, Eq.~\eqref{totddecay2}. In general, events should be generated with the photon as an explicit final-state particle and passed through the full detector simulation and event selection. An event generator implementing the radiative corrections together with the other known corrections to the spectral shape is in preparation. With the precision expected from the next generation of experiments, this will allow the $2\nu\beta\beta$ spectrum to be used as a quantitative probe of nuclear structure and of physics beyond the Standard Model.

 \section*{Acknowledgements}

We thank Ryan Plestid for comments and discussion. J.~d.~V. is funded by the Dutch Research Council (NWO) in the form of a VIDI grant and by the European Union (ERC, CRUNS, 101230525). Views and opinions expressed are, however,  those of the author(s) only and do not necessarily reflect those of the European Union or the European Research Council. Neither the European Union nor the granting authority can be held responsible for them.
E.M. and S.S. are supported by the U.S. Department of Energy Office and by the Laboratory Directed Research and Development (LDRD) program of Los Alamos National Laboratory under project numbers 20230047DR,  20250164ER and 20260246ER. Los Alamos National Laboratory is operated by Triad National Security, LLC, for the National Nuclear Security Administration of the U.S. Department of Energy (Contract No. 89233218CNA000001)
We acknowledge support from the DOE Topical Collaboration “Nuclear Theory for New Physics” award No. DE-SC0023663.

\paragraph*{Use of AI tools.} The effective field theory framework and all analytical calculations of the amplitudes and loop integrals were performed by the authors. Based on the analytical expressions, Anthropic’s Claude Opus~5.5 was used to create a separate numerical code to cross check the numerical results for the phase space integrations of the authors' own numerical calculations and those appearing in the accompanying letter \cite{deVries:2026udq}, neither of which involved AI tools. The authors have carefully reviewed and verified all results and text, and take full responsibility for the content of this paper.

\newpage
\appendix

\section{Tree-level phase-space functions}
\label{app:tree}

The tree-level phase-space functions entering Eq.~\eqref{totddecay}, beyond $\mathrm{d}G^{2\nu}_0$ of Eq.~\eqref{dGamma0}, are obtained by integrating Eq.~\eqref{dGamma} over the neutrino energy. The functions $\mathcal A_2$ and $\mathcal A_4$ give
\begin{align}
\frac{\mathrm{d} G^{2\nu}_{2}}{\mathrm{d}\epsilon\, \mathrm{d}\Delta\, \mathrm{d} y_{12} } &= \frac{\mathrm{d} G^{2\nu}_{0}}{\mathrm{d}\epsilon\, \mathrm{d}\Delta\, \mathrm{d} y_{12} } \times  \frac{(\mathcal Q-\epsilon)^2 + 28 \Delta^2}{56 m_e^2}, \nonumber\\
\frac{\mathrm{d} G^{2\nu}_{4}}{\mathrm{d}\epsilon\, \mathrm{d}\Delta\, \mathrm{d} y_{12} } &= \frac{\mathrm{d} G^{2\nu}_{0}}{\mathrm{d}\epsilon\, \mathrm{d}\Delta\, \mathrm{d} y_{12} } \times  \frac{(\mathcal Q-\epsilon)^4 + 72 (\mathcal Q-\epsilon)^2 \Delta^2 + 336 \Delta^4}{2688 m_e^4}\,.
\label{eq:G2G4}
\end{align}
$\mathcal A_{22}$ has a different dependence on $y_{12}$ and is given by
\begin{align}
    \frac{\mathrm{d} G^{2\nu}_{22}}{\mathrm{d}\epsilon\, \mathrm{d}\Delta\, \mathrm{d} y_{12} } &= \frac{1}{\ln 2}\frac{(G_F V_{ud})^4}{8 \pi^7 m_e^2}  \frac{1}{30}  E^2_{e_1}  E^2_{e_2}    \beta_1 \beta_2 (\mathcal{Q}-\epsilon)^5  F(Z_f,E_{e_1})  F(Z_f,E_{e_2}) \nonumber \\
&    \times \left[ \frac{(\mathcal Q-\epsilon)^4 + 24 (\mathcal Q-\epsilon)^2 \Delta^2 + 336 \Delta^4}{5376 m_e^4} - \beta_1 \beta_2 \frac{y_{12}}{3} \frac{(\mathcal Q-\epsilon)^4 - 88 (\mathcal Q-\epsilon)^2 \Delta^2 + 336 \Delta^4}{1792 m_e^4}  \right]
    \,.
\label{eq:G22}
\end{align}

\section{The $\mathcal O(1/\omega_n^2)$ radiative corrections}
\label{app:NLO}

At next-to-leading order in the large-$\omega_n$ expansion, the radiative corrections contain, besides the double-weak Sirlin function multiplying $\mathrm{d}G^{2\nu}_2$, the two functions appearing in Eq.~\eqref{eq:dGalpha2}. The function multiplying $\xi_{31}$ is
\begin{align}\label{largeOmegaNLO}
&    \frac{\mathrm{d} G^{2\nu}_{2\alpha}}{\mathrm{d} \epsilon\, \mathrm{d} \Delta\, \mathrm{d} y_{12}  } =  \frac{1}{\ln 2}\frac{(G_F V_{ud})^4}{8 \pi^7 m_e^2}  \frac{1}{30}  E^2_{e_1}  E^2_{e_2}    \beta_1 \beta_2 (\mathcal{Q}-\epsilon)^5  F(Z_f,E_{e_1})  F(Z_f,E_{e_2}) \frac{1}{24 E_{e_1} E_{e_2} (1-\beta_{12}^2) } \Bigg\{
\nonumber \\  &        \frac{w_{12}}{\beta_{12}}  (3 + 4 \beta_{12}^2 + 5 \beta_{12}^4) \left(  (\epsilon + 2 m_e)^2 \beta_{12}^2  - 12 \Delta^2 - \frac{4 m_e^2 \beta_{12}^2}{1-\beta_{12}^2}  \right) + \frac{1}{3} (7 - 4 \beta_{12}^2 - 15 \beta_{12}^4) (\epsilon + 2 m_e)^2
+ 8 \Delta^2 (7 + 15 \beta_{12}^2) \nonumber \\ & + \frac{4}{3} \frac{m_e^2}{1-\beta_{12}^2} (5 + 7 \beta_{12}^2+ 6 \beta_{12}^4)
+ 8 (\mathcal Q - \epsilon) \Delta (1+\beta_{12}^2) \left(\frac{w_1}{\beta_1} - \frac{w_2}{\beta_2} \right) \nonumber \\
&  + \frac{(\mathcal Q -\epsilon)^2}{49}
 \Bigg[
\frac{w_1}{E_{e_1}^2 \beta_1^3}\Big( -6 (\epsilon^2  + 4 m_e \epsilon) (1+\beta_{12}^2) + 4 \Delta^2 (22-6 \beta_{12}^2) -8 (3-4 \beta_{12}^2) \Delta (\epsilon + 2 m_e) \Big) \nonumber \\
&+\frac{w_2}{E_{e_2}^2 \beta_2^3}\Big( -6 (\epsilon^2  + 4 m_e \epsilon) (1+\beta_{12}^2) + 4 \Delta^2 (22-6 \beta_{12}^2) + 8 (3-4 \beta_{12}^2) \Delta (\epsilon + 2 m_e) \Big) + 12(1+\beta^2_{12})^2 \frac{w_{12}}{\beta_{12}} \nonumber \\
&+ \frac{1}{E_{e_1}^2 E_{e_2}^2 \beta_1^2 \beta_2^2} \left( \frac{9}{4} (1+\beta_{12}^2) \left(\epsilon^2 + 4 m_e \epsilon \right)^2  - 4 \Delta^4 (47 - 9 \beta_{12}^2) - 2 \Delta^2 (\epsilon + 2 m_e)^2 (37 - 47 \beta_{12}^2) +8 m_e^2 \Delta^2 (19 - 9 \beta_{12}^2)\right)
\Bigg] \nonumber \\
& + \frac{(\mathcal Q - \epsilon)^3 \Delta}{7} \Bigg[
\frac{w_1}{E_{e_1}^3 \beta_{1}^3}  \Big( 2\Delta + \beta_{12}^2 (\epsilon + 2 m_e)\Big) + \frac{w_2}{E_{e_2}^3 \beta_{2}^3}  \Big( 2\Delta - \beta_{12}^2 (\epsilon + 2 m_e)\Big) \nonumber \\ & + \frac{E_{e_1}^2 - E_{e_2}^2}{E^2_{e_1} E_{e_2}^2 \beta_1^2 \beta_2^2} \Big( 1 + \beta_{12}^2- \frac{1-\beta_{12}^2}{4m_e^2} \left( (\epsilon+2m_e)^2 - 4 \Delta^2\right)\Big)
\Bigg] \nonumber \\
& + \frac{(\mathcal Q - \epsilon)^4}{294} \Bigg[ \frac{w_1}{E_{e_1}^3 \beta_1^3} (\beta_{12}^2 (\epsilon+2m_e) +2 \Delta)
+ \frac{w_2}{E_{e_2}^3 \beta_2^3} (\beta_{12}^2 (\epsilon+2m_e) - 2 \Delta) \nonumber \\ & +\frac{1-\beta_{12}^2}{m_e^2 E_{e_1}^2 E_{e_2}^2 \beta_1^2 \beta_2^2} \left(
(\epsilon+2m_e)^2 (\Delta^2 + m_e^2)
- 4 (\Delta^2-m_e^2)^2 - \frac{m_e^2}{1-\beta_{12}^2} \left( \epsilon (\epsilon + 4 m_e) + 4 \Delta^2
\right)\right)
\Bigg]
\Bigg\}.
\end{align}
The function multiplying $\zeta_{31}$ originates only from the virtual corrections,
\begin{align}\label{eq:GLalpha}
    \frac{\mathrm{d} G^{2\nu}_{L\alpha}}{\mathrm{d} \epsilon\, \mathrm{d} \Delta\, \mathrm{d} y_{12}  } &= \frac{1}{\ln 2}\frac{(G_F V_{ud})^4}{8 \pi^7 m_e^2}  \frac{1}{30}  E^2_{e_1}  E^2_{e_2}    \beta_1 \beta_2 (\mathcal{Q}-\epsilon)^5  F(Z_f,E_{e_1})  F(Z_f,E_{e_2}) \frac{1}{12 E_{e_1} E_{e_2}  (1-\beta_{12}^2)^2 } \nonumber \\
& \times \Bigg[ (2-\beta_{12}^2- \beta_{12}^4) (\epsilon +2m_e)^2 - 12 \Delta^2 (1+\beta_{12}^2-2\beta_{12}^4) - 2 m_e^2 (1+ 2 \beta_{12}^2 + 3 \beta_{12}^4)\Bigg]\,.
\end{align}

\section{Loop integrals}
\label{app:Expressions}

In this appendix we collect the scalar integrals needed for the virtual corrections of Sec.~\ref{sec:virtual}. The integrals with a heavy-nucleus propagator, for general residual energy $\Delta$, are given in closed form in App.~\ref{app:scalarint}, and their expansions for large $\Delta$ in App.~\ref{app:largeDelta}. Before we begin, we give our conventions. We use the integration measure of Package-X~\cite{Patel:2015tea},
\begin{equation}
    \int \left[\mathrm{d}q\right]
    \equiv -i (4\pi)^2
    \left(\frac{\mu^2 e^{\gamma_E}}{4\pi}\right)^{\varepsilon}
    \int \frac{\mathrm{d}^d q}{(2\pi)^d} \;,
    \label{eq:integrationconvention}
\end{equation}
and the logarithms
\begin{equation}
    L_{\mu} = \ln\left(\frac{\mu^2}{m_e^2}\right),
    \qquad
    L_{\Delta} = \ln\left(\frac{m_e^2}{4\Delta^2}\right)\;.
\end{equation}
The former appears in the Sirlin functions, Eqs.~\eqref{eq:largew} and~\eqref{eq:sirlin1beta}, while the latter appears in the large-$\Delta$ expansions. We write $L_{\Delta}$ for a generic residual energy, and $L_{\omega_n}$ and $L_{\Omega_n^{+K}}$ for $\Delta = \omega_n$ and $\Delta = \Omega_n^{+K}$, respectively. We also use the rapidities $w = \tanh^{-1}\beta$, $w_{12}=\tanh^{-1}\beta_{12}$, and the notation $E_e\beta_{\pm\Delta}=\sqrt{(E_{e}\pm\Delta)^2-m_e^2}$. The quantity $\beta_{\pm\Delta}$ interpolates between $\beta$ for $\Delta \to 0$ and $\beta\tilde{\Delta}$ for $\Delta \to +\infty$, with $\tilde{\Delta}=\Delta /(\beta E_e)$.

\subsection{Scalar integrals for general $\Delta$}
\label{app:scalarint}

The ultraviolet-divergent tadpole and bubble integrals are
\begin{equation}
\begin{split}
    \bar{A}_{0}(v;\Delta)&=0\;,\\
    \bar{B}_{0}(v;\Delta, m_e) &=-2\Delta  \left(\frac{1}{\varepsilon_{UV}}+L_\mu+2-4\frac{\sqrt{\Delta ^2-m_e^2}}{\Delta} \tanh ^{-1}\left(\frac{\Delta +\sqrt{\Delta ^2-m_e^2}-m_e}{\Delta +\sqrt{\Delta ^2-m_e^2}+m_e}\right)\right)\;.
\label{eq:scalarpavebubble}
\end{split}
\end{equation}
For the triangle integrals with the electron propagator $1/(k^2+2k\cdot p_e)$, we find
\begin{equation}
\begin{split}
    \bar{C}_{0}(v,p_e;0,0,m_e)&=\frac{1}{\beta E}\left(w\left(\frac{1}{\varepsilon_{IR}}+L_{\mu}-w\right)-\textrm{Li}_2{\left(\frac{2\beta}{1+\beta }\right)}\right)\;,\\
    \bar{C}_{0}(v,p_e;\Delta,0,m_e)  &= \frac{1}{\beta E}\left(-\frac{\pi^2}{6}-\textrm{Li}_2{\left(-\frac{1-\beta}{2\beta }\right)}-\frac{1}{2}\ln^2{\left(\frac{1+\beta}{2\beta}\right)}+\frac{1}{2}\ln^2{\left(2(1-\beta)\tilde{\Delta}\right)}+\frac{1}{2}\ln^2{\left(\frac{1}{2(1+\beta)\tilde{\Delta}}\right)}\right. \\
    &-\frac{1}{2}\ln^2{\left(2(\beta+\beta_{+\Delta}+\tilde{\Delta})\right)}-\frac{1}{2}\ln^2{\left(\frac{1}{2(\beta-\beta_{+\Delta}+\tilde{\Delta})}\right)}-\textrm{Li}_2{\left(1-\tilde{\Delta}\right)}+\textrm{Li}_2{\left(1-\frac{1}{\tilde{\Delta}}\right)}\\
    &-\frac{1}{2}\ln^2{\left(\tilde{\Delta}\right)}+\frac{1}{2}\ln^2{\left(\frac{\beta-\beta_{+\Delta}-\tilde{\Delta}}{1-\beta_{+\Delta}+\beta \tilde{\Delta}}\right)}+\textrm{Li}_2{\left(\frac{(1-\beta)(1-\tilde{\Delta})}{1-\beta_{+\Delta}+\beta \tilde{\Delta}}\right)}\\
    &\left.-\textrm{Li}_2{\left(-\frac{(1-\beta)(1-\tilde{\Delta})}{\beta+\beta_{+\Delta}+ \tilde{\Delta}}\right)}\right)\;.
    \label{eq:CVpos}
\end{split}
\end{equation}
With the dimensionless $\tilde{\Delta}$ defines through $\tilde{\Delta}\beta E =\Delta$. For $-p_e$, corresponding to the electron propagator $1/(k^2-2k\cdot p_e)$, we find
\begin{equation}
    \begin{split}
    \bar{C}_{0}(v,-p_{e};0,0,m_e)&=-\bar{C}_{0}(v,p_{e};0,0,m_e)+\frac{i \pi}{\beta E}\left(\frac{1}{\varepsilon_{IR}}+\ln{\left(\frac{\mu^2}{4\vec{p}^{\,2}}\right)}+i\pi\right)\;,\\
    \bar{C}_{0}(v,-p_{e};\Delta,0,m_e)  &=-\frac{1}{\beta E}\left(-\frac{\pi^2}{3}+\frac{1}{2}\ln{\left(\frac{1-\beta}{2\beta}\right)}\left(2\pi i+\ln{\left(\frac{1-\beta}{2\beta}\right)}\right)+\frac{1}{2}\ln^2{(\tilde{\Delta})}-\frac{1}{2}\ln^2{(-2(1-\beta)\tilde{\Delta})}\right.\\
    &+\frac{1}{2}\ln^2{\left(2(\beta+\beta_{-\Delta}-\tilde{\Delta})\right)}+\frac{1}{2}\ln^2{\left(\frac{1}{2(\beta-\beta_{-\Delta}-\tilde{\Delta})}\right)}-\frac{1}{2}\ln^2{\left(\frac{-1}{2\tilde{\Delta}}\frac{1}{1+\beta}\right)}\\
    &-\frac{1}{2}\ln^2{\left(1-\frac{(1+\beta)(1-\tilde{\Delta})}{1-\beta \tilde{\Delta}+\beta_{-\Delta}\textrm{sgn}{(1-\tilde{\Delta})}}\right)}+\textrm{Li}_2{\left(\frac{1+\beta}{2\beta}\right)}+\textrm{Li}_2{\left(1-\tilde{\Delta}\right)}-\textrm{Li}_2{\left(1-\frac{1}{\tilde{\Delta}}\right)}\\
    &\left.+\textrm{Li}_2{\left(\frac{(1+\beta)(1-\tilde{\Delta})}{\beta-\tilde{\Delta}+\beta_{-\Delta}\textrm{sgn}{(1-\tilde{\Delta})}}\right)}-\textrm{Li}_2{\left(\frac{(1+\beta)(1-\tilde{\Delta})}{1-\beta \tilde{\Delta}+\beta_{-\Delta}\textrm{sgn}{(1-\tilde{\Delta})}}\right)}\right)\;,
    \end{split}
    \label{eq:CVneg}
\end{equation}
where the sign function accounts for the branch cut at $\Delta =\beta E$. The difference between $\bar{C}_{0}(v,p_e;0,0,m_e)$ and $-\bar{C}_{0}(v,-p_e;0,0,m_e)$ is the origin of the $\pi$ enhancement discussed in Sec.~\ref{sec:Zdrop_triangles}.

Diagram~7 requires two further integrals, which depend on the invariant mass $s$ of the two electrons: the triangle $\bar{C}_{0}\equiv\bar{C}_{0}(v,p_{e_1}+p_{e_2};\Delta-E_{e_1},m_e,m_e)$ 
\begin{equation}
    \begin{split}
    \bar{C}_{0}&=\frac{1}{p_t}\left(\frac{1}{2} \ln ^2\left(\frac{\beta_{12}+\tilde{E}_d}{\beta_{12}+1}\right)-\frac{1}{2} \ln ^2\left(\frac{\beta_{12}-\tilde{E}_d}{\beta_{12}+1}\right)+\frac{1}{2} \ln ^2\left(\frac{\beta_{1-\Delta} (2 \tilde{\Delta}_t+\tilde{E}_d+\tilde{E}_t)-f_1 (\tilde{E}_d \tilde{E}_t+1)}{2 \beta_{1-\Delta} \tilde{\Delta}_t+(\tilde{E}_d+\tilde{E}_t) (\beta_{1-\Delta}+f_1)}\right)\right.\\
    &+\frac{1}{2} \ln ^2\big(\tilde{E}_d(\tilde{E}_t-\beta_{2+\Delta})+\beta_{2+\Delta}(\tilde{E}_t-2\tilde{\Delta}_t)-1\big)-\frac{1}{2} \ln ^2\big(\tilde{E}_d(\tilde{E}_t+\beta_{1-\Delta})+\beta_{1-\Delta}(\tilde{E}_t+2\tilde{\Delta}_t)+1\big)\\
    &+\frac{1}{2} \ln ^2\left(\frac{1}{\tilde{E}_d(\tilde{E}_t+\beta_{2+\Delta})-\beta_{2+\Delta}(\tilde{E}_t-2\tilde{\Delta}_t)-1}\right)-\frac{1}{2} \ln ^2\left(\frac{1}{\tilde{E}_d(\tilde{E}_t-\beta_{1-\Delta})-\beta_{1-\Delta}(\tilde{E}_t+2\tilde{\Delta}_t)+1}\right)\\
    &+\textrm{Li}_2\left(\frac{(\tilde{E}_d-1) (\tilde{E}_t+1)}{\tilde{E}_d(\tilde{E}_t+f_2\beta_{2+\Delta})-f_2\beta_{2+\Delta}(\tilde{E}_t-2\tilde{\Delta}_t)-1}\right)-\textrm{Li}_2\left(\frac{(\tilde{E}_d+1) (\tilde{E}_t+1)}{\tilde{E}_d(\tilde{E}_t+f_1\beta_{1-\Delta})+f_1\beta_{1-\Delta}(\tilde{E}_t+2\tilde{\Delta}_t)+1}\right)\\
    &+\textrm{Li}_2\left(\frac{(\tilde{E}_d+1) (\tilde{E}_t+1) f_1}{(\tilde{E}_d+\tilde{E}_t) (f_1+\beta_{1-\Delta})+2 \beta_{1-\Delta} \tilde{\Delta}_t}\right)-\textrm{Li}_2\left(\frac{(\tilde{E}_d-1) (\tilde{E}_t+1) f_2}{(\tilde{E}_d-\tilde{E}_t) (f_2+\beta_{2+\Delta})+2 \beta_{2+\Delta} \tilde{\Delta}_t}\right)+\textrm{Li}_2\left(\frac{\tilde{E}_d+1}{\tilde{E}_d+\beta_{12}}\right)\\
    & -\left.\textrm{Li}_2\left(\frac{\tilde{E}_d-1}{\tilde{E}_d-\beta_{12}}\right)+\textrm{Li}_2\left(\frac{1-\tilde{E}_d}{\beta_{12}+1}\right)-\textrm{Li}_2\left(\frac{\tilde{E}_d+1}{\beta_{12}+1}\right)-\frac{1}{2} \ln ^2\left(1-\frac{(\tilde{E}_d-1) (\tilde{E}_t+1) f_2}{2 \beta_{2+\Delta} \tilde{\Delta}_t+(\tilde{E}_d-\tilde{E}_t) (\beta_{2+\Delta}+f_2)}\right)\right),
    \label{eq:diag7efttriangle}
    \end{split}
\end{equation}
with $p_t=\sqrt{(E_{e_1}+E_{e_2})^2-s}$, $\tilde{\Delta}_t=\Delta/p_t$, $p_t\tilde{E}_d=E_{e_1}-E_{e_2}-2\Delta$, $p_t \tilde{E}_t=E_{e_1}+E_{e_2}$, $f_1=\textrm{sgn}\left(1-\frac{E_{e_1}+E_{e_2}+p_t}{E_{e_1}(E_{e_2}+p_t)}\Delta\right)$, and $f_{2}=\textrm{sgn}\left(1+\frac{E_{e_1}+E_{e_2}+p_t}{E_{e_1}(E_{e_2}+p_t)}\Delta\right)$. This expression assumes $\Delta>0$, which is sufficient for our purposes, since $\omega_n>0$ and $\Omega^{\pm K,L}_{n}>0$. The infrared-divergent box $\bar{D}_{0}\equiv\bar{D}_{0}(v,-p_{e_1},p_{e_2};\Delta,0,m_e,m_e)$ is
\begin{equation}
    \begin{split}
    \bar{D}_{0}&=\frac{1}{s\beta_{12}\Delta}\left(\left(2w_{12}-i\pi\right)\left(\frac{1}{\varepsilon_{IR}}+L_\mu+2\ln{\left(\frac{m_e(1+\beta_{12})^2}{8\Delta \beta_{12}}\right)}\right)+\textrm{Li}_2{\left(\left(\frac{1-\beta_{12}}{1+\beta_{12}}\right)^2\right)}+\ln^2{\left(\frac{E_{e_2}}{m_e}\beta_2^{++}\right)}\right.\\
   &+\ln^2{\left(-\frac{E_{e_1}}{m_e}\beta_1^{--}\right)}-\frac{\pi^2}{6}+\textrm{Li}_2{\left(1-x \beta_{1}^{+-}\beta_{2}^{-+}\right)}+\textrm{Li}_2{\left(1-x \beta_{1}^{--}\beta_{2}^{-+}\right)}+\textrm{Li}_2{\left(1-x \beta_{1}^{+-}\beta_{2}^{++}\right)}\\
   &+\textrm{Li}_2{\left(1-x \beta_{1}^{--}\beta_{2}^{++}\right)}-2\pi i \Big(\Theta(-\Delta+E_{e_1}+m_e)\left(\ln\left(1-x \beta_{1}^{--} \beta_2^{-+}\right)+\ln\left(1-x \beta_{1}^{--} \beta_2^{++}\right)\right)\\
   &\left.-\Theta(-\Delta-E_{e_2}-m_e)\ln\left(1-x \beta_{1}^{+-} \beta_2^{-+}\right)\Big)\right),
    \label{eq:diag7eftbox}
    \end{split}
\end{equation}
with $m_e^2 x=E_{e_1}E_{e_2}(1-\beta_{12})/(1+\beta_{12})$, $\Theta$ the Heaviside function, $\beta_{i}^{\pm \mp}=1\pm \beta_{i \Delta} \mp\beta_{i} \tilde{\Delta}_i$, $\beta_{1\Delta}=\beta_{1-\Delta}$, and $\beta_{2\Delta}=\beta_{2+\Delta}$.

\subsection{Large-$\Delta$ expansions}
\label{app:largeDelta}

The integrals of App.~\ref{app:scalarint} are expensive to evaluate, and the resulting expressions for the diagrams are too long to be written in compact form. We therefore expand them in $1/\Delta$, using the method of regions. This expansion converges rapidly for typical values of $\Omega_n^{\pm K, L}$, except for the lowest ones, for which higher orders or the full expressions should be used. We only give the real parts, which contribute to the radiative corrections to the rate
\begin{equation}
\begin{split}
    \Re{\left(\bar{B}_0(v;\Delta, m)\right)} &\approx \Delta \left(-4-2L_{\Delta}+\frac{m^2}{\Delta^2}(L_{\Delta}-1)+\frac{m^4}{8\Delta^4}(2L_{\Delta}+1)\right),\\
    \Re{\left(\bar{C}_0(v,\pm p;\Delta, 0,m)\right)} &\approx \frac{1}{\Delta}\left(-2+L_{\Delta}\mp\frac{E}{2\Delta}(L_{\Delta}+1)+\frac{E^2}{6\Delta^2}\left(L_{\Delta}(3-\beta^2)+\frac{1}{3}(15-\beta^2)\right)\right).
    \label{eq:scalarmasterlargedelta1}
\end{split}
\end{equation}
For the integrals of diagram~7, Eqs.~\eqref{eq:diag7efttriangle} and~\eqref{eq:diag7eftbox}, we find
\begin{equation}
\begin{split}
    \Re{\left(\bar{C}_0\right)}&=\frac{1}{\Delta}\left(L_{\Delta}-2+2\beta_{12}w_{12}+\frac{E_{e_1}-E_{e_2}}{2\Delta}(L_{\Delta}+2w_{12}\beta_{12})+\frac{3(E_{e_1}-E_{e_2})^2+6m_e^2+p_t^2}{12\Delta^2}\left(L_{\Delta}-\frac{2}{3}\right.\right.\\
    &+\left.\left.2\beta_{12}w_{12}+\frac{6E_{e_1}^2-8E_{e_1}E_{e_2}+3E_{e_2}^2+2\frac{m_e^2}{s}(-p_t^2(1-2w_{12}\beta_{12})+(E_{e_1}+E_{e_2})^2(5-6 w_{12}\beta_{12}))}{3(E_{e_1}-E_{e_2})^2+6m_e^2+p_t^2}\right)\right),\\
    \Re{\left(\bar{D}_0\right)}&=\frac{1}{s\beta_{12}\Delta}\left(\frac{2\pi^2}{3}-2w_{12}\left(L_{\Delta}+w_{12}-\frac{1}{w_{12}}\textrm{Li}_2\left(\frac{1-\beta_{12}}{1+\beta_{12}}\right)+2\ln{\left(\frac{2\beta_{12}}{1+\beta_{12}}\right)}-\frac{1}{\varepsilon_{IR}}\right)-2w_{12}\frac{E_{e_1}-E_{e_2}}{\Delta}\right.\\
    &\left.+\frac{s\beta_{12}}{4\Delta^2}\left(L_{\Delta}-1+\frac{2}{s}(E_{e_1}+E_{e_2})^2+\frac{2w_{12}}{s^2 \beta_{12}}\left(4(E_{e_1}+E_{e_2})^2m_e^2-2(E_{e_1}^2+E_{e_2}^2+2m_e^2)s+s^2\right)\right)\right).
    \label{eq:scalarmasterlargedelta2}
\end{split}
\end{equation}

\subsection{Relativistic integrals}
\label{app:relint}

The standard relativistic integrals that appear in the diagrams, written in the same variables, are
\begin{equation}
\begin{split}
    \int [\mathrm{d}k] \frac{1}{k^2}\frac{1}{k^2-2k\cdot p_{e_1}}\frac{1}{k^2+2k\cdot p_{e_2}}&=\frac{1}{s\beta_{12}}\left((i\pi-2w_{12})\left(\frac{1}{\varepsilon_{IR}}+L_{\mu}-2w_{12}+i\pi\right)+2\textrm{Li}_2\left({\frac{2\beta_{12}}{1+\beta_{12}}}\right)\right.\\
    &+\left.i\pi\left(w_{12}-\ln{\left(\frac{4\beta_{12}^2}{1+\beta_{12}^2}\right)}\right)\right)\;,\\
    \int [\mathrm{d}k] \frac{1}{k^2-2k\cdot p_{e_1}}\frac{1}{k^2+2k\cdot p_{e_2}}&=2+\frac{1}{\varepsilon_{UV}}+L_{\mu}+\beta_{12}(i\pi-2w_{12})\;,\\
    \int [\mathrm{d}k] \frac{1}{k^2}\frac{1}{k^2\pm2k\cdot p_{e}}&=2+\frac{1}{\varepsilon_{UV}}+L_{\mu}\;.
    \label{eq:noneeftpave}
\end{split}
\end{equation}

\section{Virtual diagrams and counterterms}
\label{app:virtual}

The result for diagrams~8--11 is given in Eq.~\eqref{eq:M8to11_final}, and the leading term of the large-$\omega_n$ expansion of diagrams~1--6 in Sec.~\ref{sec:virtual}. For the first lepton ordering, the full result for diagrams~1--3 is
\begin{equation}
\begin{split}
    i\mathcal{M}_{1-3}&=\frac{\alpha}{4\pi}\left(i\mathcal{M}_{0}^{+K}\left(\frac{1}{\varepsilon_{UV}}+2+L_{\mu}+\frac{4w_1}{\beta_1}\left(L_{\mu}+\frac{1}{\varepsilon_{IR}}-w_1-\frac{1}{w_1}\textrm{Li}_2{\left(\frac{2\beta_1}{1+\beta_1}\right)}\right)\right.\right.\\
    &+\left.\left(2E_{e_1}-\Omega_n^{+K}\right)\bar{C}_{0}(v,-p_{e_1};\Omega_n^{+K},0,m_e)\right)+\frac{\Omega_n^{+K}}{E_{e_1}\beta_1^2}\left(i\mathcal{M}_0^{+K}-\frac{m_e}{E_{e_1}}i\mathcal{M}_{0}^{+K}[\gamma^0,1]\right)\\
    & \times\left(\frac{4E_{e_1}\beta_1 w_1}{\Omega_n^{+K}}\left(1+\frac{\beta_{1-\Omega_n^{+K}}}{\beta_1}\frac{1}{w_1}\tanh^{-1}{\left(\frac{\frac{\Omega_n^{+K}-m_e}{E_{e_1}}+\beta_{1-\Omega_n^{+K}}-1}{\frac{\Omega_n^{+K}+m_e}{E_{e_1}}+\beta_{1-\Omega_n^{+K}}-1}\right)}\right)\left.+L_{\Omega_n^{+K}}-E_{e_1}\bar{C}_{0}(v,-p_{e_1};\Omega_n^{+K},0,m_e)\right)\right)+\ldots\;,
    \label{eq:diag13full}
\end{split}
\end{equation}
where the integrals $\bar{C}_{0}(v,\pm p_{e};\Omega_n^{\pm K,L},0,m_e)$ are given in Eqs.~\eqref{eq:CVpos} and~\eqref{eq:CVneg}. For diagrams~4--6 we find
\begin{equation}
\begin{split}
    i\mathcal{M}_{4-6}&=\frac{\alpha}{4\pi}\left(i\mathcal{M}_{0}^{+K}\left(\frac{1}{\varepsilon_{UV}}+2+L_{\mu}+\frac{4w_2}{\beta_2}\left(L_{\mu}+\frac{1}{\varepsilon_{IR}}-w_2-\frac{1}{w_2}\textrm{Li}_2{\left(\frac{2\beta_2}{1+\beta_2}\right)}\right)\right.\right.\\
    &-\left.\left(2E_{e_2}+\Omega_n^{+K}\right)\bar{C}_{0}(v,p_{e_2};\Omega_n^{+K},0,m_e)\right)+\frac{\Omega_n^{+K}}{E_{e_2}\beta_2^2}\left(i\mathcal{M}_0^{+K}-\frac{m_e}{E_{e_2}}i\mathcal{M}_{0}^{+K}[1,\gamma^0]\right)\\
    & \times\left.\left(\frac{4E_{e_2}\beta_2 w_2}{\Omega_n^{+K}}\left(1-\frac{\beta_{2+\Omega_n^{+K}}}{\beta_2}\frac{1}{w_2}\tanh^{-1}{\left(\frac{\frac{\Omega_n^{+K}-m_e}{E_{e_2}}+\beta_{2+\Omega_n^{+K}}+1}{\frac{\Omega_n^{+K}+m_e}{E_{e_2}}+\beta_{2+\Omega_n^{+K}}+1}\right)}\right)+L_{\Omega_n^{+K}}-E_{e_2}\bar{C}_{0}(v,p_{e_2};\Omega_n^{+K},0,m_e)\right)\right)+\ldots\;.
    \label{eq:diag46full}
\end{split}
\end{equation}
Using the large-$\omega_n$ expansion of the integrals, diagram~7 becomes
\begin{equation}
\begin{split}
    i\mathcal{M}_7&\approx \frac{\alpha}{4\pi}\frac{-1}{s\beta_{12}}\left(i\mathcal{M}_0^{+K}2(s-2m_e^2)\left(\frac{\pi^2}{6}+\left(2w_{12}-i\pi\right)\left(\frac{1}{\varepsilon_{IR}}+L_{\mu}+\frac{i\pi}{2}-w_{12}-2\ln{\left(\frac{2\beta_{12}}{1+\beta_{12}}\right)}\right)\right.\right.\\
    &+\left.\left.2\textrm{Li}_2{\left(\frac{1-\beta_{12}}{1+\beta_{12}}\right)} \right)+2\left(i\mathcal{M}_0^{+K}[\slashed{p}_{e_2}\gamma_\mu,1]-i\mathcal{M}_0^{+K}[1,\slashed{p}_{e_1}\gamma_\mu]\right)(2w_{12}-i\pi)\left(p_{e_1}^\mu-p_{e_2}^\mu\right)\right.\\
    &+\left.i\mathcal{M}_0^{+K}[\gamma^\mu \gamma^\nu,\gamma_\mu \gamma_\nu]\frac{s\beta_{12}}{4}(2-L_{\Delta}-\beta_{12}(2w_{12}-i\pi))\right.+\frac{m_e^2}{s \beta_{12}}(1-\beta_{12}(2w_{12}-i \pi))\Big(i\mathcal{M}_0^{+K}[\gamma^{\mu}\slashed{p}_{e_2},\gamma_{\mu}\slashed{p}_{e_1}]\\
    &- 2m_e\left(i\mathcal{M}_0^{+K}[\slashed{p}_{e_2},1]+i\mathcal{M}_0^{+K}[1,\slashed{p}_{e_1}]\right)+\frac{s-2m_e^2}{2m_e}\left(i\mathcal{M}_0^{+K}[\gamma^{\mu}\slashed{p}_{e_2},\gamma_{\mu}]+i\mathcal{M}_0^{+K}[\gamma^{\mu},\gamma_{\mu}\slashed{p}_{e_1}]\right)\\
    &+m_e^2\,  i\mathcal{M}_0^{+K}[\gamma^{\mu},\gamma_{\mu}]\Big)+\ldots\;.
\label{eq:M7largeomegaexpansion}
\end{split}
\end{equation}

The counterterms are $Z$ independent once $\varepsilon_{IR}=\varepsilon_{UV}$ is taken at the end. We split the corrections into an ultraviolet part ($Z$) and an infrared and finite part ($R$):
\begin{equation}
    \begin{split}
        \delta Z_e&=\frac{\alpha}{2\pi}\times  \left(-\frac{1}{2\varepsilon_{UV}}\right),\quad
        \delta Z_{A_i}=\frac{\alpha}{2\pi} \times Z^2\frac{1}{\varepsilon_{UV}},\quad
        \delta Z_{B_n}=\frac{\alpha}{2\pi} \times (Z+1)^2\frac{1}{\varepsilon_{UV}},\quad
        \delta Z_{A_f}=\frac{\alpha}{2\pi} \times (Z+2)^2\frac{1}{\varepsilon_{UV}},\\
        \delta R_e&=\frac{\alpha}{2\pi} \left(-\frac{1}{\varepsilon_{IR}}-2-\frac{3}{2}L_{\mu}\right),\quad
        \delta R_{A_i}=\frac{\alpha}{2\pi} \times \left(-Z^2\frac{1}{\varepsilon_{IR}}\right),\quad
        \delta R_{B_n}=\frac{\alpha}{2\pi} \times \left(-(Z+1)^2\frac{1}{\varepsilon_{IR}}\right),\\
        \delta R_{A_f}&=\frac{\alpha}{2\pi} \times \left(-(Z+2)^2\frac{1}{\varepsilon_{IR}}\right),\quad
        \delta g_{A}^{in}=\frac{\alpha}{2\pi} \left(Z(Z+1)-\frac{1}{2}\right)\frac{1}{\varepsilon_{UV}},\quad
        \delta g_{A}^{fn}=\frac{\alpha}{2\pi} \left((Z+2)(Z+1)-\frac{1}{2}\right)\frac{1}{\varepsilon_{UV}}\;.
    \label{eq:CTterms}
    \end{split}
\end{equation}
Unlike that of the electron, the wave-function renormalization of the heavy nuclei contains no logarithm of the renormalization scale, because the integrals with a single linear propagator are scaleless. The counterterm contribution to the amplitude is
\begin{equation}
    i\mathcal{M}_{CT} = i\mathcal{M}_0\, \frac{\alpha}{4\pi} \left(2Z (Z+2)\left(\frac{1}{\varepsilon_{UV}}-\frac{1}{\varepsilon_{IR}}\right)-\frac{6}{\varepsilon_{IR}}-3L_\mu-4\right).
    \label{eq:CTmatrixelement}
\end{equation}

\section{Real emission}
\label{app:real}

The real-emission contributions of Sec.~\ref{sec:real} are evaluated with the same tensor-reduction methods, and require the angular integrals over the photon direction. These are the same as in the derivation of the single-$\beta$ Sirlin function, except for $I_{1,1}$, which depends on $s$
\begin{equation}
\begin{split}
    I_{0}&=\int\frac{\mathrm{d}\Omega}{4\pi}=1+2\varepsilon,
    \\I_{1}&=\int\frac{\mathrm{d}\Omega}{4\pi} \frac{1}{1-\beta \hat{\beta}\cdot \hat{k}}= \frac{1}{\beta}\left(w+\varepsilon \left(w^2+\textrm{Li}_{2}\left(\frac{2 \beta}{1+\beta}\right)\right)\right),\\
    I_{2}&=\int\frac{\mathrm{d}\Omega}{4\pi} \frac{1}{(1-\beta \hat{\beta}\cdot \hat{k})^2}=\frac{1}{1-\beta^2}\left(1+\varepsilon\frac{2w}{\beta}\right),\; \\
    I_{1,1}&=\int\frac{\mathrm{d}\Omega}{4\pi} \frac{1}{1-\beta_1\hat{\beta}_1\cdot \hat{k}}\frac{1}{1-\beta_2 \hat{\beta}_2\cdot \hat{k}}=\frac{2E_{e_1}E_{e_2}}{s(1+\beta_{12}^2)}\left(\frac{2w_{12}}{\beta_{12}}\left(1+\beta_{12}^2\right)+\varepsilon I_{\varepsilon}\right)\;.
    \label{eq:photonanglemasters}
\end{split}
\end{equation}
The function $I_\varepsilon$ can be reconstructed from Ref.~\cite{Alioli:2010xd}. With the auxiliary quantities
\begin{align}
a &= \beta_1^2 + \beta_2^2 - 2 \vec\beta_1 \cdot \vec\beta_2\,, &\;    b &= \frac{\beta_1^2 \beta_2^2 - (\vec\beta_1 \cdot \vec \beta_2)^2}{a}\,, &\;  c &= \sqrt{\frac{b}{4a}}\,, &\; x_1 &= \frac{\beta_1^2 - \vec\beta_1 \cdot \vec\beta_2}{a}\,, \nonumber\\
x_2 &= \frac{\beta_2^2 - \vec\beta_1 \cdot \vec \beta_2}{a}\,, &\; z_1 &= \frac{\sqrt{x_1^2 + 4 c^2} - x_1}{2c}\,, &\; z_2 &= \frac{\sqrt{x_2^2 + 4 c^2} + x_2}{2c}\,, &\;    z_\pm &= \frac{1\pm \sqrt{1-b}}{\sqrt{b}}\,,
\end{align}
it reads
\begin{align}\label{eq:Ieps}
I_\varepsilon = \left(K(z_2) - K(z_1) \right) \frac{1 - \vec \beta_1 \cdot \vec \beta_2}{\sqrt{a (1-b)}}\,,
\end{align}
with
\begin{align}
K(z) &=  -\frac{1}{2} \ln^2 \frac{(z - z_-)(z_+ - z)}{(z_+ + z)(z_- + z)}  - 2\, {\rm Li}_2\left(\frac{2 z_- (z_+ - z)}{(z_+ - z_-) (z_- + z)}\right) - 2\, {\rm Li}_2\left(-\frac{2 z_+ (z_- + z)}{(z_+ - z_-) (z_+ - z)}\right).
\end{align}

The remaining real-emission contribution follows the standard procedure, simplified by the fact that factors of $v\cdot k$ introduce no additional angular dependence, since the nucleus does not recoil. The terms beyond the soft limit can be identified through their factors of $(\mathcal{Q}-\epsilon)^{n}$. At leading order in the large-$\omega_n$ expansion we find
\begin{align}
  &  |\mathcal R|^2= |\mathcal R|^2_{\text{soft}} + |\mathcal M_0|^2 \frac{\alpha}{2\pi} \times 4 \Bigg\{ \left[ \frac{3}{2}  - \frac{w_1}{\beta_1}  - \frac{w_2}{\beta_2}
 +   \frac{1+\beta_{12}^2}{2} \frac{w_{12}}{\beta_{12}} \right] \frac{137}{30}+\frac{(\mathcal{Q}-\epsilon)m_e}{6(s-2m_e^2)}
\nonumber \\
& \times\left[\frac{E_{e_1}}{m_e}\left(-3\left(1-y_{12}\frac{\beta_1}{\beta_2}\right)+\frac{w_1}{\beta_1}(1-\beta_1^2)+\frac{w_2}{\beta_2}\left(2+y_{12}\frac{\beta_{1}}{\beta_2}(\beta_2-3)\right)\right)\right.\nonumber\\
 &\left.+\frac{E_{e_2}}{m_e}\left(-3\left(1-y_{12}\frac{\beta_2}{\beta_1}\right)+\frac{w_2}{\beta_2}(1-\beta_2^2)+\frac{w_1}{\beta_1}\left(2+y_{12}\frac{\beta_{2}}{\beta_1}(\beta_1-3)\right)\right)\right]+\frac{(\mathcal{Q}-\epsilon)^2}{42(s-2m_e^2)^2} \nonumber\\
 &\times \left[-2+\frac{w_1}{\beta_1}\frac{E_{e_2}}{E_{e_1}}+\frac{w_2}{\beta_2}\frac{E_{e_1}}{E_{e_2}}-y_{12}\left(\frac{E_{e_2}\beta_{2}}{E_{e_1}\beta_{1}}\left(\frac{w_1}{\beta_1}-1\right)+\frac{E_{e_1}\beta_{1}}{E_{e_2}\beta_{2}}\left(\frac{w_2}{\beta_2}-1\right)\right)\right]
\Bigg\}.
\label{eq:r2largeomega}
\end{align}

\section{The single-$\beta$ Sirlin function and the Fermi function
}
\label{app:sirlin}

Many of the structures in the double-weak Sirlin functions for $2\nu\beta\beta$ and $0\nu\beta\beta$ decay also appear in the single-$\beta$ Sirlin function,
\begin{equation}
      g(E_e,\mathcal Q) =
      \frac{3}{2}L_\mu+8
      -\frac{4}{3}\frac{\bar{E}}{E_e}
      +2\ln{\left(\frac{m_e^2}{4\bar{E}^2}\right)}
      -\frac{4}{\beta}
      \left(
          \textrm{Li}_2{\left(\frac{2\beta}{1+\beta}\right)}
          +w^2
          -\frac{w}{4}
          \left(
              \frac{\bar{E}^2}{6E_e^2}
              +\frac{4\bar{E}}{3E_e}
              +2\beta^2-4
              -2\ln{\left(\frac{m_e^2}{4\bar{E}^2}\right)}
          \right)
      \right)\;,
      \label{eq:sirlin1beta}
\end{equation}
with $\bar{E}=  \mathcal Q_\beta+m_e-E_e$ the maximal photon energy, whose analogue in our case is $\mathcal{Q}-\epsilon$. This expression differs from the traditional one~\cite{Sirlin:1967zza} by a constant and by the replacement $\ln m_p\to \ln\mu$, since the latter was not derived in dimensional regularization.

The single-$\beta$ and double-weak Sirlin functions share the terms involving $w$, ${\rm Li}_2(2\beta/(1+\beta))$, and $w^2$, which arise from master integrals common to both processes, such as $\bar C_0(v,\pm p_e;0,0,m_e)$ and $I_1$. However, their coefficients differ: for example, the terms $w_1^2/\beta_1$ and ${\rm Li}_2(2\beta_1/(1+\beta_1))/\beta_1$ enter $f_{2\nu}$ with coefficient $-8$, compared with $-4$ in the single-$\beta$ function. In addition, $f_{2\nu}$ contains terms that depend on the invariant mass $s$ of the two electrons, which have no analogue in single-$\beta$ decay. The double-weak Sirlin function is therefore not simply the sum of two single-$\beta$ functions with a correction for the repulsion between the electrons.

Ref.~\cite{Nitescu:2024tvj} proposed to use the Sirlin function to account for radiative corrections. This would be equivalent to replacing $\dd G_\alpha$ with
\begin{align}
\label{eq:sirlin0}
\frac{\mathrm{d} \hat{G}^{2\nu}_\alpha}{\mathrm{d}\epsilon\, \mathrm{d}\Delta} =   \frac{\mathrm{d} {G}^{2\nu}_0}{\mathrm{d}\epsilon\, \mathrm{d}\Delta} \left( g(E_{e_1}, \mathcal Q + m_e) + g(E_{e_2}, \mathcal Q + 2 m_e - E_{e_1}) \right)\,.
\end{align}

For the Fermi function, we use the traditional expression \cite{Hayen:2017pwg}
\begin{eqnarray}\label{eq:FFtrad}
F(Z,E_e) = 4 (2 E_e \beta R )^{2(\gamma_0 - 1)} \frac{|\Gamma(\gamma_0 + i y)|^2}{|\Gamma(2\gamma_0+1)|^2} e^{\pi y}, \quad \gamma_0 = \sqrt{1- (\alpha Z)^2}, \quad y = \frac{\alpha Z }{\beta}.
\label{eq:Fermi_function}
\end{eqnarray}
At $\mathcal O(\alpha^2)$, this expression depends on the nuclear radius $R$, which, for a uniform charge distribution, is related to the charge radius of the daughter nucleus by
\begin{equation}
    R = \sqrt{\frac{5}{3}} R_{\text{ch}}. 
\end{equation}
In the numerical evaluations we use \cite{Angeli:2013epw}
\begin{equation}
    R_{\text{ch}}(^{76}\text{Se}) = 4.1395 \, \text{fm}, \quad 
R_{\text{ch}}(^{100}\text{Ru}) = 4.4531 \, \text{fm}, \quad
        R_{\text{ch}}(^{130}\text{Xe}) = 4.7818\, \text{fm}, \quad 
        R_{\text{ch}}(^{136}\text{Ba}) = 4.8334 \, \text{fm}.
\end{equation}
For extensions of the EFT framework to higher order in $\alpha$, it will be necessary to replace the traditional Fermi function with the one calculated in dimensional regularization \cite{Hill:2023acw,Hill:2023bfh}. 
This is given by
\begin{align}\label{eq:FF}
\bar F(Z,E_e,\mu)  &= 4\left(\frac{2 E_e \beta}{\mu}e^{-\gamma_{E}}\right)^{2(\gamma_0-1)} \frac{2\gamma_0}{1+\gamma_0}\frac{|\Gamma(\gamma_0+iy)|^2}{\Gamma(2\gamma_0+1)^2} e^{\pi y},
\end{align}    
with $\mu$ the $\overline{\text{MS}}$ renormalization scale. The renormalization scale dependence in Eq.~\eqref{eq:FF} is compensated by nuclear matrix elements that depend logarithmically on the nucleon distances \cite{Cirigliano:2024msg}
and provide an operator definition
of the radius to be used in Eq.~\eqref{eq:FFtrad}.
Eq.~\eqref{eq:FF} agrees with the traditional Fermi function for the choice $\mu^{-1} = R e^{\gamma_E}$, apart from a factor of $(1+\gamma_0)/2$, with in the $\beta$ decay literature is typically included in the finite size correction $L_0$, and the residual factor
\begin{equation}
    \frac{4 \gamma_0}{(1+\gamma_0)^2} = 1 - \frac{\alpha^4 Z^4}{16} + \mathcal O((\alpha Z)^6).
\end{equation}

\bibliography{biblio}

\end{document}